\documentclass[twocolumn]{aastex631}
\usepackage{graphicx}
\newcounter{subfigure}

\newcommand{\panelimage}[4]{%
  \begin{minipage}[c]{#2}\centering
  \includegraphics[width=\linewidth,height=#3,keepaspectratio]{#4}\par
  \vspace{2pt}{\footnotesize\textbf{(#1)}}%
  \end{minipage}%
}\usepackage{CJK}

\usepackage{color}
\usepackage{natbib}

\begin{document}
\begin{CJK*}{UTF8}{gbsn}

\title{On the Orbital Evolution of Giant Planet Pair Embedded in a Gaseous Disk. III. simultaneous gas accretion and the final mass ratio}
	
\correspondingauthor{Hui Zhang}
\affiliation{Shanghai Astronomical Observatory, Chinese Academy of Sciences, Shanghai 200030, China}
	\email{zhangh@shao.ac.cn}

\author[0000-0002-1007-1991]{Bixuan Zhao (赵碧轩)}
\affil{Shanghai Astronomical Observatory, Chinese Academy of Sciences, Shanghai 200030, China}
    
\author[0000-0003-3491-6394]{Hui Zhang (张辉)}
\affil{Shanghai Astronomical Observatory, Chinese Academy of Sciences, Shanghai 200030, China}

\author[0000-0003-3728-8231]{Yinhao Wu (吴寅昊)}
\thanks{EACOA Fellow}
\affiliation{Shanghai Astronomical Observatory, Chinese Academy of Sciences, Shanghai 200030, China}
\affiliation{National Astronomical Observatory of Japan, 2-21-1 Osawa, Mitaka, Tokyo 181-8588, Japan}
    
\author[0000-0002-7329-9344]{Ya-Ping Li (李亚平)}
\affil{Shanghai Astronomical Observatory, Chinese Academy of Sciences, Shanghai 200030, China}

\shorttitle{Gas accretion and orbital evolution of giant planet pair}
	
\shortauthors{Zhao et al.}
	
\begin{abstract}

Recent exoplanet surveys reveal a systematic trend in the mass ratios of giant planet pairs ($> 30 M_\oplus$). Although the $M_{outer}/M_{inner}$ is broadly centered near unity, it exhibits a dependence on orbital separation. Compact pairs tend to have comparable masses, whereas widely separated pairs show outer-heavy configurations. To investigate the origin of this trend, we perform two-dimensional hydrodynamic simulations that couple gas accretion and orbital migration of two giant planet embryos with initial masses of $15-30 M_\oplus$. We explore the effects of disk viscosity, surface density slope, and initial mass ratio on the coupled evolution of planetary growth and migration. We find that convergent migration drives the planets to open a common gap, restructuring the gas flow. This configuration induces asymmetric accretion, in which the inner planet becomes starved as gap clearing suppresses its gas supply, while the outer planet continues to accrete from the high-density region at the outer gap edge. Consequently, the system naturally evolves toward mass-ratio convergence, and in many cases produces an inversion of the initial core-mass hierarchy. We further show that migration outcomes are highly sensitive to initial conditions. Systems with initially more massive inner cores tend to diverge, whereas steeper surface density profiles and more massive outer cores favor convergent migration and subsequent resonance capture. Overall, our results identify common-gap-driven asymmetric accretion as a unified mechanism linking orbital architecture, resonant capture, and the observed mass-ratio demographics of giant planet pairs.

		
		
\end{abstract}
	
\section{Introduction}
	
\label{sec:intro}
Recent exoplanet surveys have found many multiple-planet systems and revealed that planetary systems hosting multiple gas giants, like Jupiter and Saturn in the solar system, are not rare \citep{Zhu-Dong-2021}. Previous hydrodynamic simulations have shown that two giant planets usually undergo the convergent migration if the outer planet is litter than the inner one (i.e., $M_{outer}/M_{inner}<1$) in a disk with a negative surface density profile (i.e., $\Sigma \propto r^{-1}$ or $r^{-3/2}$) \citep{Masset-Snellgrove-2001, Morbidelli-Crida-2007, Zhang-Zhou-2010a, Baruteau-Papaloizou-2013, Bitsch-etal.2015, Pichierri-etal.2018}. However, many giant planet pairs close to the MMRs have a reversed mass ratio (i.e., $M_{outer}/M_{inner}>1$). For example, the two giant planets in Kepler-553, Kepler-58, HD 217107 and others. With an initial reversed mass ratio, the two planets systems usually undergo divergent migration and the odds of forming an MMR are low \citep{Zhang-Zhou-2010b}. 
    
Several solutions have been proposed to resolve this discrepancy. One possible solution is outward migration of the inner or both giant planets, which is triggered by various disk setups \citep{Kley2000, Peplinski-etal.2008, Crida-etal.2009, Bitsch-Kley-2011, Lin-Papaloizou-2012, Chen2020,Kimmig-etal.2020,Laune2024,Wu-Chen-2025,Pan2026, Liu2026}. Another explanation is to consider the reverse of the mass ratio after the two giant planets have locked into MMRs. The gaps opened by the two giant planets will merge and form a common gap enclosing both planets when their orbital configuration becomes compact. At this stage, gas accretion onto each planet no longer occurs in isolation but is strongly influenced by the dynamical and hydrodynamical effects of the companion planet. Which planet ultimately accretes more gas and thus whether the initial mass hierarchy is preserved or inverted becomes a key question for explaining the observed diversity of giant planet pairs.
	


Previous studies have explored several aspects of giant planet migration and accretion. Early hydrodynamical simulations demonstrated that giant planet pairs can undergo joint migration when embedded in a common gap, and in some cases, when an outer planet captures an inner planet through migration within the protoplanetary disk to an MMR or forms a co-gap \citep{Kanagawa-etal.2020}, the inner planet may migrate outward rather than simply inward toward the host star \citep{Masset-Snellgrove-2001, Morbidelli-Crida-2007, Andre-Papaloizou-2016}. The physics of gap opening has been studied extensively, with analytic criteria developed to predict when a planet is able to carve a deep gap depending on disk scale height and viscosity \citep{Crida-etal.2006, Kanagawa-etal.2018}. Some works have emphasized the importance of disk viscosity and thermodynamics in setting accretion rates and migration pathways \citep{Machida2010, Morbidelli-etal.2014, Ziampras-etal.2020,Li2021, Choksi2023,Li2023,Li2024, Wu2025}. However, most of these models assume fixed planetary masses during migration or consider gas accretion in isolation, without coupling it to orbital evolution. 

\cite{Bergez-Casalou-etal.2023} modeled simultaneous gas accretion by two giant planets embedded in the same disk using long-term two-dimensional (2D) hydrodynamical simulations with \texttt{FARGO-2D1D} code \citep{Crida-etal.2007}. Keeping the planets on fixed orbits, they found that when both planets start runaway accretion at the same time, their masses tend to converge (typically $0.9<M_{out}/M_{in}<1.1$ after 0.5 Myr) and a wide range of observed mass ratios is reproduced only if gas accretion begins asynchronously. Their results emphasize the crucial role of gap formation and disk viscosity in shaping how multiple planets share a common gas reservoir. In contrast, we extend these investigations by allowing the planets to migrate and capture into resonance, rather than fixing their orbits. In this work, our hydrodynamical simulations followed the runaway gas accretion and orbital evolution of planetary cores that have already reached critical mass ($\sim 30 M_{\oplus}$). We find that the inner planet's growth is self-limited by early gap opening and inner-disk depletion, while the outer planet continues to accrete efficiently from the outer disk. Consequently, the outer planet often becomes equal to or even more massive than the inner one. This mechanism suggests a plausible pathway for the reversed mass ratio observed in resonant giant planet systems such as HD 128311 \citep{Sandor-Kley-2006, Rein-Papaloizou-2009} and Kepler-27 \citep{Steffen-etal.2012, Hadden-Lithwick-2014}, complementing and extending the findings of \cite{Bergez-Casalou-etal.2023}.

In this work, we carry out a suite of 2D hydrodynamical simulations to study the simultaneous gas accretion and orbital migration of giant planet pairs. Our goal is to investigate how disk properties such as surface density slope and viscosity and planet properties including initial core-mass and orbital position determine the final orbital configuration and mass ratio. We particularly focus on the physical conditions that lead to mass-ratio reversal, a phenomenon with direct implications for the architecture of observed giant planet systems. We believe that the planet cores should accrete gas when they are migrating. The density profile changed by their growth will affect their accretion rates and alter their final mass ratios. By including the gas accretion process, we perform a series of 2D hydrodynamical simulations with one or two planets cores. Our results show that the mass configuration of the planet pair changes during the evolution of their orbits. Additionally, the preferred mass configuration is the reversed one, where the outer planet becomes more massive than the inner planet. This implies that the mass ratio we observed now is not the original value. It may reverse during the convergent migration or even after the compact orbital configuration had formed. The architecture in which the inner planet is more massive than the outer one, by contrast, should be the result of a different process, in which the outer planet forms and is captured long after the inner giant planet has already formed.
	
This paper is organized as follows: In Section \ref{sec:observation facts}, we present some observational evidence for the similarity in mass between neighboring giant planets. Section \ref{sec:Simulation methology} describes our physical model and simulation setup. Results are presented in Section \ref{sec:results}. The tests of a single accreting planet and two orbit-fixed accreting planets are presented in Section \ref{sec:single-planet} and Section \ref{sec:two fixed planets}. The investigations on two orbit-free accreting planets are presented in Section \ref{sec:two free planets}. Finally, Section \ref{sec:discuss} summarizes our main conclusions and Section \ref{sec:conclusion} discusses the broader implications for the formation and evolution of giant planet systems.

\section{The Similarity and Dispersion in Mass of Neighboring Giant Planets}\label{sec:observation facts}
The study of planet similarity within multi-planet systems has traditionally focused on low-mass planets, where the planetary radii can be used as a reliable proxy for mass \citep{Weiss2014, Weiss2018, Millholland2017}. For giant planets, however, the situation is more complicated: planetary radii are relatively insensitive to mass beyond the Neptune-Jupiter regime \citep{Militzer2013, Laughlin2011}. As a result, the mass ratio between adjacent giant planets provides a more meaningful diagnostic of their similarity \citep{Otegi2022}.

\begin{figure*}[!htbp]
\centering
\includegraphics[width=0.7\textwidth]{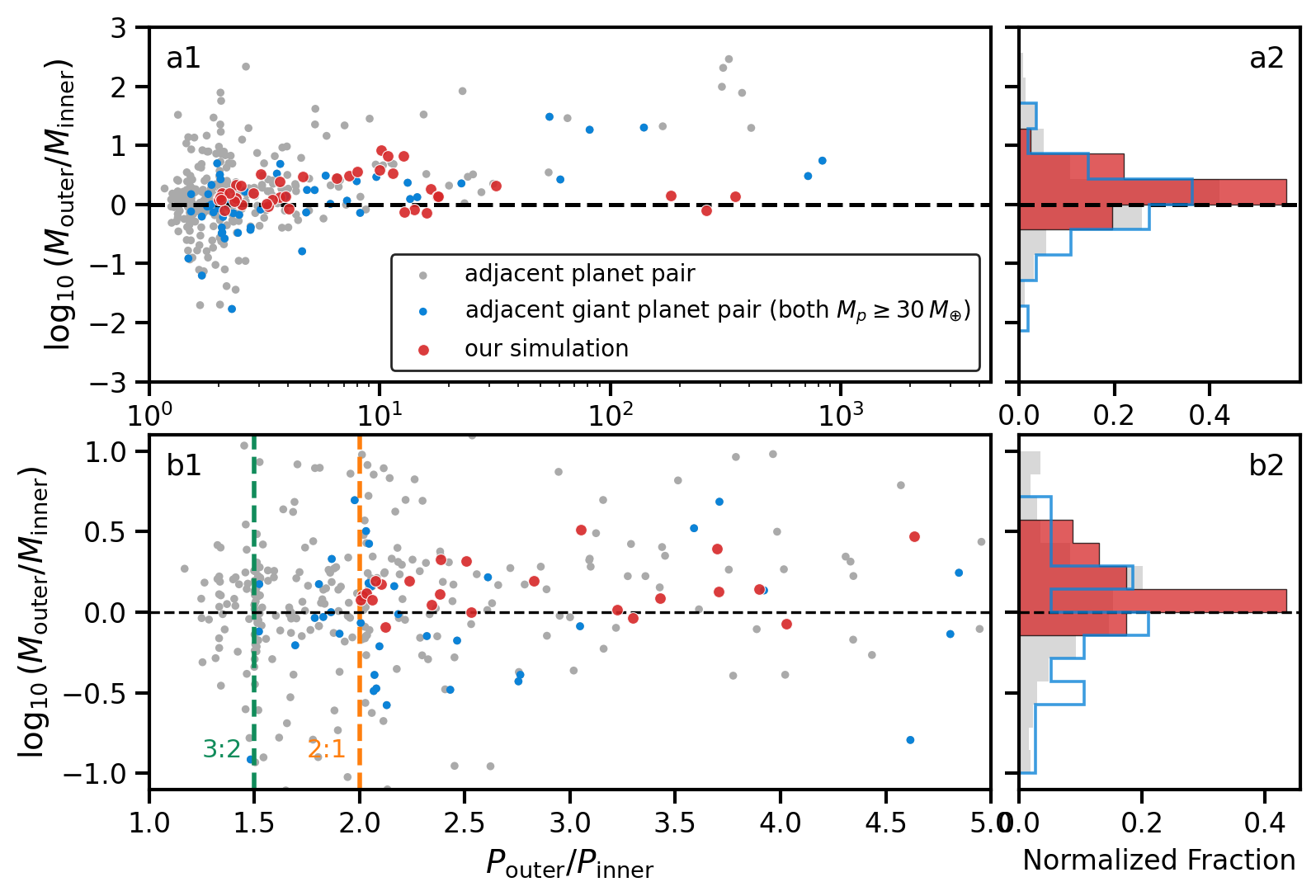}
\caption{Adjacent exoplanet pairs exhibit a systematic trend between the planetary mass ratio ($M_{\mathrm{outer}}/M_{\mathrm{inner}}$) and the orbital period ratio ($P_{\mathrm{outer}}/P_{\mathrm{inner}}$). Grey points represent all adjacent pairs in the observational sample. Blue points denote adjacent giant pairs with both planets $M_p\geq30\,M_\oplus$. Red points show the final states of our hydrodynamical runs (Tables~\ref{tab1}--\ref{tab3}), with each run shown once and given equal weight. The right-hand panels show normalized histograms of $\log_{10}(M_{\mathrm{outer}}/M_{\mathrm{inner}})$.}\label{fig1}
\end{figure*}
    
To date, most measurements of planetary mass are provided by the radial velocity (RV) method \citep{Mayor1995, Fischer2016, Perryman2018}. Therefore, the reliability of any single-planet mass or the mass distribution of a sample depends strongly on how well we know the host star's mass and the planetary orbital inclination. However, if we take the mass ratio of two adjacent planets within the same system instead, we can cancel out the dependence on the host star's properties. This diagnostic is particularly valuable for giant planets, whose radii are poor tracers of mass, but whose formation likely involved substantial gas accretion and strong dynamical interactions with their natal environment.

Figure \ref{fig1} presents the distribution of planetary mass ratios ($M_{p,\mathrm{outer}}/M_{p,\mathrm{inner}}$) as a function of orbital period ratios ($P_{\mathrm{outer}}/P_{\mathrm{inner}}$) for adjacent planet pairs. A weak but noticeable trend emerges whereby systems with wider orbital separations tend to host more massive outer planets than their inner companions, with the effect becoming more apparent at $P_{\mathrm{outer}}/P_{\mathrm{inner}} > 3$. However, this apparent trend may be influenced by observational selection effects. In radial velocity surveys, lower-mass or longer-period planets are intrinsically more difficult to detect, which can bias the inferred mass hierarchy within widely separated pairs. Moreover, large orbital gaps increase the likelihood of undetected intermediate planets, potentially leading to an artificial enhancement of the apparent mass asymmetry between the observed giants. As illustrated in panels b1 and b2, the underlying initial conditions and evolutionary pathways of real planetary systems are likely significantly more diverse than those assumed in our simulations.

When focusing on more compact systems ($P_{outer}/P_{inner}\leq5$), the picture changes. As shown in the lower panels of Figure \ref{fig1}, adjacent planets in close proximity tend to be more similar in mass, with the scatter in $M_{outer}/M_{inner}$ significantly reduced. This effect is most pronounced for systems where both planets exceed 30 $M_{\oplus}$ (i.e. in the giant-planet regime). In these cases, the observed mass ratios cluster more tightly around unity, suggesting that some mechanism during the disk phase tends to equalize the final masses of neighboring giant planets.

\begin{figure*}[t]
\centering
\includegraphics[width=0.6\textwidth]{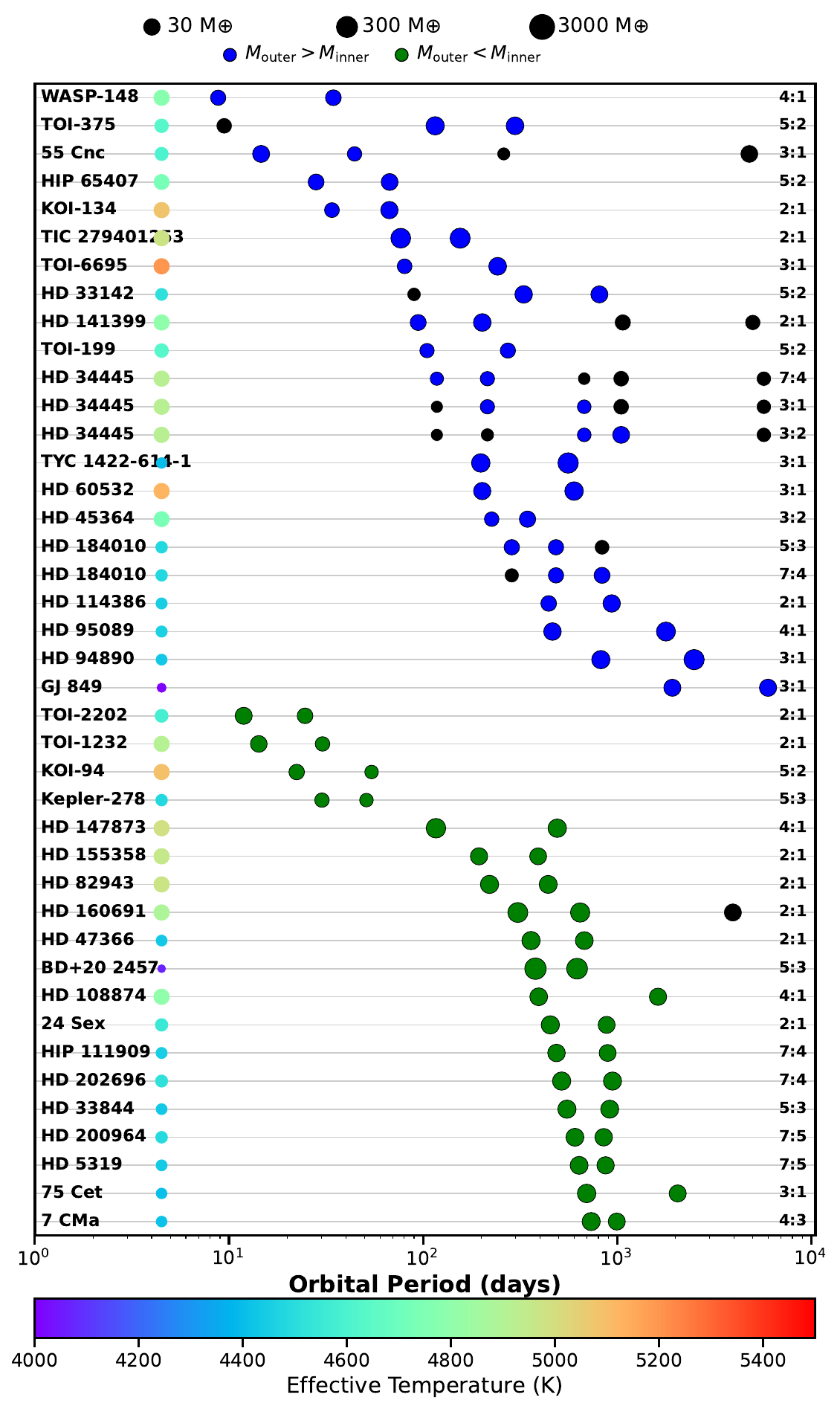}
\caption{A subset of giant exoplanet pairs close to MMRs. We only include giant planet pairs where both planets are more massive than $30\,M_\oplus$. The color of the host star represents its effective temperature, while its size represents its radius. The size of the planets represents their mass, and the approximate period ratio of the two planets in each pair is listed on the right. The color of the planet pairs represents the mass ratio (i.e., $M_{\mathrm{outer}}/M_{\mathrm{inner}}$) being greater than 1 (blue) or less than 1 (green). Regardless of whether the mass ratio of near-resonant planet pairs is greater than or less than 1, their distribution range in orbital period is similar. In particular, there are giant planet pairs with mass ratios $M_{\mathrm{outer}}/M_{\mathrm{inner}} > 1$ in the period range of $10-100$ days, which is difficult to explain using the outward migration model. The more likely scenario is that the two planets migrate inward and their mass ratio is reversed after they are trapped in MMRs.}
\label{fig2}
\end{figure*}

Figure \ref{fig2} highlights a subset of giant planet pairs that lie near MMRs. Interestingly, these resonant or near-resonant systems exhibit a wide range of mass ratios: in some cases the outer planet is heavier ($M_{outer}/M_{inner} > 1$, blue), while in others the inner planet dominates ($M_{outer}/M_{inner} < 1$, green). The distribution of orbital periods, however, does not show a strong dependence on which planet is more massive. Notably, several systems with $M_{outer}/M_{inner} > 1$ are located at relatively short orbital periods ($10-100$ days), where standard outward migration models struggle to explain their configuration. This provides strong motivation for exploring the possibility that the mass ratio of giant planet pairs may reverse after they have been captured into resonance.

As mentioned above, these observational statistics suggest two important points. First, while widely separated giant-planet pairs show diverse mass ratios partly driven by detection biases, compact and resonant giant-planet pairs are often more similar in mass than naive expectations would suggest. Second, the existence of systems with reversed mass ratios challenges the conventional picture of convergent migration, and calls for detailed hydrodynamical modeling of simultaneous gas accretion. These considerations form the basis for our numerical study presented in the following sections.

\section{Physical Model and Numerical Simulation} \label{sec:Simulation methology}

Our physical model and numerical method are the same as our previous works \citep{Zhang-Zhou-2010a, Zhang-Zhou-2010b}. The gas is treated as a locally isothermal fluid, with the equation of state given by $P = c_s^2 \Sigma$, where $P$ is the vertically integrated pressure, $\Sigma$ is the gas surface density, and $c_s$ is the local sound speed. The sound speed is prescribed as $c_s = h r \Omega_K$, with $h \equiv H/r = 0.04$ being the constant aspect ratio adopted throughout this work, and $\Omega_K$ the Keplerian angular frequency. The evolution of the planets is integrated using a fifth-order Runge--Kutta integrator, which handles the gravitational interactions between the planets and the gas disk self-consistently. The disk self-gravity is neglected. The code employs a softening length $\epsilon$ for the planetary gravitational potential, which is set to a fraction of the local scale height to avoid singularities and to properly resolve the planet--disk interaction. For the boundary conditions, we adopt wave-killing zones in the inner and outer parts of the disk to prevent artificial wave reflections that could affect the long-term evolution of the system. The inner and outer radial boundaries are set at $r_{\rm in}$ and $r_{\rm out}$, which are chosen to be sufficiently far from the planet orbits to minimize boundary effects on the planetary migration. The masses of the cores are set to be $15$ or $30 M_{\oplus}$, which is believed to be the critical mass to trigger the runaway accretion. In this runaway accretion stage, the mass of the growing planet is only regulated by the disk density around it. Details can be found in \cite{ang2008}. Thus, the gas accretion process is described in a relatively simple way: at every time step, we remove some mass from the disk around the planet and add it onto the planet, adjusting the planet's velocity to conserve total angular momentum. The amount of accreted mass is determined by a half-emptying timescale prescription commonly adopted in hydrodynamical simulations of gas accretion onto planets \citep{ang2008}. Many studies have examined the migration and/or gas accretion process of planet cores embedded in the proto-stellar disk. In this article we will explore the joint effects of migration and gas accretion processes in a multiple-planet system, and compare the results to the observation facts.

\subsection{Model Set-ups}	\label{subsec:Model set-ups}
To investigate the coupled processes of orbital migration and gas accretion in giant planet pairs, we perform a series of 2D hydrodynamical simulations with the 2nd-order Godunov code \texttt{Antares} which has been described and tested by \cite{zhang2008}. This code is in the Cartesian coordinates whose origin locates at the central star and covers the whole star-planet-disk interaction range without a cavity around the central star to ensure the global consistency of the dynamics evolution of the system \citep{Zhang-Zhou-2010a}. It also adopts a corner-flux correction method and a global viscous description to reduce the grid effect and ensure the consistency of the disk's viscous evolution \citep{Zhang-Zhou-2010b}. For computational accuracy and convenience, we set the mass of the central star, $M_\ast = M_\odot$, as the mass unit and the initial position of the inner planet, $r^0_{inner} = 5.2 AU$, as the length unit. Then the  time unit is $P/2\pi$, where $P=1$ is the orbital period at $r^0_{inner}$ when the gravitational constant $G=1$. The surface density of the gaseous disk follows a power law:
\begin{equation}
\Sigma(r) = \Sigma_0 r^\alpha,
\end{equation}
where $\Sigma_0=6.4\times10^{-3}$ is the surface density at $r=1$ under our unit normalization. To investigate the influences of the disk slope we adopt different profile indexes, where $\alpha=-1/2, -1, -3/2$. We fixed the disk aspect ratio $h=H/r=0.04$ for all cases and investigate different kinematic viscosities $\nu$ from $10^{-6}$ to $10^{-4}$. The computational domain of our disks are from $-4$ to $4$ in both x- and y- directions, which encloses a total disk mass $M_{disk}\sim 0.01-0.02 M_\ast$ depending on different $\alpha$. The whole disk is resolved by a grid with $N_x\times N_y = 1024\times1024$, which delivers an overall spatial resolution of $\Delta x=\Delta y\approx0.008$. A planet core locating at $r_p=1$ with the initial mass of $30M_e$ has a Hill radius of $0.032$ in our unit. This means its Hill region is resolved by $\sim8$ cells in diameter. As the planet is getting more massive by accreting gas, its Hill region would be resolved by more cells.

The initial locations of the two planets cores are $r^0_{inner}=1$ and $r^0_{outer}=2$. Instead of using grown-up giant planets, we start with two tiny embryos whose initial masses are both set to be $0.001M_\oplus$. During the first $100$ orbits we hold the embryos on their initial circular orbits and increase their masses smoothly to the critical mass. Then we stop the mass growth and hold the planets on fixed circular orbits for another 100 orbits to allow the local gas flow around their orbits to adjust to the planetary potential. This preparation phase reduces the artificial response associated with the initialization before the planets are released to migrate and accrete. By doing so, we avoid any artificial initial jerk to the disk and feed backs to the accreting planet cores. When the two cores are released, they start to grow by accreting gas from the disk. The detailed gas accretion model will be described in the next section. The major goal of this work is to investigate the mass and period ratios of the two interacting gas giants. According to \cite{Pollack1996}, the cross-over mass of Jupiter is $10-30M_\oplus$ depending on the disk properties. Since the solid accretion process and the trigger of the runaway accretion are beyond the scope of this work, we focus only on the subsequent evolution of gas-accreting planet cores. To ensure that the planet cores are already in the runaway gas accretion regime when the artificial orbital constraints are removed, we set the critical mass to be $30M_\oplus$ in most cases. This setup allows us to focus on the relative growth and migration of two interacting planets in a shared disk environment, while neglecting the earlier stages of solid accretion and envelope contraction. Such simplifications are necessary to isolate the dynamical competition between neighboring giants and the role of disk parameters in shaping their final architecture.

Each run begins with two low-mass embryos held on circular orbits. Their masses are smoothly increased to the target core masses over the first 100 orbits and are then held fixed for another 100 orbits ($t\leq200P_{\rm in,0}$) to allow the local gas flow to adjust before the planets are released to migrate and accrete. We do not require the disk to reach a fully steady gap configuration before release. In particular, previous simulations have shown that formally reaching a steady state can require much longer integrations in very low-viscosity disks, e.g., $\alpha\sim10^{-4}$ in \cite{Wu-etal.2023}. Our adopted viscosities, $\nu=10^{-6}-10^{-4}$ with $h=0.04$, correspond to effective $\alpha_{\rm SS}\simeq6.3\times10^{-4}-6.3\times10^{-2}$, and therefore lie above this extremely low-viscosity regime. Nevertheless, because the planets are released before deep gaps are fully established, an early transient in both $\dot M$ and $da/dt$ is expected. We therefore distinguish this initial transient from the subsequent evolution over thousands of orbits, including gap opening, convergent or divergent migration, and possible MMR capture, on which our conclusions are based. The subsequent long-term evolution, including gap opening, convergent or divergent migration, and possible MMR capture, represents the physically relevant outcome of each run. Integrations are terminated at a prescribed total time $t_{\mathrm{f}}$ (typically a few thousand $P_{\mathrm{in},0}$, set by the input parameter \texttt{Norbs}), rather than by a mass-convergence criterion. Consequently, the final masses reported in Tables~\ref{tab1}--\ref{tab3} are the values at $t_{\mathrm{f}}$ rather than guaranteed asymptotic masses in an undepleted disk.

\subsection{Gas Accretion} 	\label{subsec:GasAcc}

We adopt a modified version of the commonly used sink–cell prescription for gas accretion described by \cite{Kley1999}. The basic model is simple: for each accreting planet, gas within its vicinity is taken away and added to the planet. Meanwhile, the velocity of the planet is adjusted to ensure the conservation of the angular momentum of the accreting planet and the accreted gas. The accretion vicinity is described by a radius $r_{acc}$ which is the minimum value of Bondi radius $R_B$ (distance beyond which the thermal energy of the gas is larger than the gravitational energy that binds the gas to the planet) and Hill radius $R_H$ of the accreting planet:

\begin{eqnarray}
r_{acc} \equiv \min(R_{B},R_{H}), \\
R_{B} = \frac{2Gm_p}{v^2_{rel}+c^2_s}, \\
R_{H} \simeq r_p (\frac{m_p}{3M_\ast})^{\frac{1}{3}},
\end{eqnarray}
where $G$ is the gravitational constant, $v^2_{rel}$ is the relative velocity between the accreting planet and gas, $c_s$ is the speed of sound at the planet's position, $M_\ast$, $m_p$ and $r_p$ are the stellar mass, planet's mass and position, respectively.

Two important adjustments are implemented to improve realism. First, we include a geometrical reduction factor $f_{3d}$ to account for the $3D$ effect of gas accretion. The formula of $f_{3d}$ follows the fact that the actual accretion volume is a sphere in a $3D$ disk instead of a cylinder which has been integrated over vertical direction in a $2D$ simulation:

\begin{equation}
f_{3d} \equiv \frac{\frac{4}{3}r^3_{acc}}{r^2_{acc} \cdot 2H_p} = \frac{2}{3}\frac{r_{acc}}{h r_p} ,
\end{equation}

where $H_p$ is the scale height of the disk at the position of an accreting planet, $h=H/r$ is the disk's aspect ratio.  During the runaway accretion regime in our simulations, the typical values after normalization are $G=1$, $M_\ast=1$, $m_p=0.0001$ ($\approx 30 M_{\oplus}$), $r_p=1$ and $h=0.04$, then we have $f_{3d}\sim 0.1 $.

The second adjustment is that the accretion rates in grids within the accretion area are not identical. Instead, a Gaussian-like smoothing function $f_{sm}$ is adopted and it is set to be $1$ at the position of the planet and smoothly vanishes to $0$ outside the accreting radius :
\begin{equation}
f_{sm} = e^{-r^2_d/m},
\end{equation}
where $r_d$ is the distance from the grid center to the planet and it is in the unit of $r_{acc}$. m denotes the slow variation of the accretion rate and based on our experience, its value is 0.2. The gravitational potential of the star-planet-disk system is given by $\Phi = \Phi_s + \Phi_p + \Phi_d + \Phi_i$, where $\Phi_s$, $\Phi_p$, $\Phi_d$, and $\Phi_i$ denote the softened potentials of the central star and the planet, the disk self-gravity, and the indirect potential arising from the acceleration of the coordinate origin by the planet and the disk, respectively. The softened stellar and planetary potentials are $\Phi_s = -[G M_\odot / \left(x^2 + y^2 + \epsilon_{\rm star}^2\right)^{1/2}]$, $\Phi_p = -(G M_p / \left[(x - x_p)^2 + (y - y_p)^2 + \epsilon_p^2\right]^{1/2})$, where $\epsilon_{\rm star}$ and $\epsilon_p$ are the corresponding softening lengths. In our simulations, we adopt $\epsilon_p = 0.5\,R_{\rm Roche}$ and $\epsilon_{\rm star} = 0.05$ in units where the planet's initial semimajor axis is unity \citep{zhang2008}. Affected by this smoothing function, the overall accretion rate drops to $30\%$ of our original (unsmoothed) simulation result. Combined with the effect of $f_{3d}$, the typical accretion rate in our $2D$ simulations decreases to $\sim 0.03$ of the standard $2D$ results. This turns out to be consistent with the results obtained by high-resolution $3D$ simulations that the accretion rate in $3D$ simulations is typically two orders of magnitude smaller than that in $2D$ simulations \citep{Machida2010}. In particular, our tests for a single planet with $\nu=10^{-5}$ and $h=0.04$ yield an accretion rate slightly smaller than that reported by \citet{Machida2010,Tanigawa2016}. In the sub-thermal regime ($M_{\rm p}/M_{*}\lesssim h^{3}$), the $\dot{M}_{\rm p}$-$M_{\rm p}$ scaling is broadly consistent with those earlier results. In the super-thermal regime ($M_{\rm p}/M_{*}\gg h^{3}$), gap opening strongly suppresses accretion, placing our rates in rough agreement with \citet{DAngelo2003,Bodenheimer2013,Li2021,Li2023,Choksi2023}. Some of the remaining discrepancy likely arises from the fact that \citep{Machida2010,Tanigawa2016} adopted a larger disk aspect ratio of $h=0.05$, and that \citet{Machida2010} simulated inviscid disks. Since a higher $h$ increases the planetary accretion rate \citep{Li2023,Choksi2023}, adopting that parameter choice would naturally bring their results into better agreement with ours. 
Nevertheless, these refinements allow our simulations to capture both the global evolution of the disk and the relative growth of planets with more physical fidelity than a standard $2D$ sink approach, while remaining computationally tractable for long-term integrations.

In this work we adopt a globally constant aspect ratio $h\equiv H/r=0.04$, a conventional choice in two-dimensional planet--disk simulations that isolates the effects of varying the surface density slope, viscosity, and core-mass ratio at fixed thermal structure \citep[e.g.,][]{Li2023,Li2024}. As the planets migrate inward and grow, however, $r_{\mathrm{acc}}=\min(R_B,R_H)$ often exceeds the local scale height $H=h\,r_p$, and therefore vertical structure effects must be accounted for. We therefore recompute $f_{3\mathrm{D}}$ from the instantaneous $r_{\mathrm{acc}}$ and $H_p$, progressively suppressing the effective accretion rate in the regime where $r_{\mathrm{acc}}/H\gtrsim1$. Real disks are expected to exhibit flaring (e.g.\ $h\propto r^{1/4}$), which could modestly alter the gap depths and the relative Type~II migration rates of the two planets. The qualitative competition between gap opening, gas supply, and convergent and divergent migration is expected to remain robust.

\begin{figure*}[!htbp]
\centering
\includegraphics[width=\textwidth]{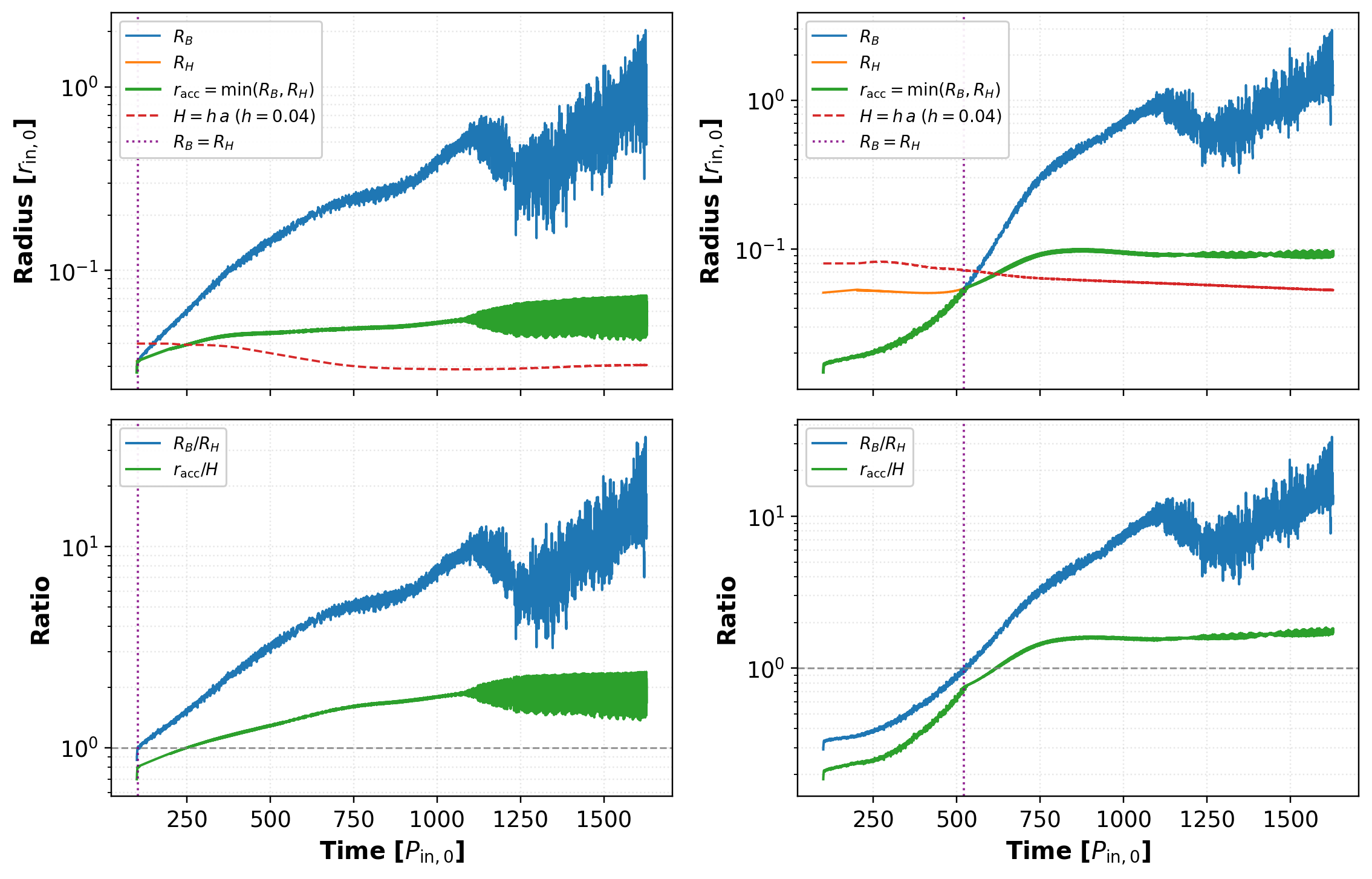}
\caption{Evolution of the Bondi radius $R_B$, Hill radius $R_H$, accretion radius $r_{\mathrm{acc}}=\min(R_B,R_H)$, and local disk scale height $H=h\,a$ ($h=0.04$) for a representative convergent run ($\Sigma \propto r^{-1.0}$, $\nu=10^{-5}$, core masses $30/15\,M_\oplus$). Left and right columns show the inner and outer planets, respectively. Top panels: absolute radii in units of $r_{\mathrm{in},0}$. Bottom panels: ratios $R_B/R_H$ and $r_{\mathrm{acc}}/H$; the horizontal dashed line marks unity. The vertical dotted line indicates the first crossing $R_B=R_H$. After early growth, $r_{\mathrm{acc}}$ is set by $R_H$ and typically satisfies $r_{\mathrm{acc}}\gtrsim H$, so that the geometrical factor $f_{3\mathrm{D}}$ strongly regulates accretion.}\label{rb_rh_evolution}
\end{figure*}

We also examine the evolution of the Bondi-to-Hill radius ratio during planetary growth. In a representative convergent run with $\Sigma \propto r^{-1.0}$, $\nu=10^{-5}$, and $M_{\mathrm{inner}}/M_{\mathrm{outer}}=30/15$, $R_B/R_H$ rises from order unity shortly after release to $\gg1$ at late times. After the early growth phase, $r_{\mathrm{acc}}$ is set by $R_H$ for nearly the entire subsequent evolution of the inner planet and for most of that of the outer planet, with $r_{\mathrm{acc}}\gtrsim H$ throughout much of the runaway/gap-opening stage. Thus, even with a fixed $h$, our model already captures the transition from a Bondi-limited to a Hill-limited accretion regime (Figure \ref{rb_rh_evolution}) through the factor $f_{3\mathrm{D}}(r_{\mathrm{acc}}/H)$.

\section{Results} \label{sec:results}

\subsection{Single-Planet Tests}\label{sec:single-planet}
\begin{figure*}[!htbp]
\centering
\includegraphics[width=\textwidth]{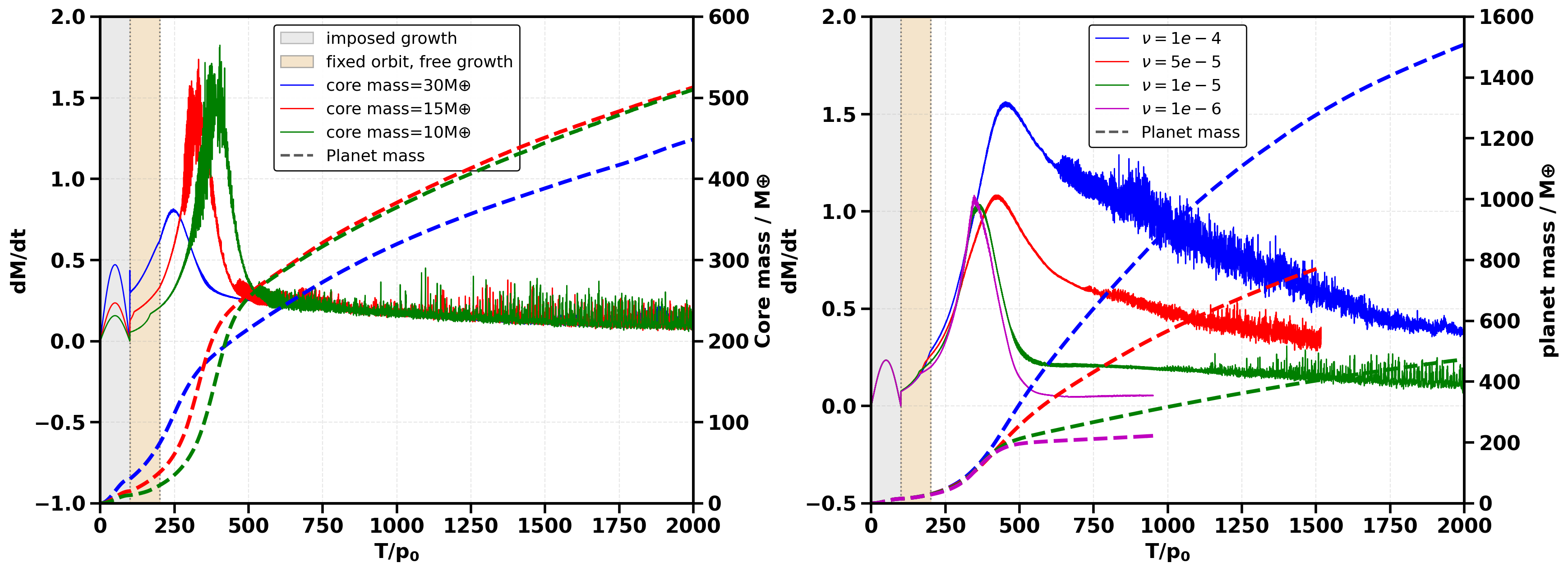}
\caption{Gas accretion onto a single planet with varying parameters. Solid lines show the accretion rate $\mathrm{d}M/\mathrm{d}t$ (left axis). Dashed lines show the planetary mass (right axis). The gray shaded interval at t = 0 to 100$P_0$ marks the imposed-growth phase and the orange shaded interval at 100$P_0$ to 200$P_0$ marks the orbit is held fixed while the planet is allowed to accrete freely. After $t=200\,P_0$, the planet evolves freely in both orbit and mass. Left panel: runs with different initial core masses at fixed viscosity. Right panel: runs with different disk viscosities at fixed core mass. A more massive core tends to trigger earlier gap opening and thus earlier saturation of growth, whereas higher viscosity sustains a larger late-time accretion rate and a larger final mass.}\label{fig3}
\end{figure*}

To validate our accretion prescription and to provide a benchmark for the two planets cases, we start with a free single-planet core. Results are shown in Figure \ref{fig3}. It is interesting that a higher core-mass can result in a lower final planet mass. One possible reason is that a high-mass core grows more quickly and opens a gap earlier and more rapidly, so that gas is tidally pushed away instead of accreting onto the planet. Once the gap is cleared, the planet's accretion rate drops dramatically. We also notice that the disk viscosity dominates the final mass of an accreting planet. Viscosity controls the gap opening time and the diffusion of gas at the gap boundaries. Although the accretion rate becomes very small after the gap opens, the majority of the planet's mass is accreted after the gap is cleared. These results provide useful context for understanding the behavior of an accreting planet pair. 

\begin{figure*}[t]
\centering
\noindent\makebox[\textwidth][c]{%
\begin{minipage}[c]{0.33\textwidth}
\centering\includegraphics[width=\linewidth]{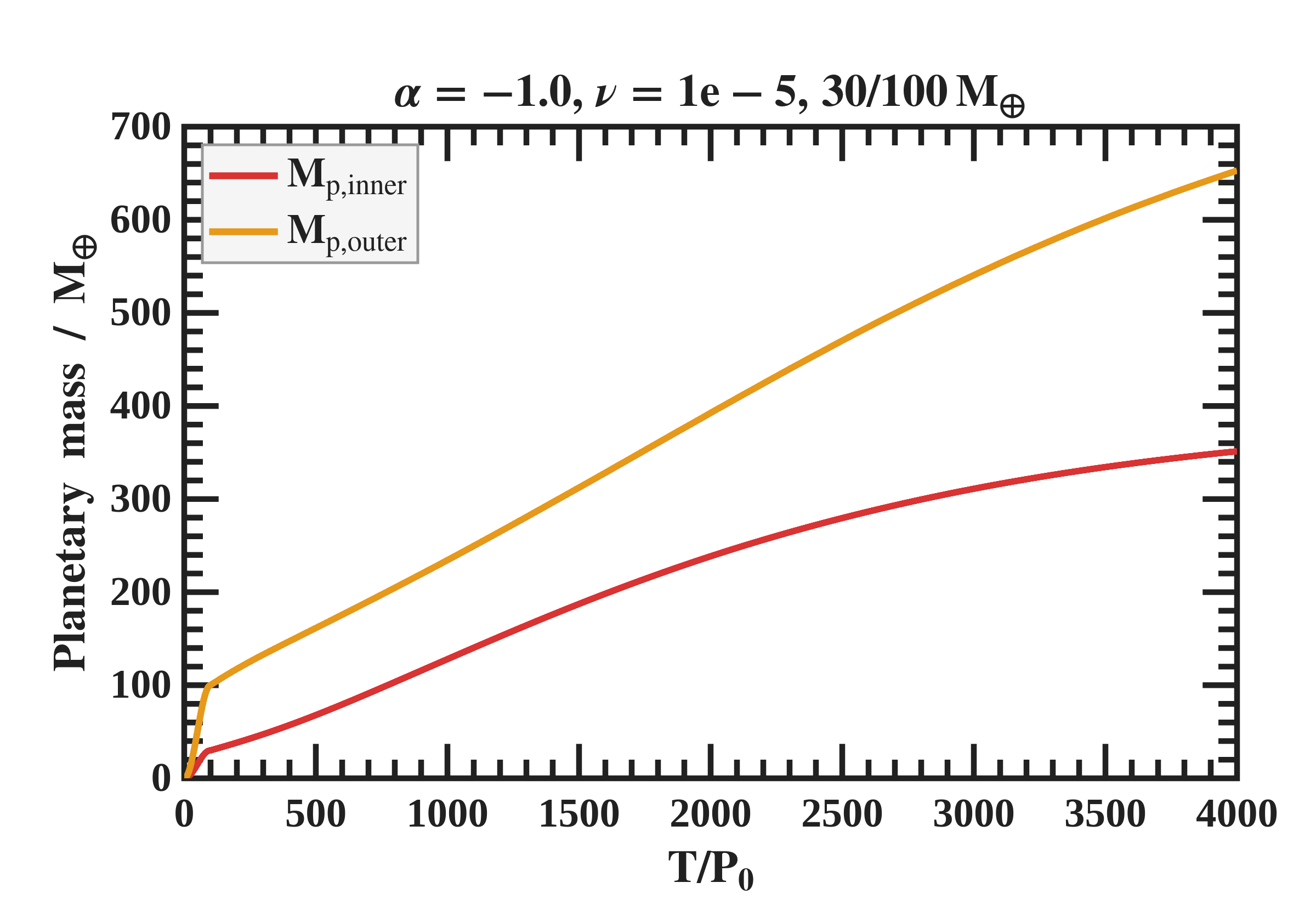}
\par\vspace{1mm}{\footnotesize (a)}
\end{minipage}%
\hspace{0.018\textwidth}%
\begin{minipage}[c]{0.33\textwidth}
\centering\includegraphics[width=\linewidth]{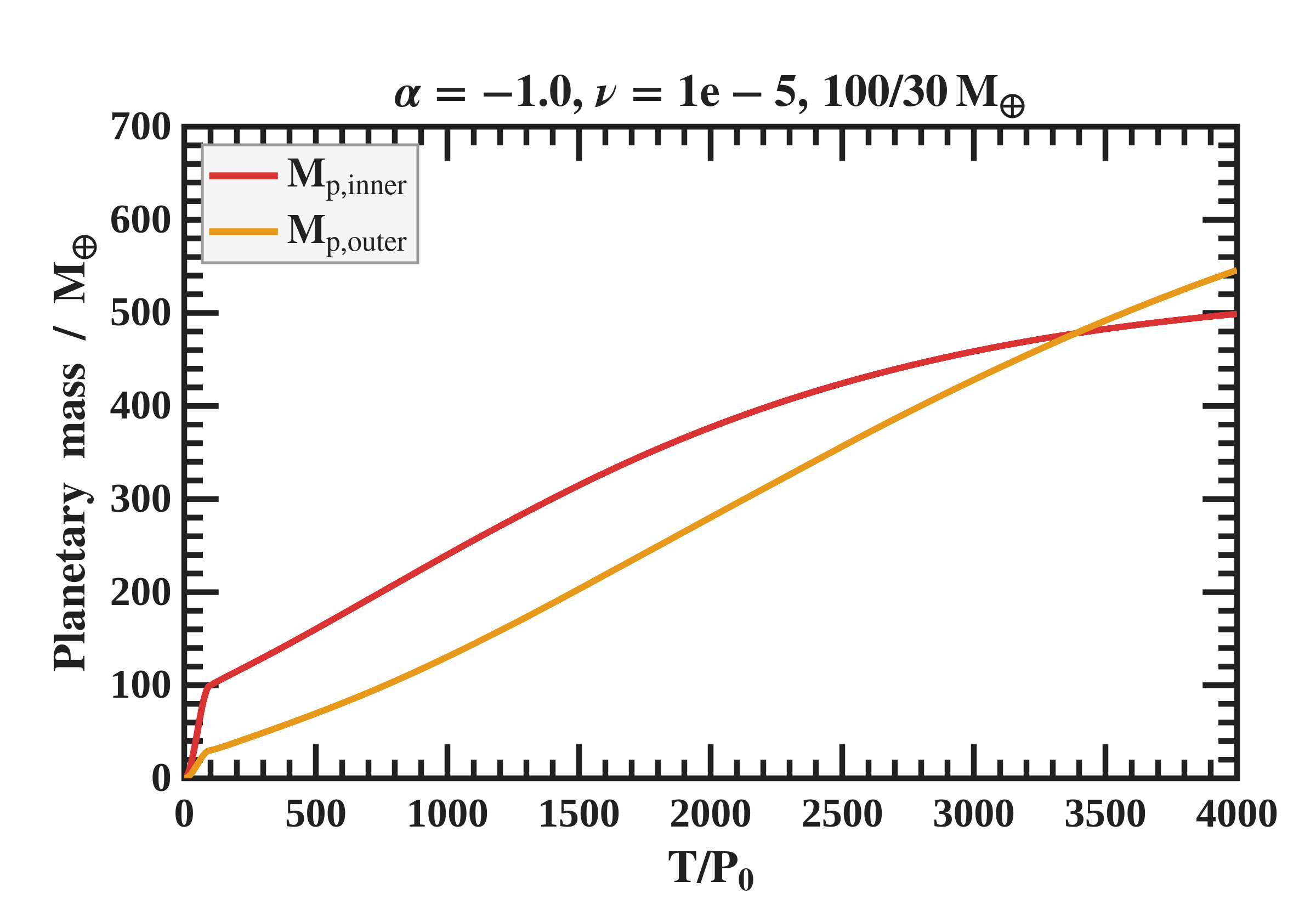}
\par\vspace{1mm}{\footnotesize (b)}
\end{minipage}%
\hspace{0.018\textwidth}%
\begin{minipage}[c]{0.33\textwidth}
\centering\includegraphics[width=\linewidth]{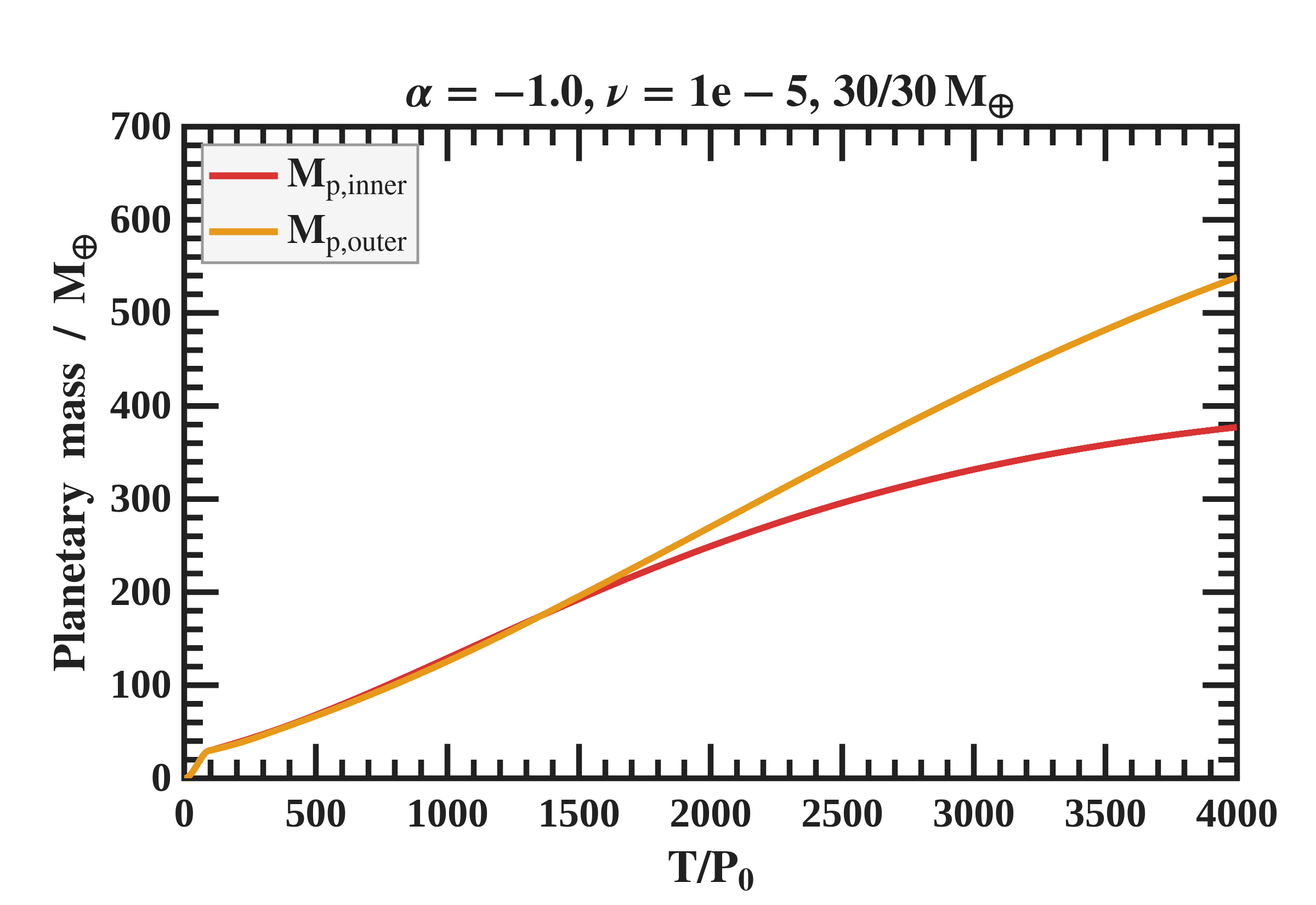}
\par\vspace{1mm}{\footnotesize (c)}
\end{minipage}%
}\par
\caption{Two-planet runs with fixed semi-major axes and varying initial core-mass combinations ($\alpha=-1.0$, $\nu=10^{-5}$). Panels (a)--(c) show the mass evolution of the inner (red) and outer (orange) planets for core pairs $30/100$, $100/30$, and $30/30\,M_\oplus$, respectively.The migration is suppressed, the panels isolate differential gas accretion that even when the cores are unequal, the outer planet often overtakes the inner one once a common gap forms and the outer feeding zone remains better supplied.}
\label{2fixed-mass}
\end{figure*}

\begin{figure*}[t]
\centering
\noindent\makebox[\textwidth][c]{%
\begin{minipage}[c]{0.305\textwidth}
\centering\includegraphics[width=\linewidth]{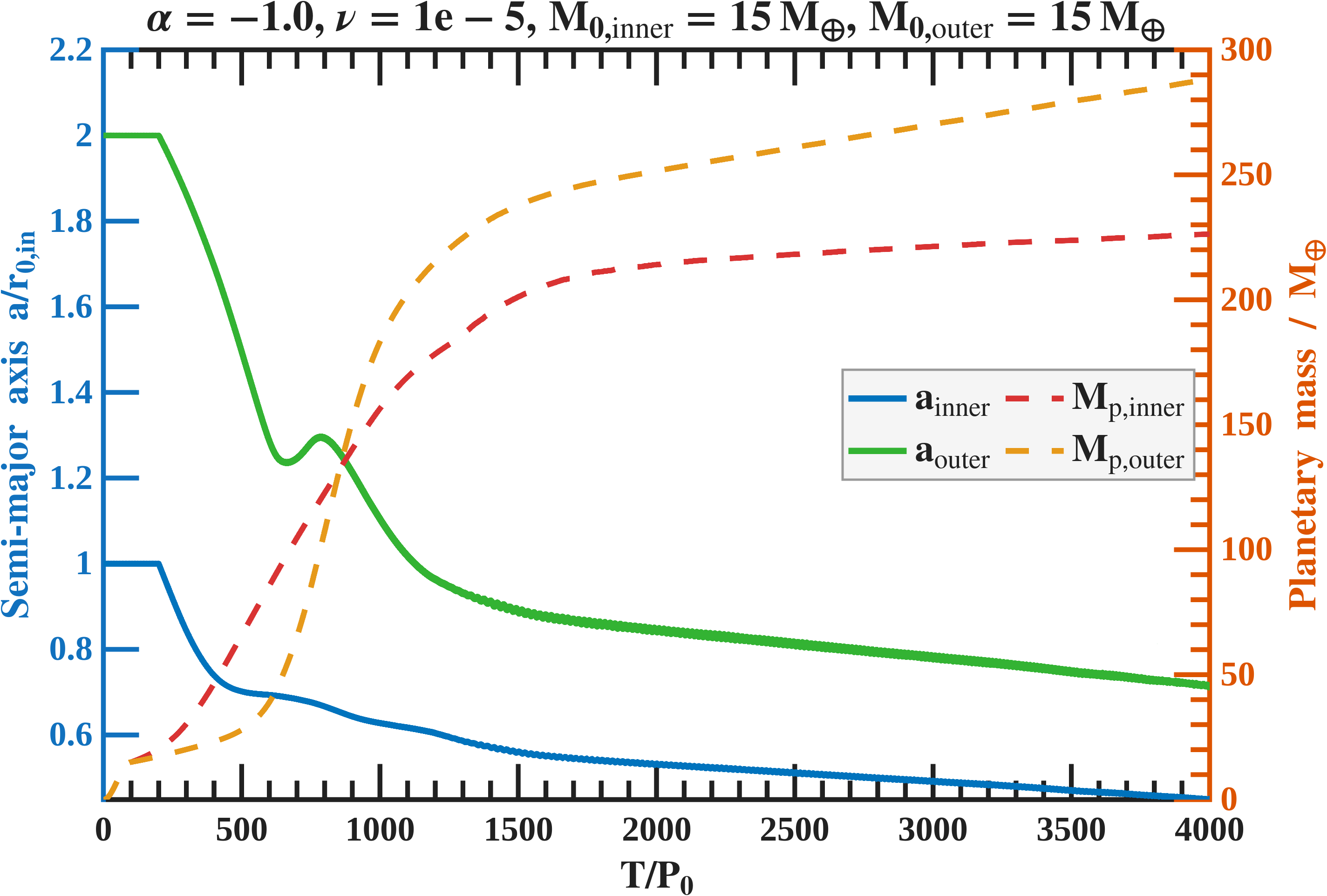}
\par\vspace{1mm}{\footnotesize (a)}
\end{minipage}%
\hspace{0.018\textwidth}%
\begin{minipage}[c]{0.305\textwidth}
\centering\includegraphics[width=\linewidth]{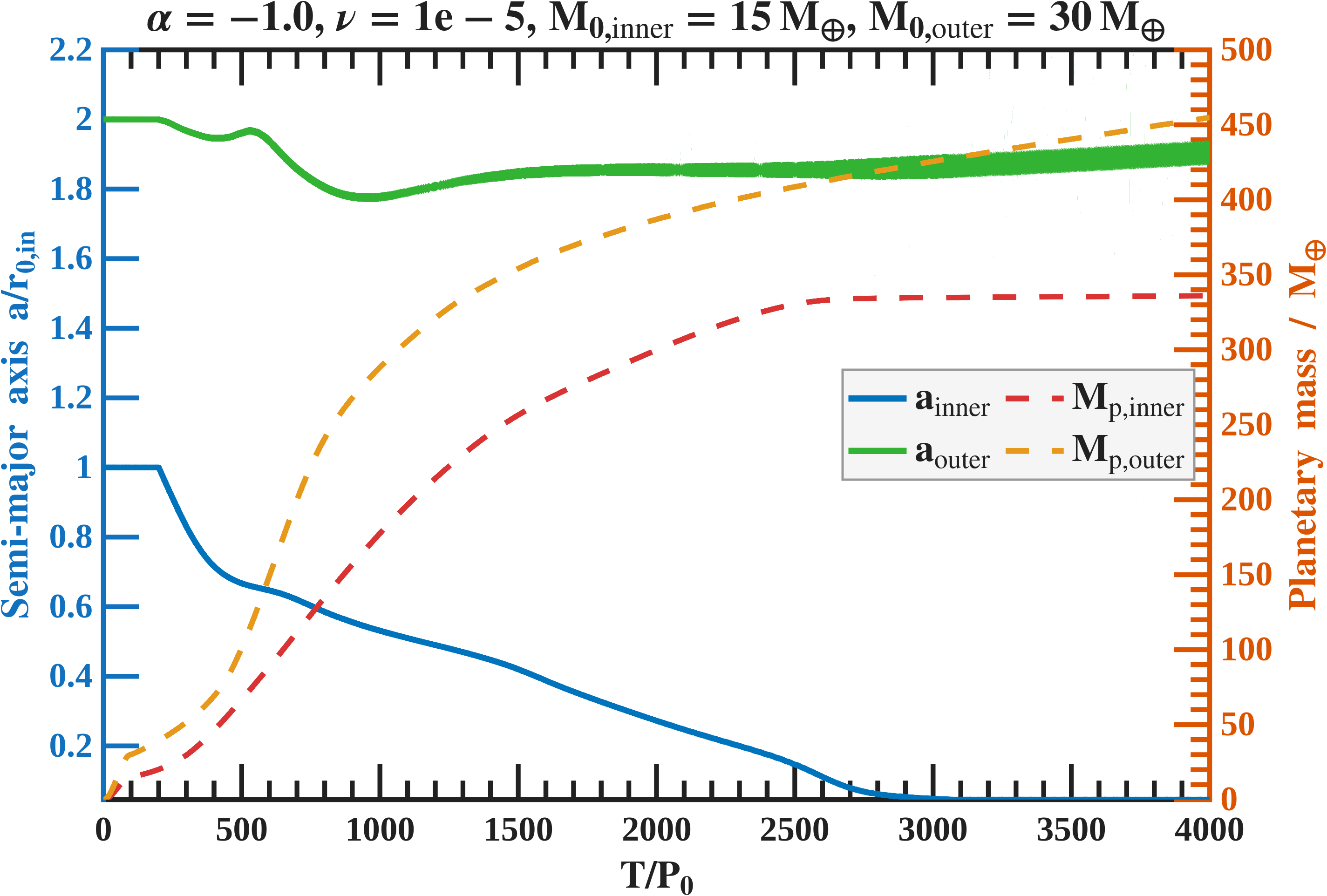}
\par\vspace{1mm}{\footnotesize (b)}
\end{minipage}%
\hspace{0.018\textwidth}%
\begin{minipage}[c]{0.305\textwidth}
\centering\includegraphics[width=\linewidth]{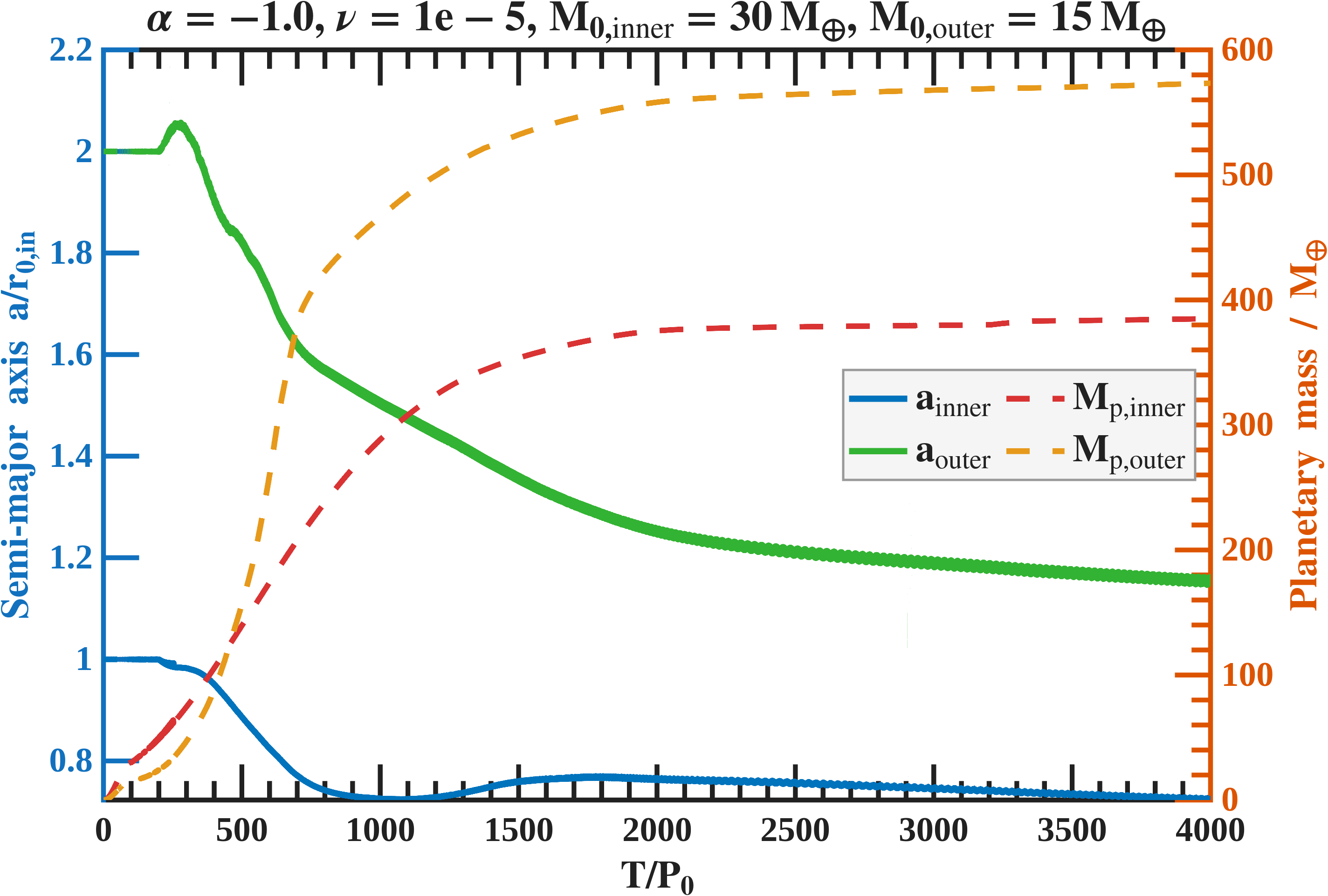}
\par\vspace{1mm}{\footnotesize (c)}
\end{minipage}%
}\par
\vspace{2mm}
\noindent\makebox[\textwidth][c]{%
\begin{minipage}[c]{0.305\textwidth}
\centering\includegraphics[width=\linewidth]{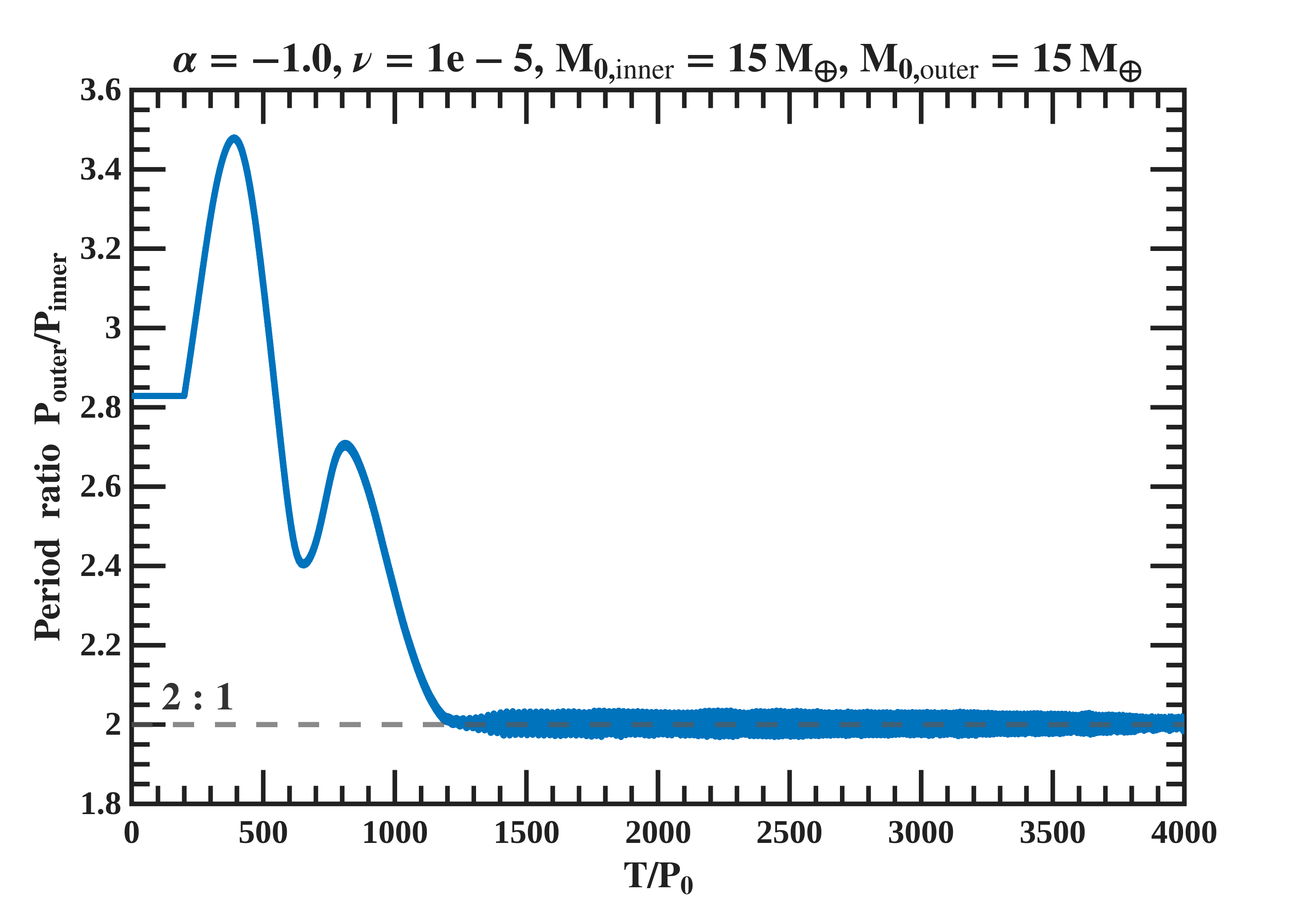}
\par\vspace{1mm}{\footnotesize (d)}
\end{minipage}%
\hspace{0.018\textwidth}%
\begin{minipage}[c]{0.305\textwidth}
\centering\includegraphics[width=\linewidth]{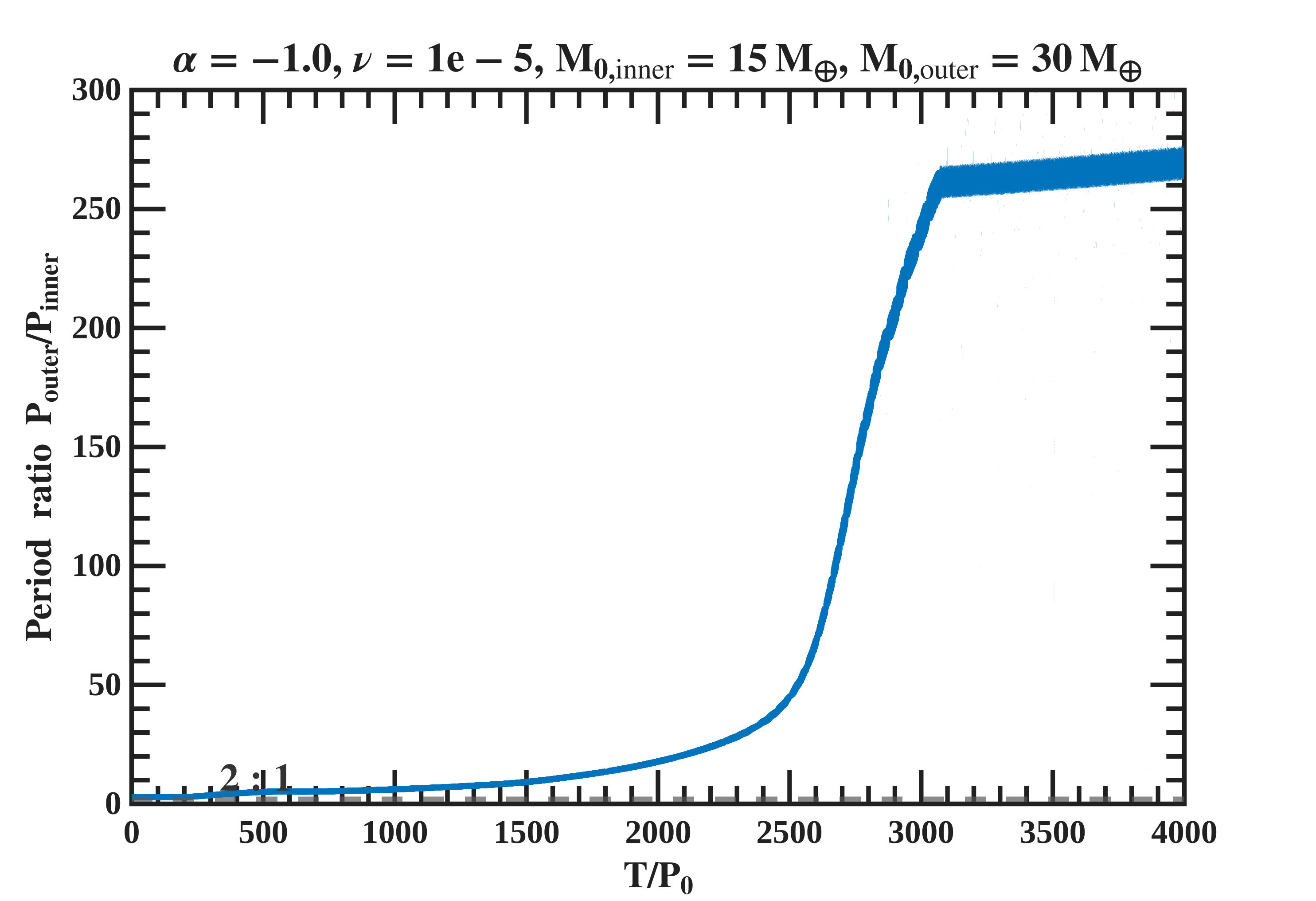}
\par\vspace{1mm}{\footnotesize (e)}
\end{minipage}%
\hspace{0.018\textwidth}%
\begin{minipage}[c]{0.305\textwidth}
\centering\includegraphics[width=\linewidth]{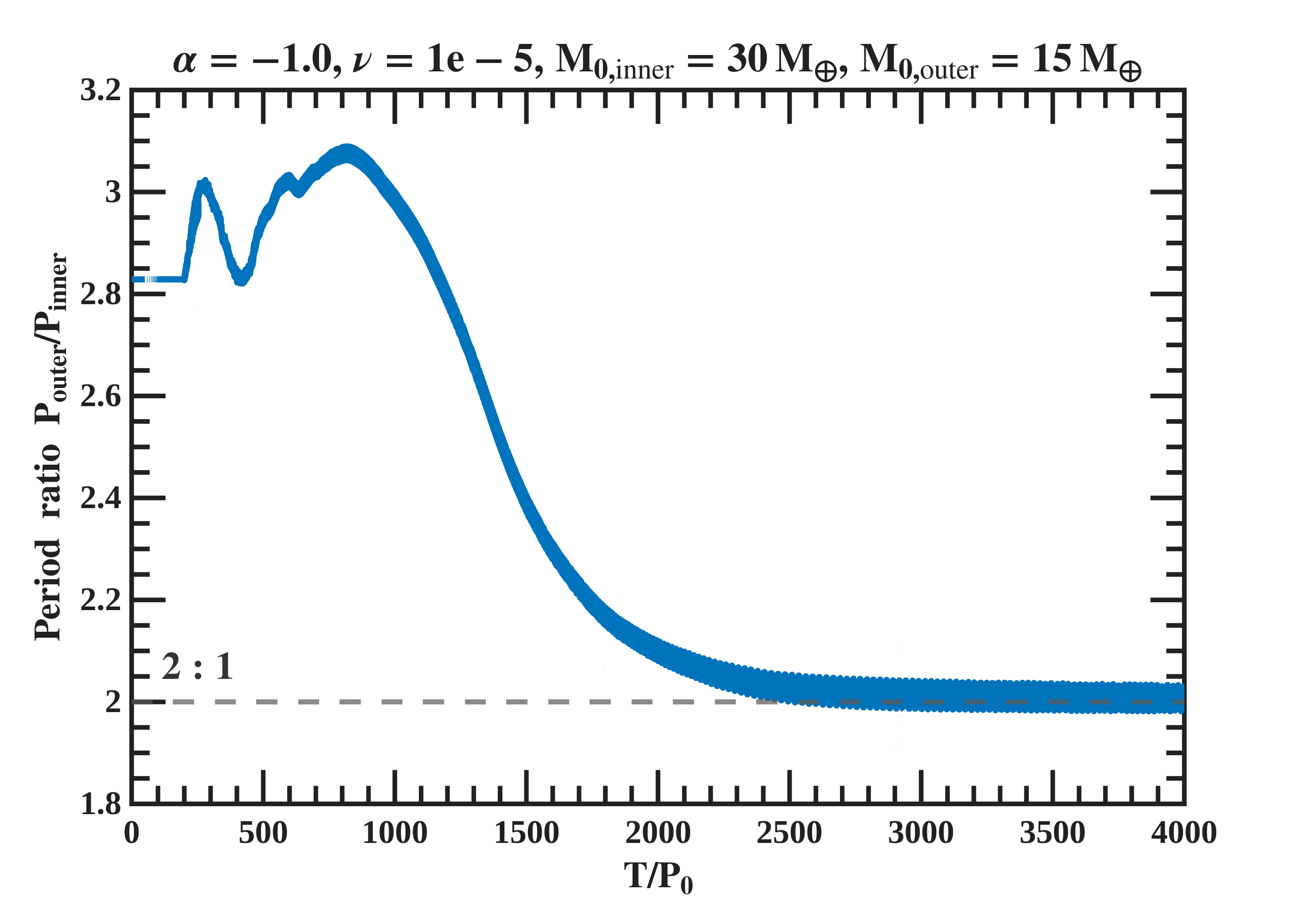}
\par\vspace{1mm}{\footnotesize (f)}
\end{minipage}%
}\par
\caption{Interacting planet-pair planet simulations with different initial core-mass combinations for the inner and outer planets ($\alpha=-1.0$, $\nu=10^{-5}$). Top panels: semi-major axes (solid) and masses (dashed) for the inner and outer planets. Bottom panels: period ratio $P_{\rm outer}/P_{\rm inner}$, with the horizontal dashed line marking the 2:1 MMR. Depending on the core-mass pairing, the system may converge and lock near 2:1 (e.g., panels d, f) or diverge with a rapidly increasing period ratio (panel e), illustrating the coupling between differential accretion and orbital evolution.}
\label{fig4}
\end{figure*}

\begin{figure*}[t]
\centering
\noindent\makebox[\textwidth][c]{%
\panelimage{a}{.315\textwidth}{0.20\textheight}{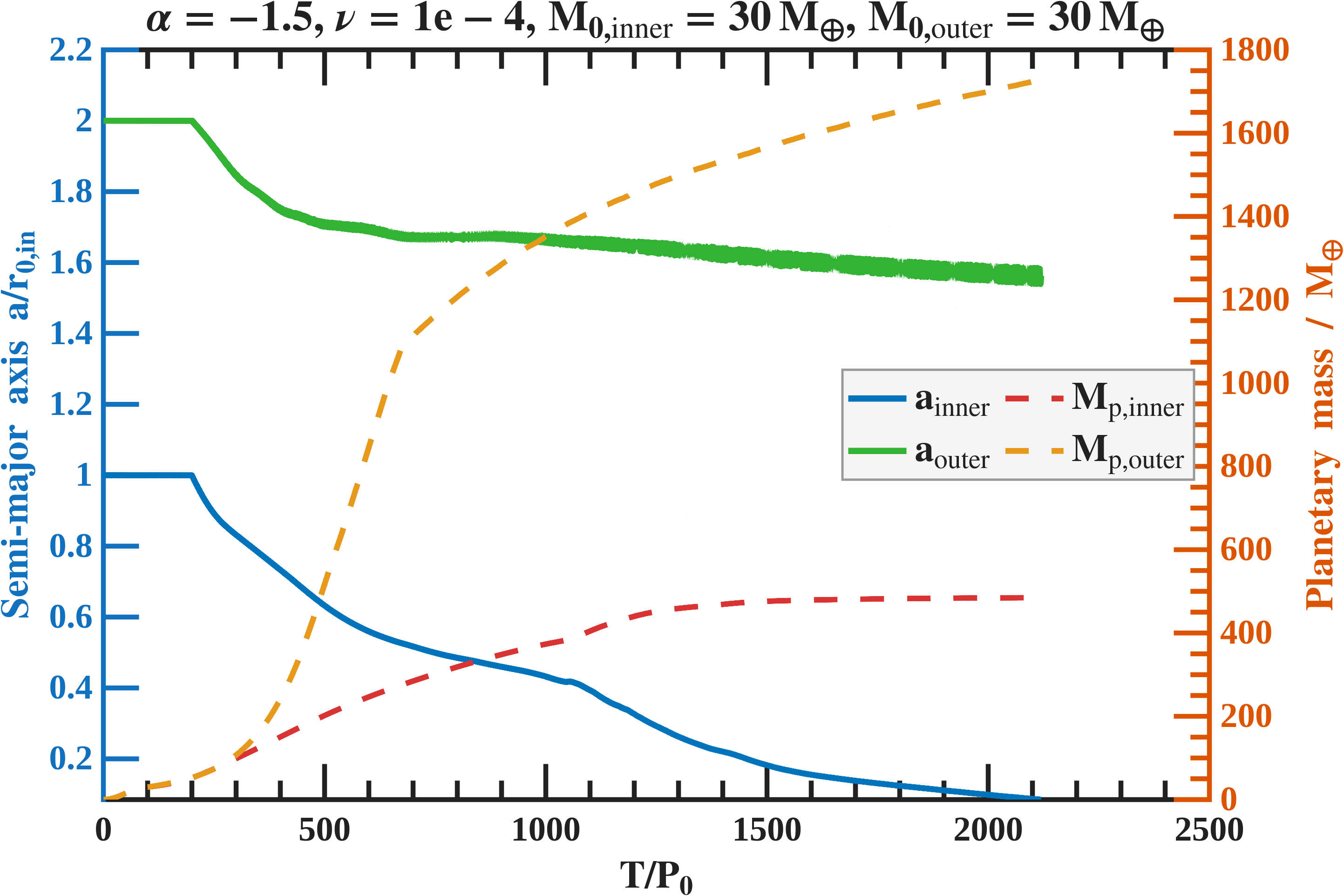}\hfill
\panelimage{b}{.315\textwidth}{0.20\textheight}{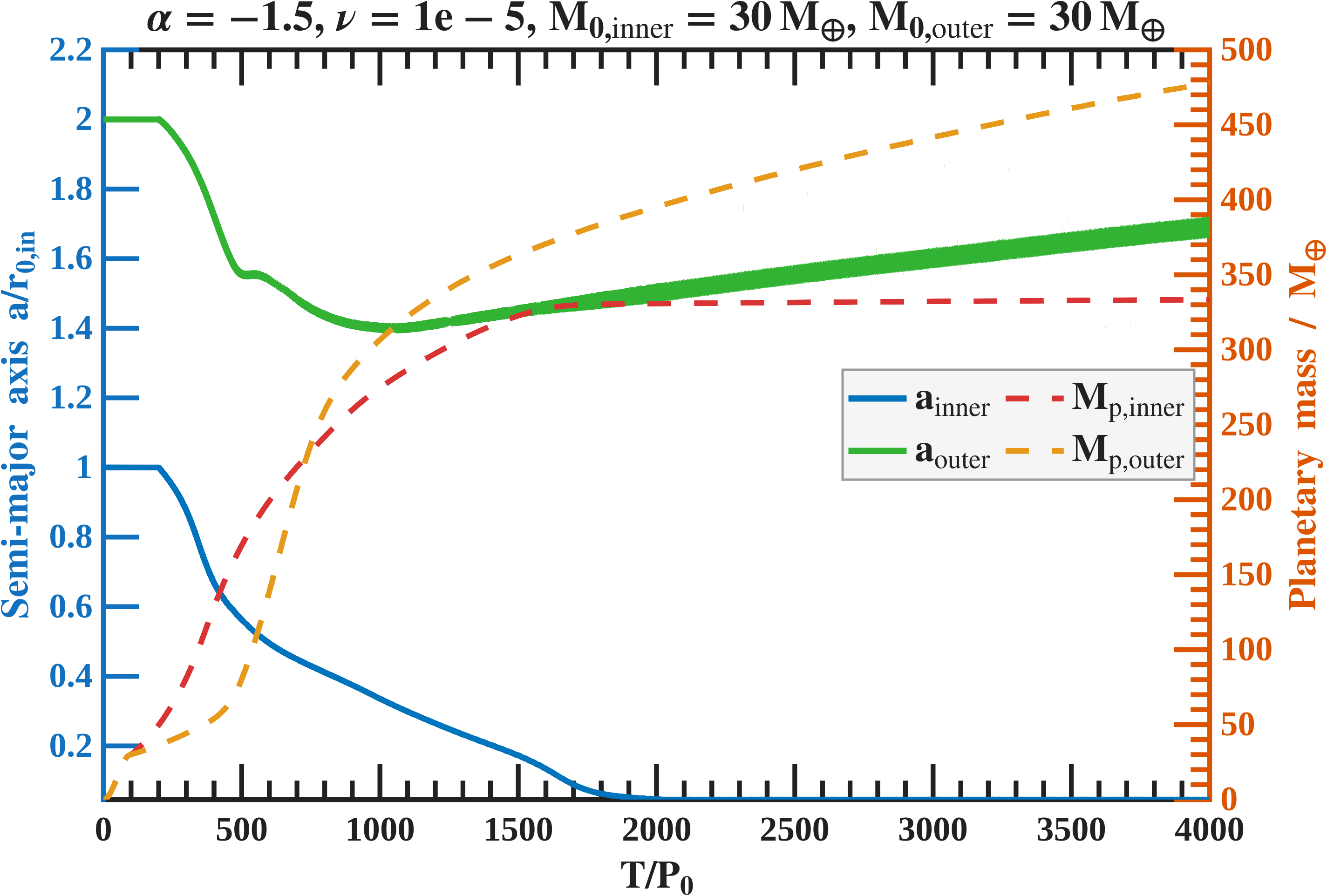}\hfill
\panelimage{c}{.315\textwidth}{0.20\textheight}{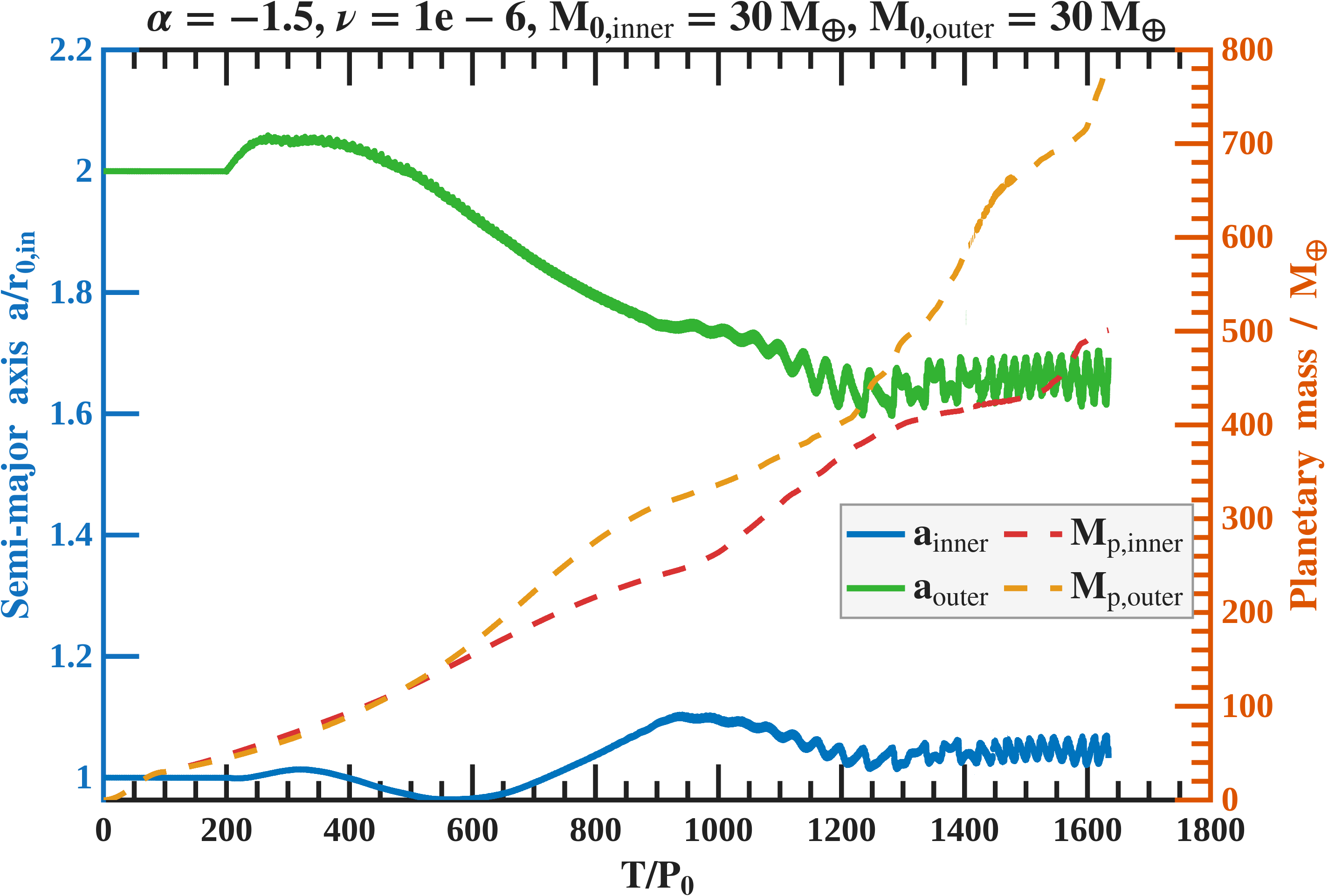}}\par\vspace{0.16cm}
\noindent\makebox[\textwidth][c]{%
\panelimage{d}{.315\textwidth}{0.20\textheight}{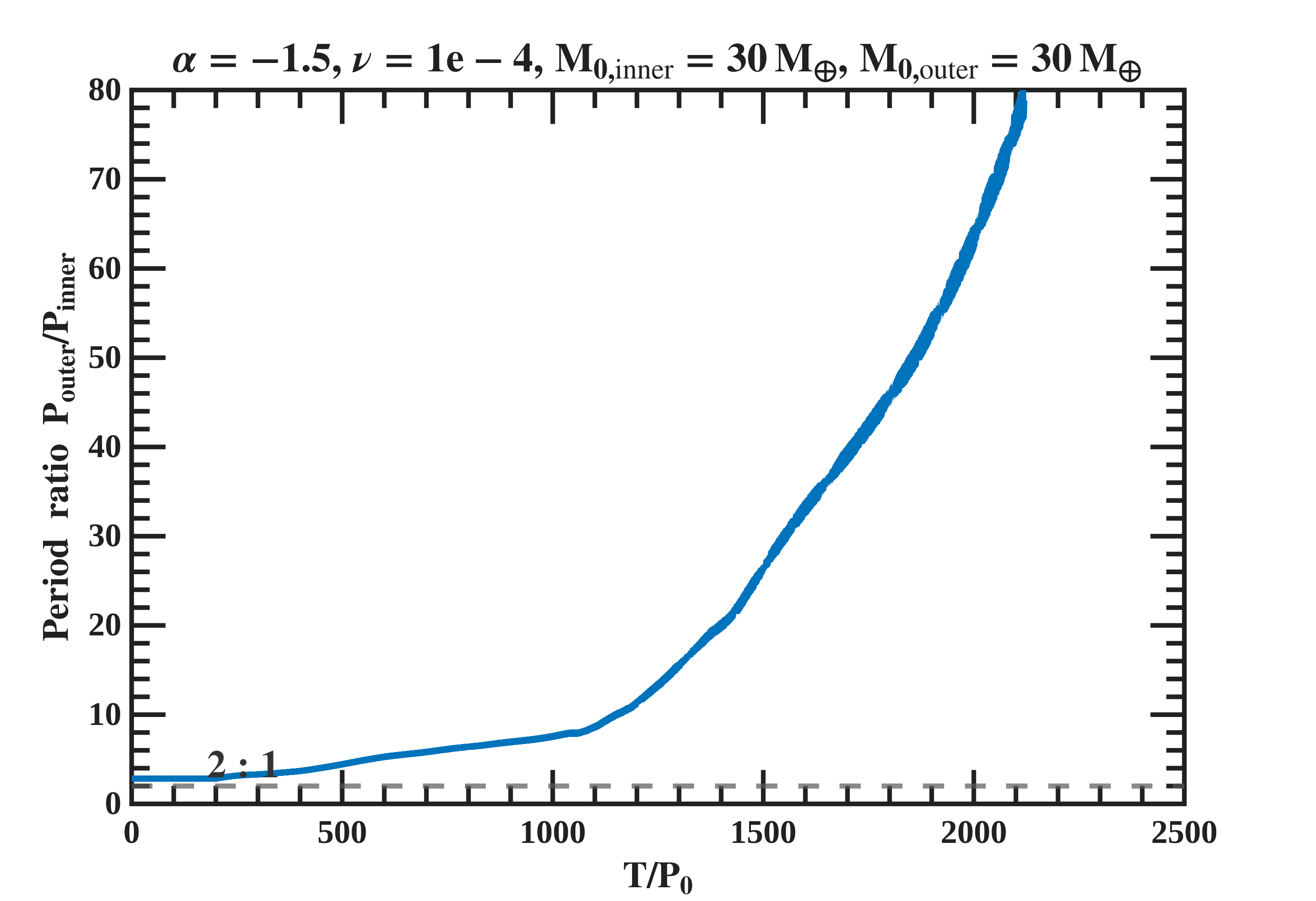}\hfill
\panelimage{e}{.315\textwidth}{0.20\textheight}{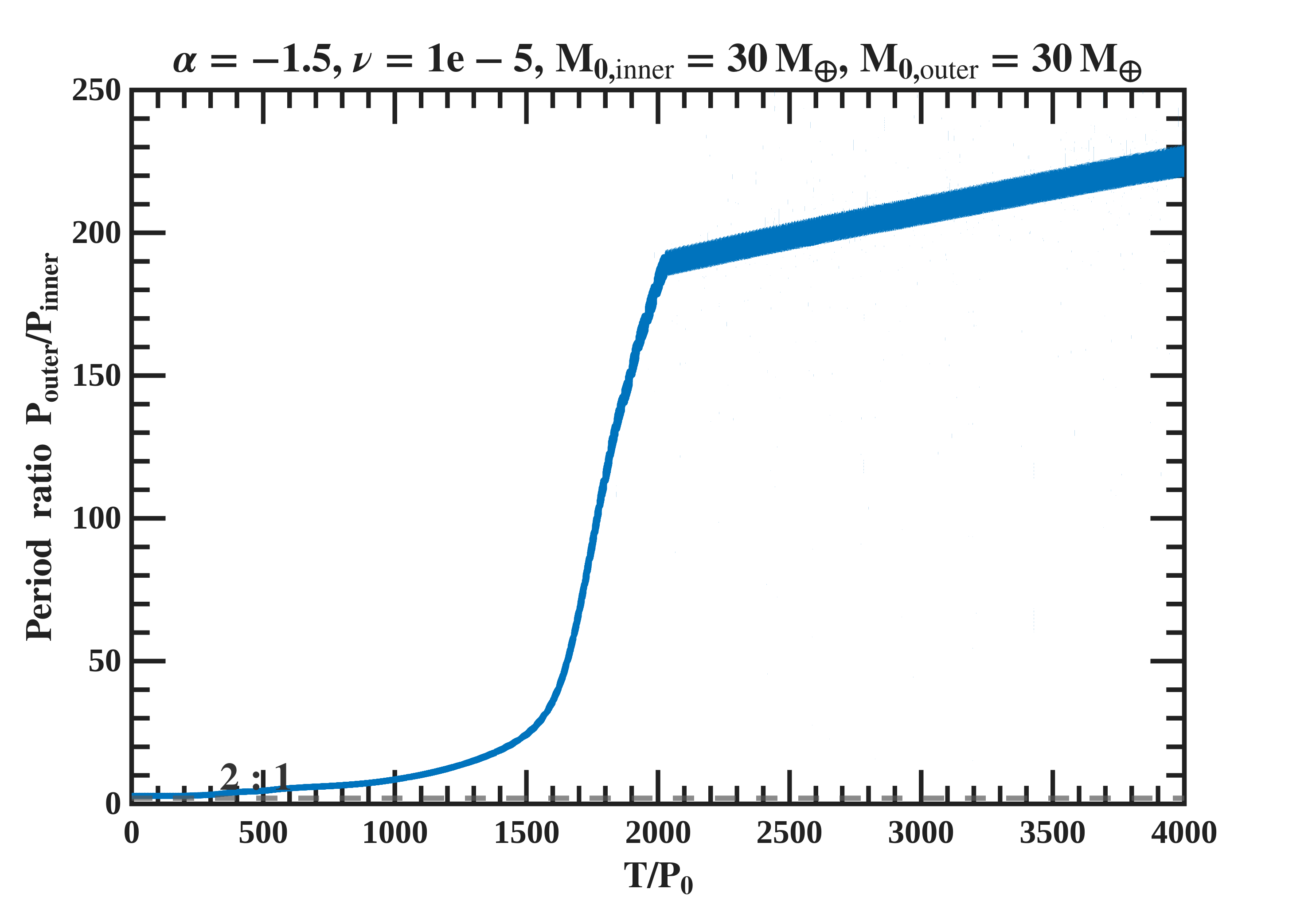}\hfill
\panelimage{f}{.315\textwidth}{0.20\textheight}{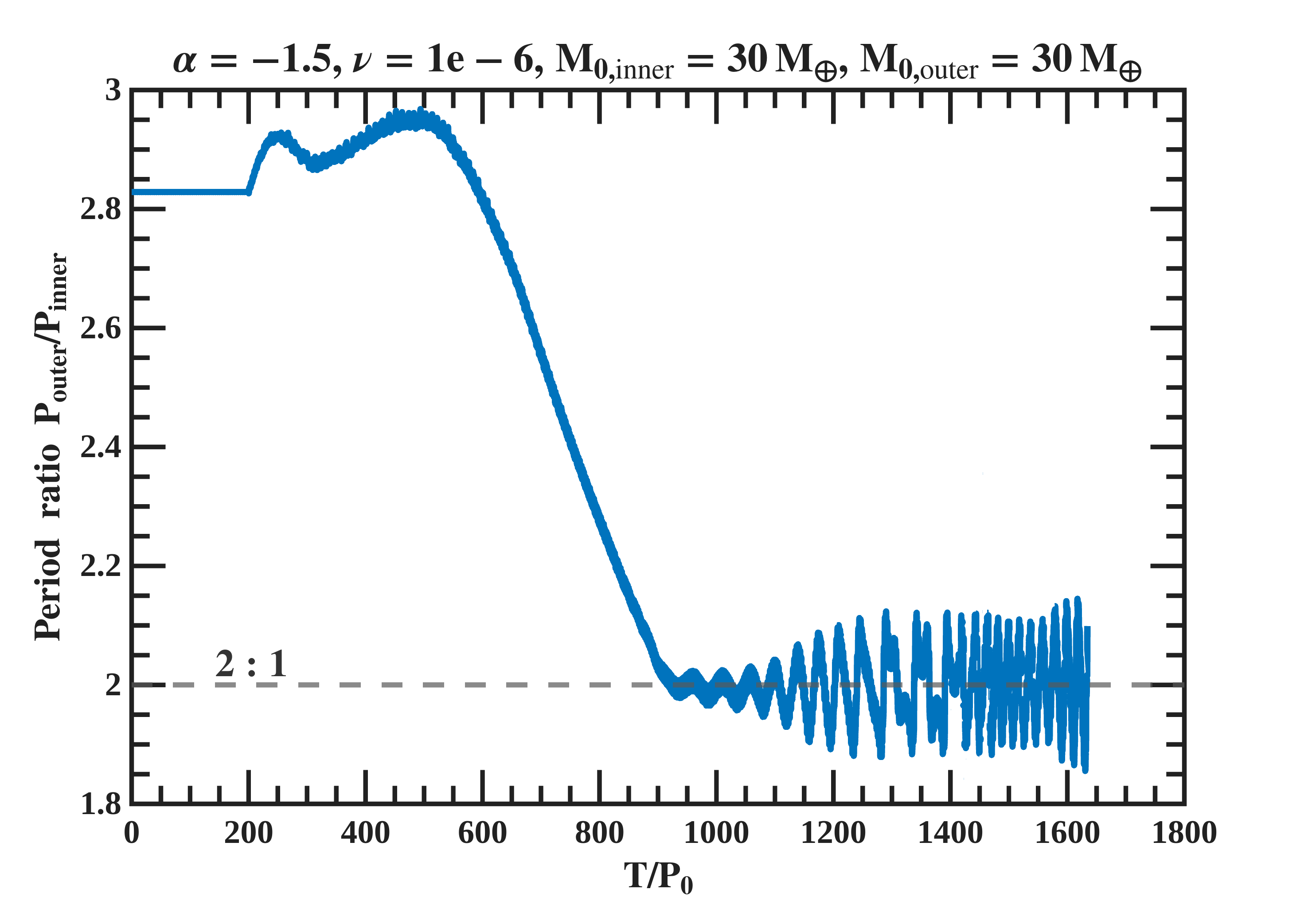}}
\caption{Interacting planet-pair planet simulations with different disk viscosity ($30/30\,M_\oplus$, $\alpha=-1.5$) at three viscosities. Top panels: semi-major axes (solid) and masses (dashed). Bottom panels: period-ratio evolution, with the 2:1 MMR indicated. Higher viscosity favors stronger differential growth and more divergent migration (large period ratios), whereas the lowest viscosity yields convergent migration and capture near 2:1, highlighting viscosity as a control on both accretion asymmetry and resonant outcome.}\label{fig5}
\end{figure*}

Left panel of Figure \ref{fig3} shows the gas accretion histories of single-planet with different initial core-masses with $\nu =10^{-5}$. A clear trend emerges that a more massive core tends to trigger gap opening earlier, which paradoxically suppresses its long-term growth. For example, a 15 $M_{\oplus}$ core undergoes rapid runaway accretion within the first few hundred orbits, but the resulting deep gap strongly reduces the gas supply from the surrounding disk. Consequently, its final mass ends up lower than that of a 10 $M_{\oplus}$ core, which accretes more steadily over a longer timescale. This counter–intuitive outcome highlights the self-limiting nature of rapid gap formation. Similar results have been reported in high–resolution studies of giant planet growth, where gas depletion near the orbit becomes the bottleneck for accretion.

Disk viscosity also plays a decisive role, as shown in  the right panel of Figure \ref{fig3} (core-mass = 15 $M_{\oplus}$). In disks with higher viscosity ($\nu \geq 10^{-5}$), the replenishment of gas into the gap is more efficient, allowing the planet to continue accreting for longer timescales. By contrast, in nearly inviscid disks ($\nu \leq 10^{-6}$), the gap edges become sharp and gas diffusion is strongly suppressed, so the planet's growth halts at Neptune-like masses. These trends are broadly consistent with previous high-resolution simulations and with analytic estimates of the gap opening criterion. We find that the majority of the final mass is acquired after the gap formation stage, highlighting the importance of disk viscosity in regulating long-term accretion efficiency.

In summary, our single-planet experiments confirm the robustness of the modified accretion prescription and clarify the dominant parameters controlling giant planet growth. core-mass determines the timing of gap opening, while disk viscosity regulates the eventual mass reservoir accessible to the planet. These insights provide the necessary baseline for interpreting the more complex behavior of interacting planet pairs discussed in Section \ref{sec:two fixed planets} and Section \ref{sec:two free planets}.

\subsection{Two Fixed Planets Test}\label{sec:two fixed planets}
In order to establish how two neighboring planets interact during their simultaneous growth, we next consider systems where both embryos are released into the disk together. Compared to the single-planet tests, these runs allow us to study not only how each planet accretes from the common disk reservoir, but also how the presence of a companion modifies gap opening and mass evolution.

\subsubsection{Core Mass}
We test the outcomes of simulations with different initial core-masses assigned to the inner and outer planets (Figure \ref{2fixed-mass}). When both cores are identical (e.g., 30 $M_{\oplus}$ each), their growth is relatively symmetric at early times. However, even in these symmetric cases, the outer planet often overtakes the inner one in mass once a common gap forms. This reversal occurs because the inner planet depletes the local inner disk rapidly, while the outer planet continues to accrete fresh material from the outer disk. When the cores start with unequal masses, the asymmetry becomes even more pronounced. A more massive inner core quickly opens a gap and thereby suppresses its own further growth, while the lighter outer core benefits from both the higher gas density at the outer gap edge and the tidal perturbations exerted by its companion. In several runs, we find that the initially lighter outer planet ends up significantly more massive than the inner one, providing a natural route to the reversed mass ratios observed in many resonant giant planet systems. These results suggest that the initial mass hierarchy of planet pairs does not uniquely determine the final outcome. Instead, the competition for gas in the shared disk tends to favor the outer planet, especially once the system enters the resonant and common–gap regime.

\subsubsection{Viscosity}

Since viscous diffusion is the only mechanism that can transport gas across the gap in the absence of migration, this correlation demonstrates that viscosity regulates the coupling between the two planets.

\begin{figure*}[t]
\centering

\noindent
\begin{minipage}[t]{0.32\textwidth}
\centering\includegraphics[width=\linewidth]{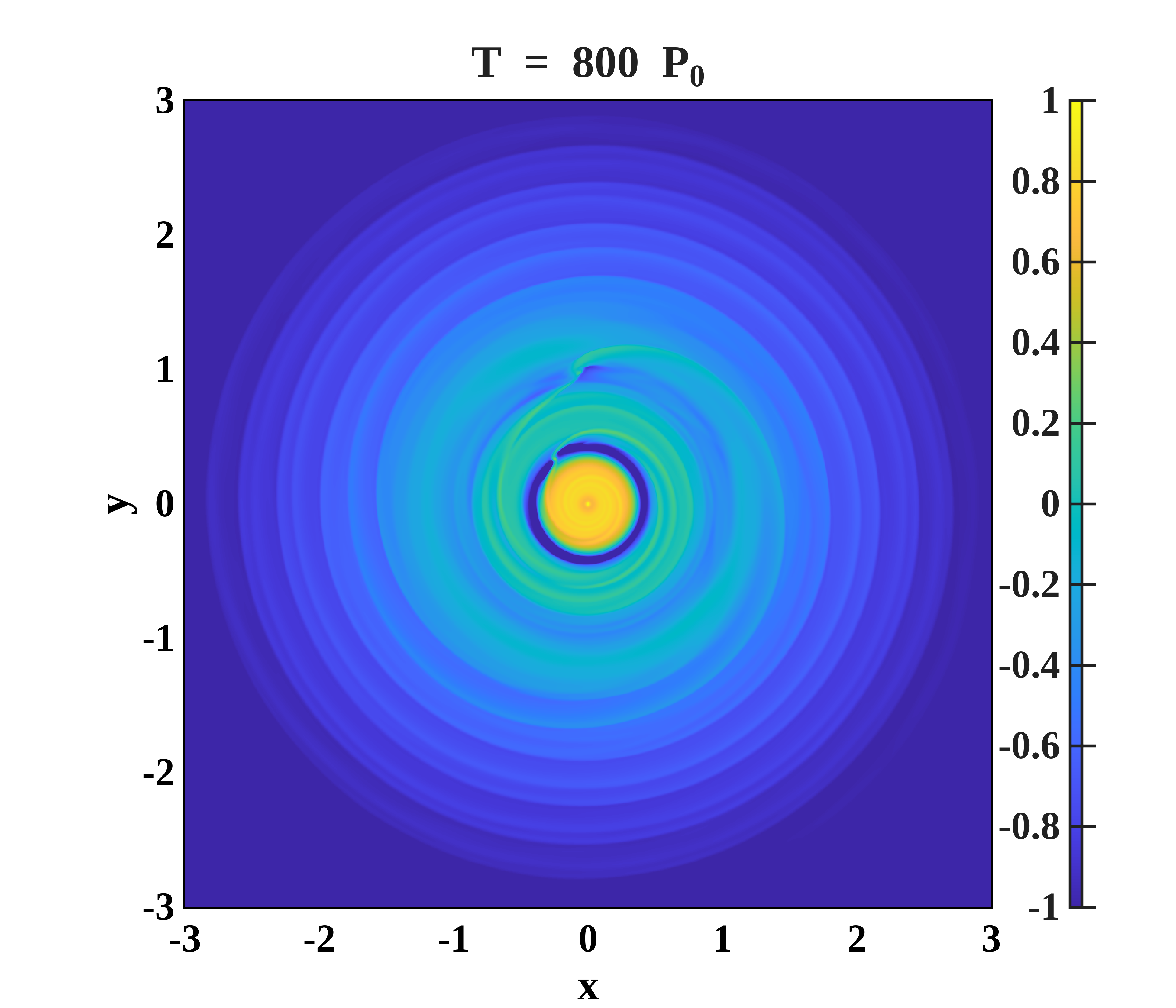}
{\footnotesize\bfseries (a1)}
\end{minipage}
\hfill
\begin{minipage}[t]{0.32\textwidth}
\centering\includegraphics[width=\linewidth]{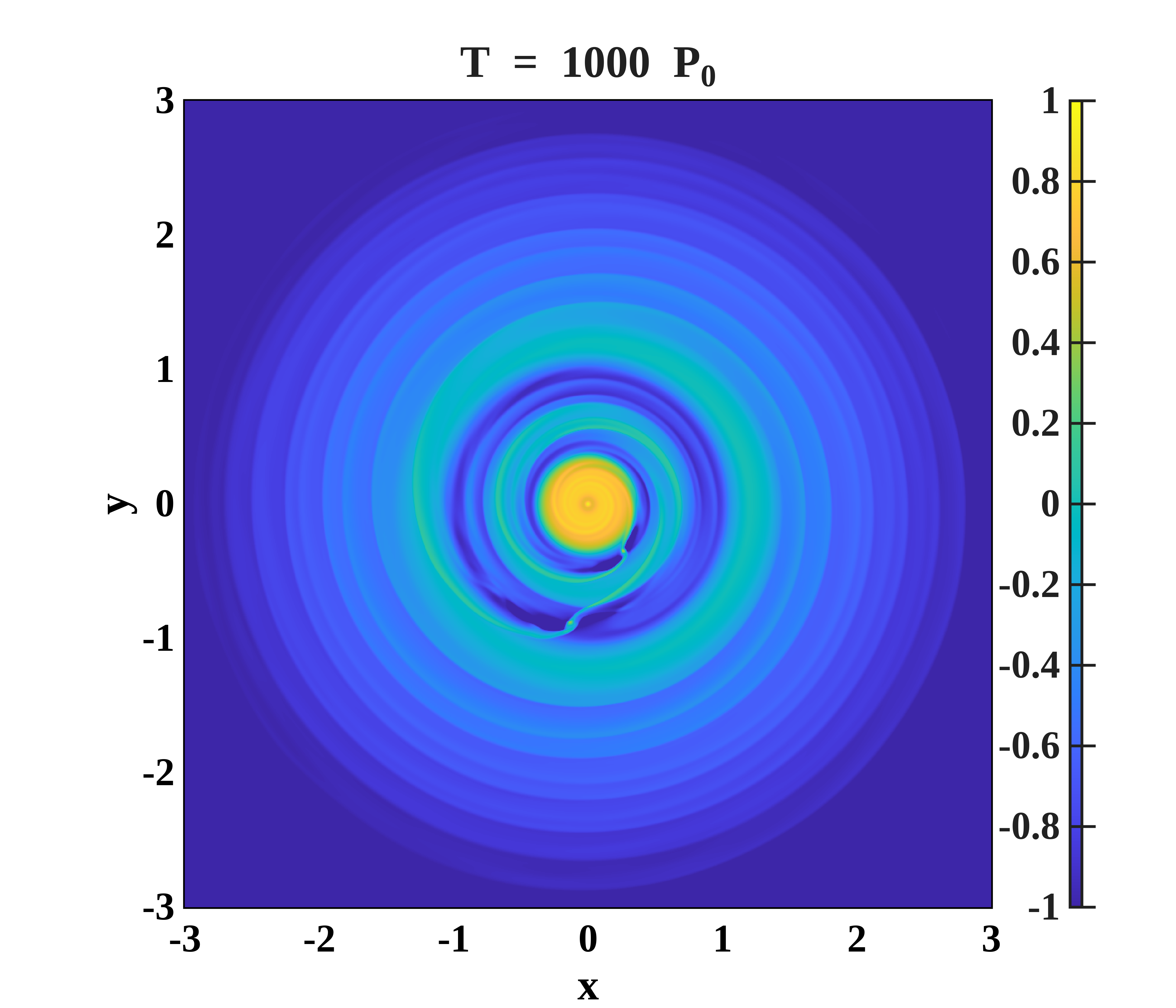}
{\footnotesize\bfseries (b1)}
\end{minipage}
\hfill
\begin{minipage}[t]{0.32\textwidth}
\centering\includegraphics[width=\linewidth]{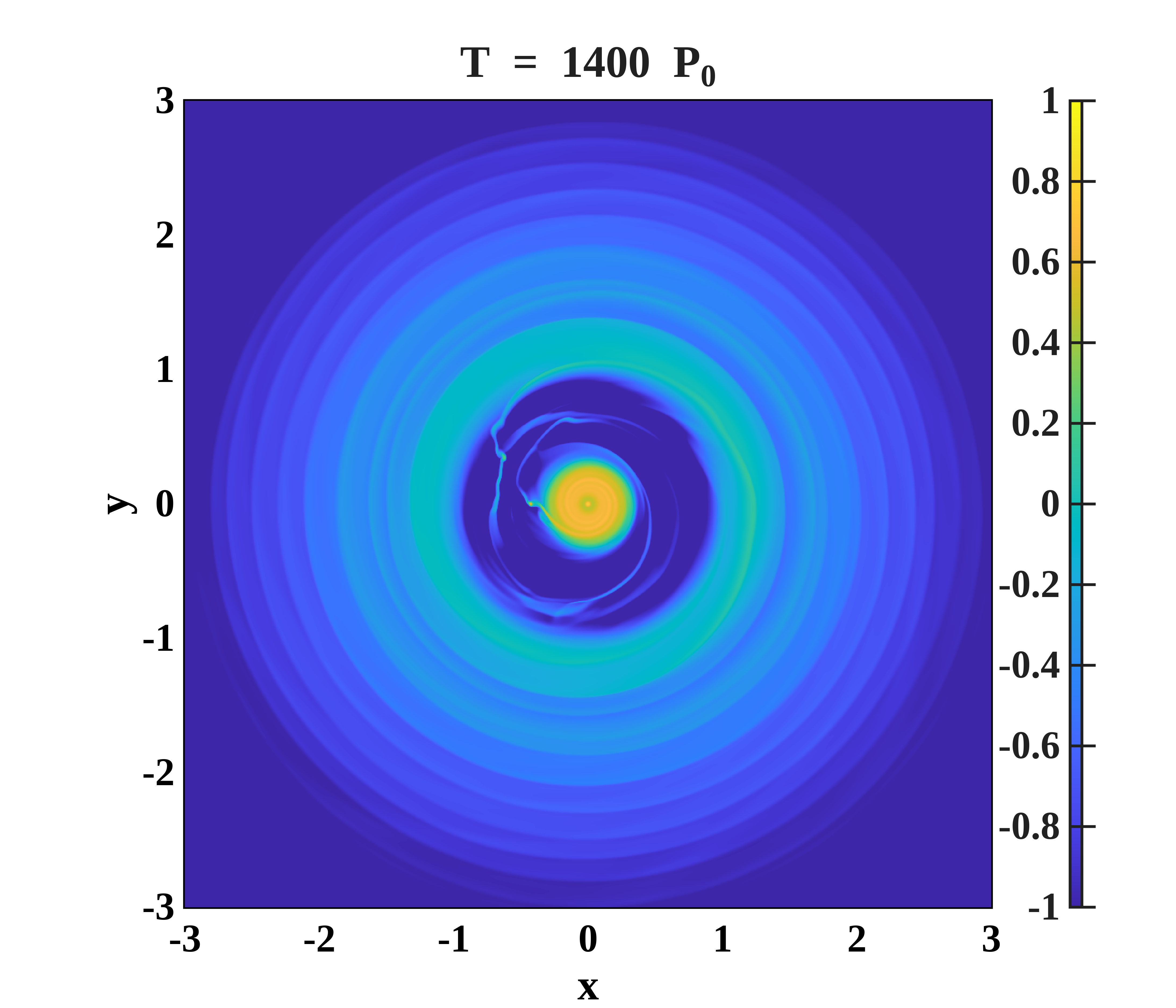}
{\footnotesize\bfseries (c1)}
\end{minipage}
\par
\vspace{3mm}

\noindent
\begin{minipage}[t]{0.32\textwidth}
\centering\includegraphics[width=\linewidth]{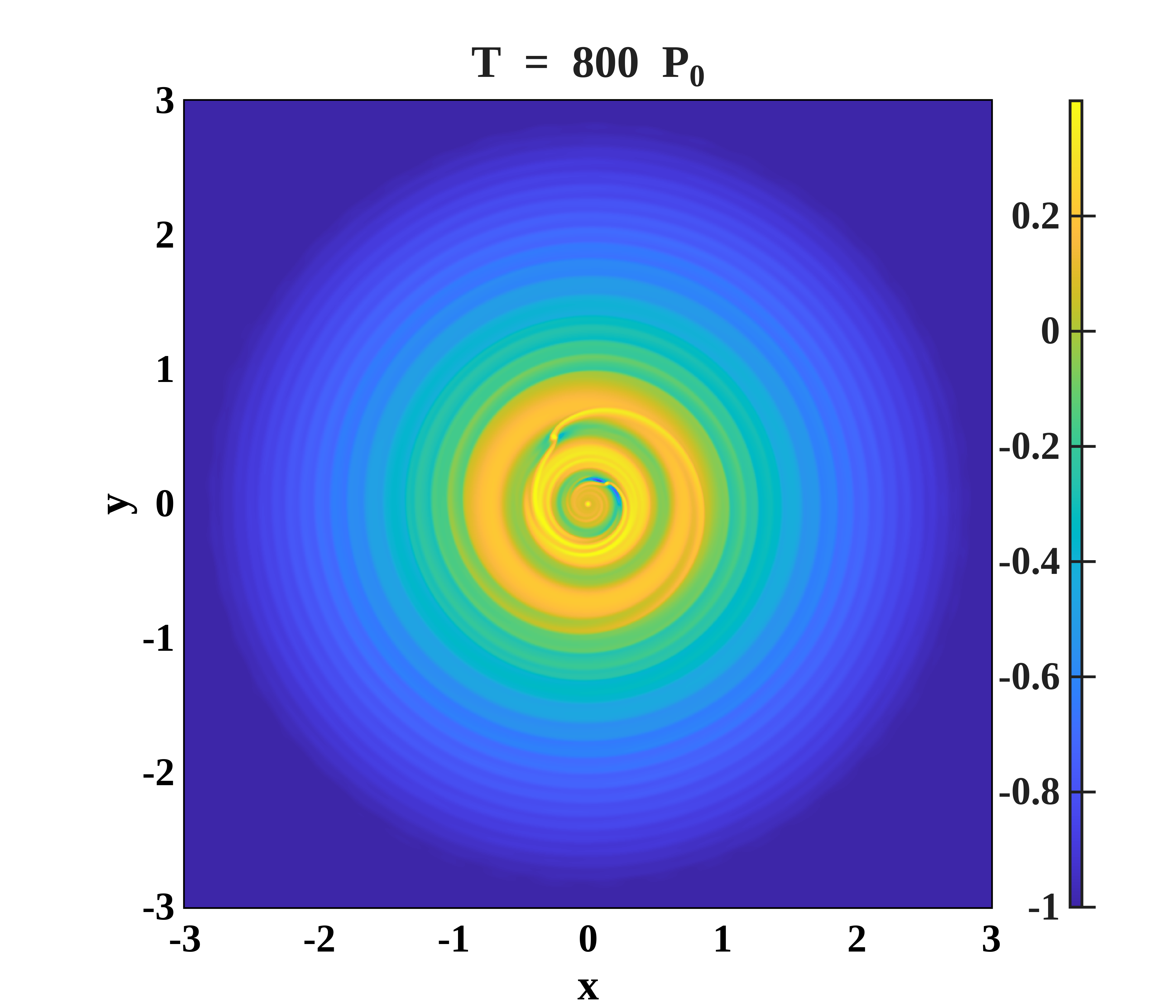}
{\footnotesize\bfseries (a2)}
\end{minipage}
\hfill
\begin{minipage}[t]{0.32\textwidth}
\centering\includegraphics[width=\linewidth]{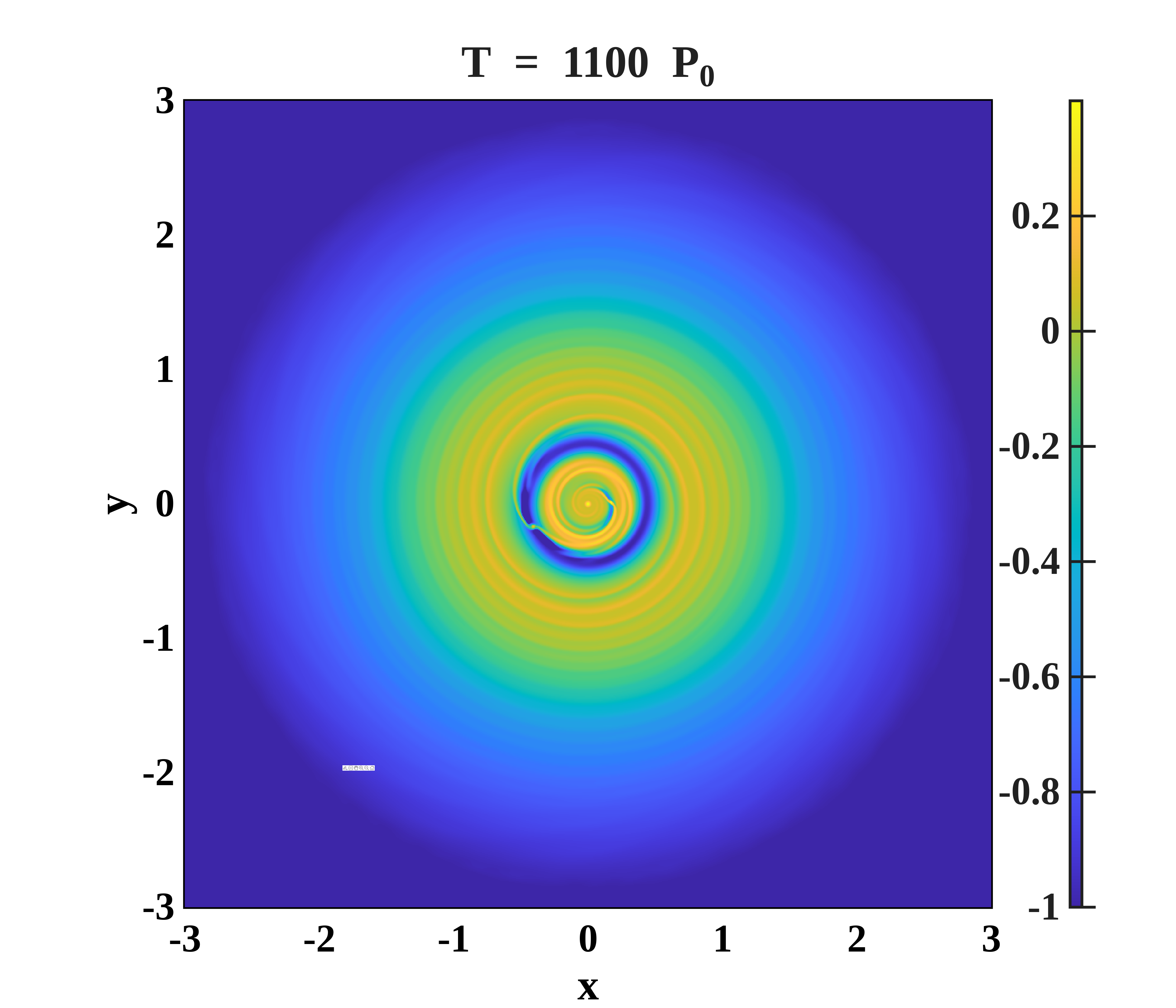}
{\footnotesize\bfseries (b2)}
\end{minipage}
\hfill
\begin{minipage}[t]{0.32\textwidth}
\centering\includegraphics[width=\linewidth]{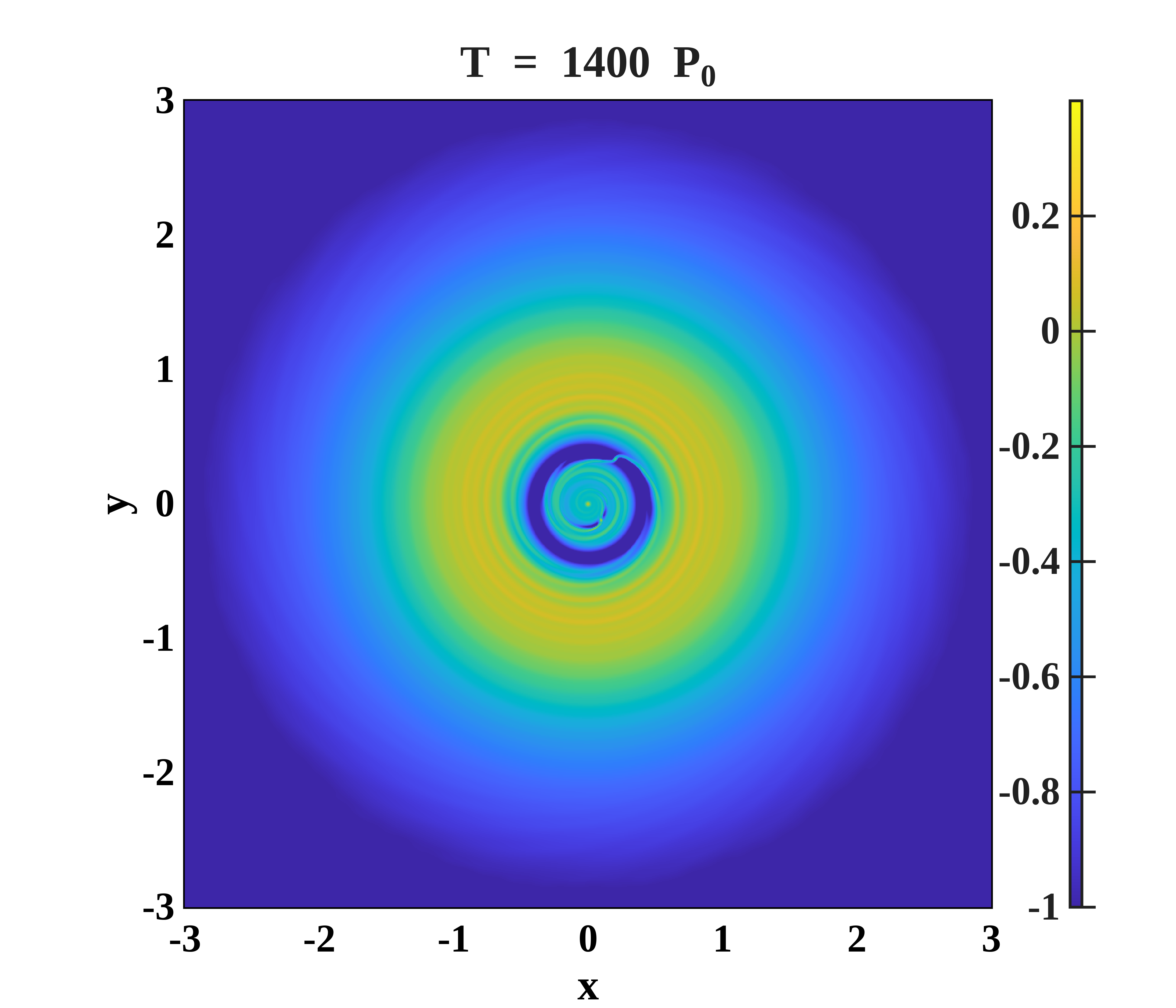}
{\footnotesize\bfseries (c2)}
\end{minipage}
\par
\vspace{3mm}

\noindent
\begin{minipage}[t]{0.46\textwidth}
\centering\includegraphics[width=\linewidth]{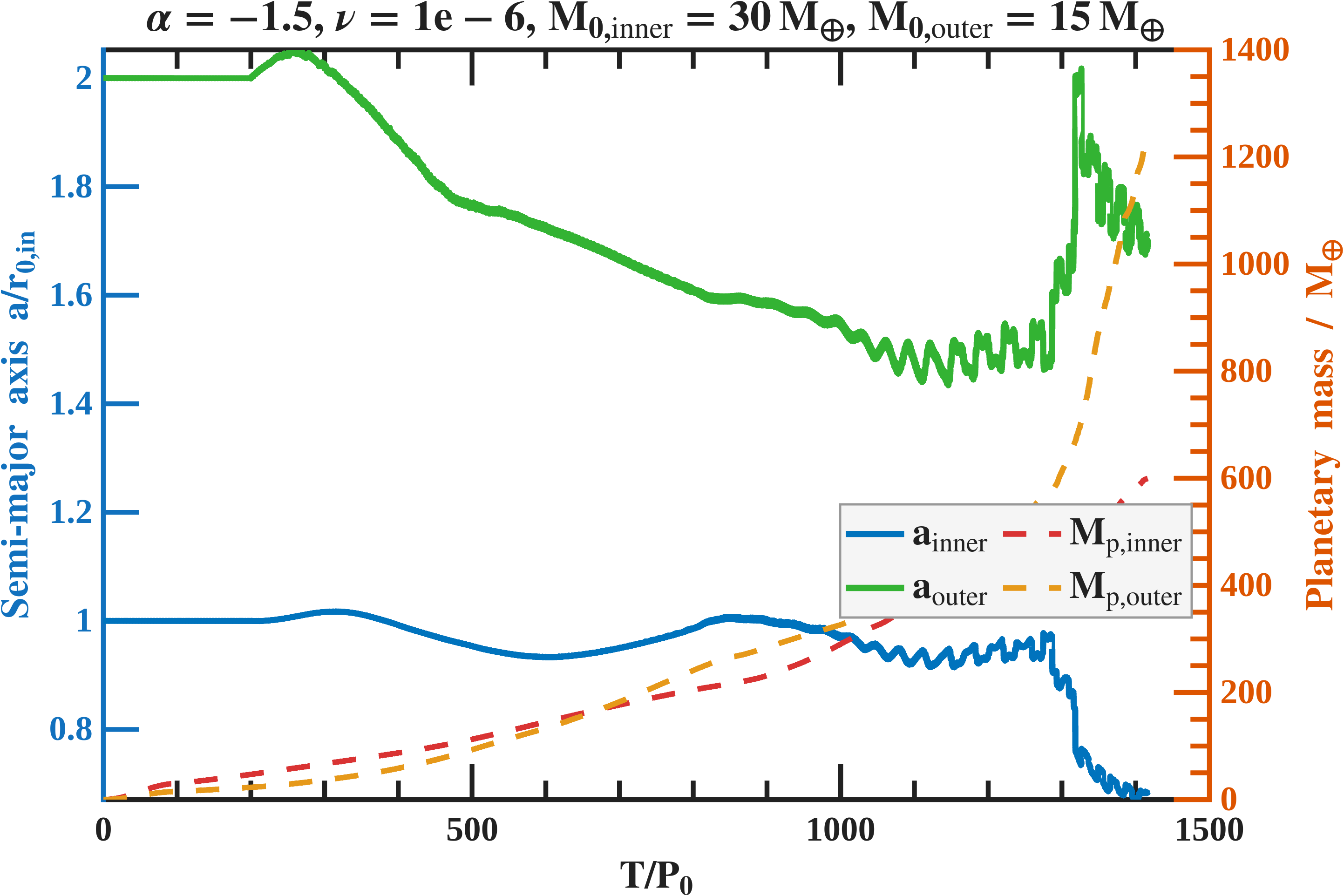}
{\footnotesize\bfseries (d1)}
\end{minipage}
\hfill
\begin{minipage}[t]{0.46\textwidth}
\centering\includegraphics[width=\linewidth]{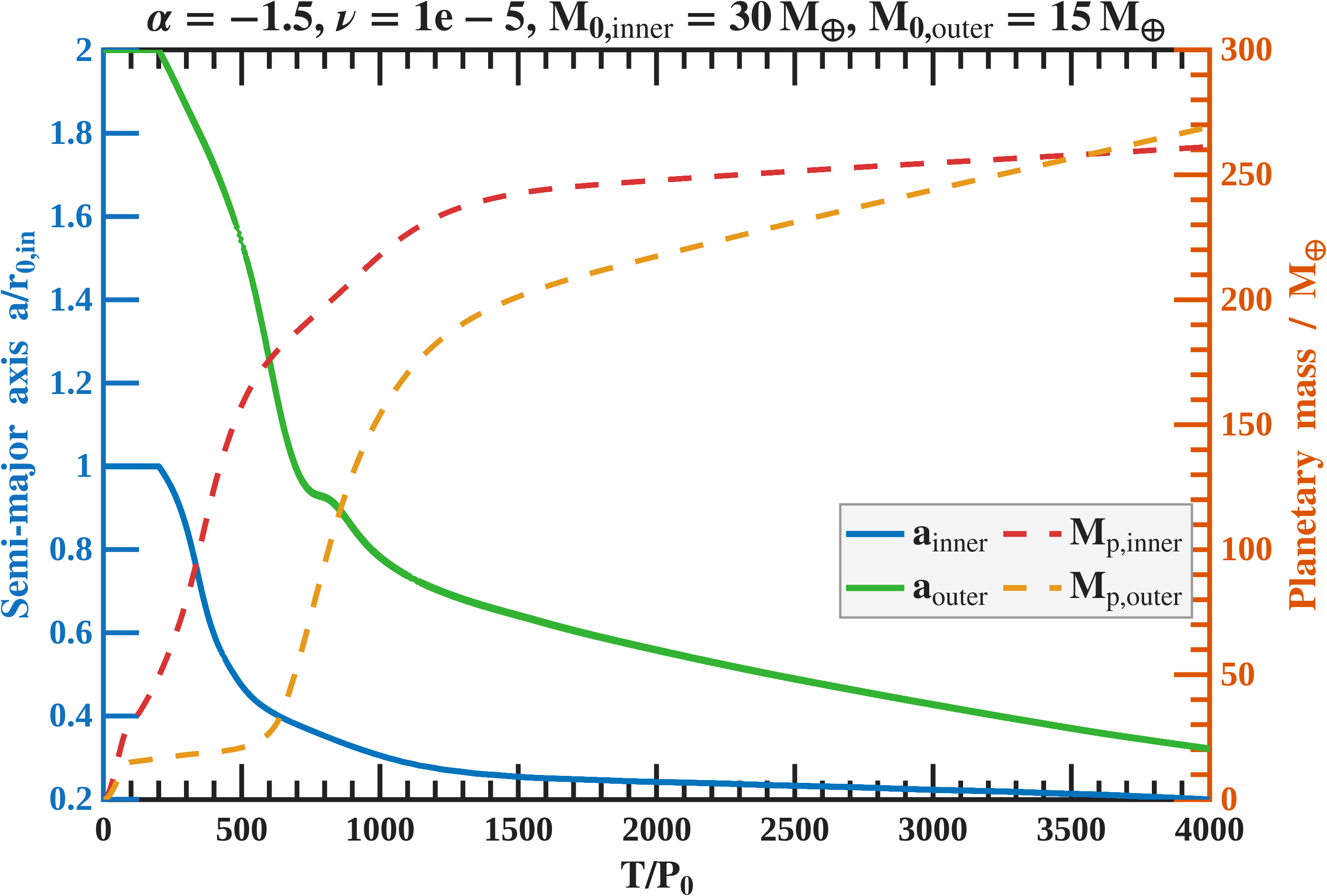}
{\footnotesize\bfseries (d2)}
\end{minipage}
\par
\caption{Orbital evolution and gas accretion of a planet pair embedded in a protoplanetary disk with an initial surface density profile $\Sigma \propto r^{-1.5}$. Panels $a1-c1$ show the gas surface density map for $\nu=10^{-6}$ and $a2-c2$ show the same sequence for $\nu=10^{-5}$, with an inner core of $30 M_\oplus$ and an outer core of $15 M_\oplus$. Panel $d1$ and $d2$ shows the corresponding orbital migration and mass growth of the inner and outer planets. Solid lines indicate the evolution of the semi-major axes, while dashed lines denote the planetary masses. Higher viscosity drives faster inward migration and earlier saturation of gas accretion, whereas lower viscosity preserves sharper spiral structure and can allow a late phase of rapid mass growth for the outer planet.} \label{fig6}
\end{figure*}

\begin{figure*}[!htbp]
\centering
\includegraphics[width=0.78\textwidth,height=0.55\textheight,keepaspectratio]{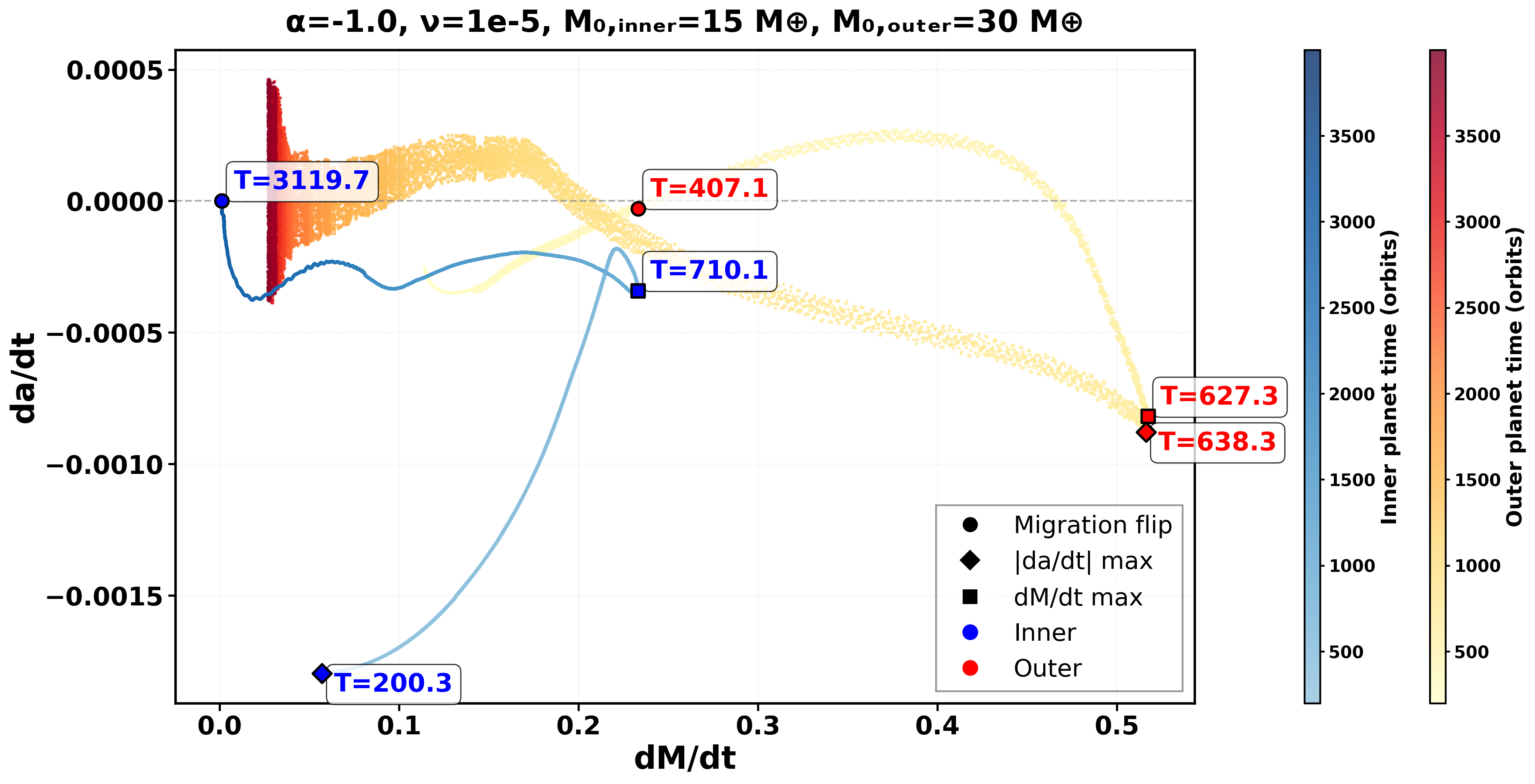}
\caption{Evolution of planetary accretion and migration for the inner (blue) and outer (red) planets. The trajectories are color-coded by time, illustrating the temporal progression of the system. Key evolutionary events are marked, including the migration reversal ($T_{\rm flip}$), the peak migration rate ($T_{\rm peak(a)}$) and the peak accretion rate ($T_{\rm peak(M)}$). The early stage is characterized by inward Type I migration with low accretion rates. As the outer planet undergoes runaway gas accretion, the enhanced corotation torque leads to a reversal of migration direction, producing a phase of outward migration. At later times, the weakening of positive torques and the onset of gap formation drive the system back to inward migration, consistent with a transition toward Type II evolution.}\label{fig7}
\end{figure*}

\begin{figure*}[!htbp]
\centering
\includegraphics[width=0.75\textwidth]{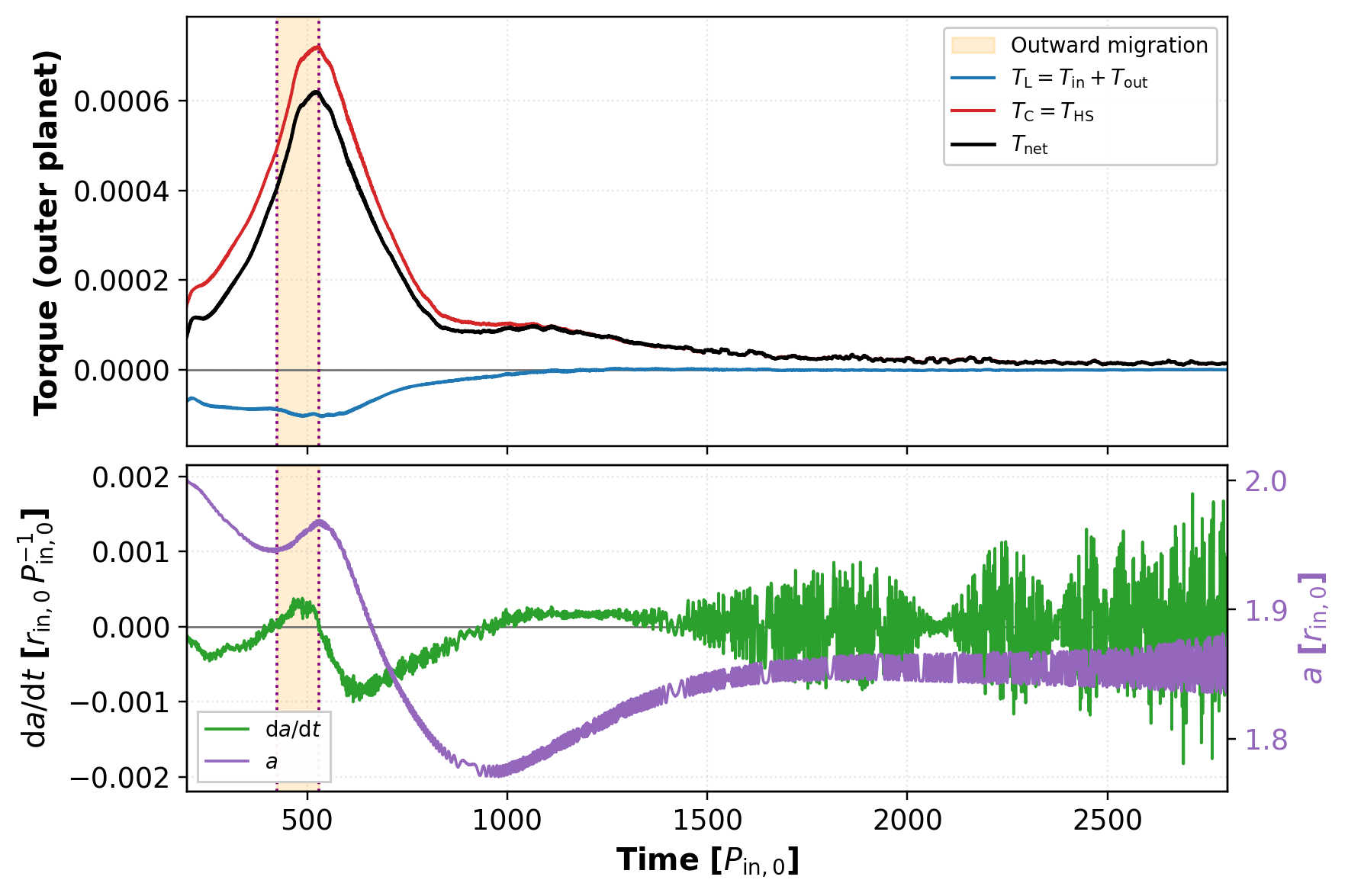}
\caption{Torque decomposition for the outer planet in a representative run ($r^{-1.0}$, $\nu=10^{-5}$, cores $15/30\,M_\oplus$). Torques are split radially about the planet. The horseshoe contribution
$T_{\mathrm{C}}=T_{\mathrm{HS}}$ is integrated over $|r-R_p|<2.5\,R_H$, while $T_{\mathrm{L}}=T_{\mathrm{in}}+T_{\mathrm{out}}$ sums the torque from gas inside and outside this band (a proxy for Lindblad torques). Top: smoothed $T_{\mathrm{L}}$, $T_{\mathrm{C}}$, and $T_{\mathrm{net}}=T_{\mathrm{L}}+T_{\mathrm{C}}$. Bottom: smoothed $\mathrm{d}a/\mathrm{d}t$ and semimajor axis $a$. The shaded interval marks outward migration; during this phase $T_{\mathrm{C}}$ dominates $T_{\mathrm{L}}$, consistent with a corotation-driven migration reversal.}\label{fig:torque_decomp}
\end{figure*}	

The two planets tests highlight two key mechanisms that differ from the single-planet scenario. Firstly, once the two planets gaps merge into a common gap, the accretion geometry is fundamentally altered. Secondly, the outer planet enjoys a systematic advantage in gas supply. The relative importance of these effects depends on the disk viscosity, which determines whether accretion proceeds in a cooperative or competitive fashion. These insights are important for understanding the full orbital and mass evolution of freely migrating planet pairs, as discussed next in Section \ref{sec:two free planets}.

\subsection{Interacting Planet-Pair}\label{sec:two free planets}
We now examine the fully interacting case in which both planets migrate and accrete simultaneously (following the single-planet and fixed-orbit two planets tests above). This scenario provides the most direct comparison to observed resonant giant planet pairs.

\subsubsection{Orbital Evolutions}
We observe inward convergent migration of the two planets, ultimately resulting in capture into the 2:1 MMR (see Figure \ref{fig4} and Figure \ref{fig5}). The convergent nature of the migration arises because the outer planet migrates inward more rapidly than the inner one, as a result of differential migration rates driven by torques. Such convergence is a generic outcome of disk-driven migration and has been widely observed in hydrodynamical simulations of planet--disk interactions \citep{Kley2000, LeePeale2002, Kley2004, PapaloizouTerquem2006}, which can occur under either Type I or Type II, depending on the planet masses and the local disk structure. When the outer planet catches up with the inner one, their mutual gravitational interactions may lead to MMRs capture or further dynamical evolution. Once the resonance is established, the two planets migrate inward as a pair, maintaining the commensurability throughout the remainder of the disk lifetime. Figure \ref{fig6} illustrates the coupled growth and migration of a planet pair embedded in a proto planetary disk. As the planets accrete material, they carve gaps in the surrounding gas, which progressively deepen and eventually merge. The spiral density waves launched by the two planets overlap at later times, highlighting the strong dynamical interactions with the disk. Meanwhile, the semi-major axes of both planets decrease due to disk-driven migration, while their masses steadily increase (see Figure \ref{fig4} panel b).

The migration of the planet pair is closely coupled to their gas accretion. Figure \ref{fig7} shows the relationship between the planetary migration rate and the gas accretion rate measured in our simulations. The trajectories reveal a sequence of evolutionary stages. At early times, both planets reside in the low mass regime and undergo inward migration ($\mathrm{d}a/\mathrm{d}t < 0$), consistent with Type I migration. During this stage, accretion remains relatively inefficient, and the torque balance is only weakly perturbed by mass growth. The inner planet quickly settles into a quasi-steady state, while the outer planet begins to experience a gradual increase in accretion rate as it interacts more strongly with the surrounding disk gas. As the outer planet grows, its accretion rate rises sharply and reaches a maximum at $T_{\rm peak(M)}$, marking the onset of a runaway gas accretion phase. This rapid increase in mass modifies the local disk structure, enhancing the corotation torque and partially saturating the Lindblad torque. As a result, the net torque acting on the planet changes sign, leading to a migration reversal at $T_{\rm flip}$, where $\mathrm{d}a/\mathrm{d}t$ becomes positive \citep{Li2024, Pan2026, Ida2026}. This transition reflects a shift from standard inward Type I migration to a regime dominated by positive corotation torque. Following this transition, the system enters a migration-dominated phase in which the outer planet undergoes sustained outward migration while continuing to accrete mass at a decreasing rate. The trajectory in phase space bends toward lower $\mathrm{d}a/\mathrm{d}t$, indicating that the positive torque gradually weakens, likely due to the depletion of co-orbital gas and the onset of partial gap formation. At later times, the accretion rate declines significantly, and both planets migrate inward again ($\mathrm{d}a/\mathrm{d}t < 0$). This behavior is consistent with the transition toward gap-opening conditions, where the torque balance becomes dominated by disk viscosity and global angular momentum transport. In this regime, the outer planet exhibits a sharp inward migration, resembling Type II migration, while the inner planet remains in a slower, torque-regulated evolution. Overall, the phase-space evolution highlights the tight coupling between gas accretion and migration torques.

Figure~\ref{fig:torque_decomp} quantifies the torque by showing a decomposition acting on the outer planet for the same case as in Figure~\ref{fig7} ($r^{-1.0}$, $\nu=10^{-5}$, cores $15/30\,M_\oplus$). Following the diagnostic implemented in our code, we integrate the torque separately within a horseshoe band of half-width $2.5\,R_H$ ($T_{\mathrm{C}}=T_{\mathrm{HS}}$) and outside that band ($T_{\mathrm{L}}=T_{\mathrm{in}}+T_{\mathrm{out}}$). We also compare their sum with the migration rate $\mathrm{d}a/\mathrm{d}t$. During the temporary outward-migration interval, the horseshoe contribution
dominates the exterior Lindblad-like torque ($\langle T_{\mathrm{C}}\rangle/|\langle T_{\mathrm{L}}\rangle|\sim6.6$), so that $T_{\mathrm{net}}$ remains positive while $a$ increases.
This supports the interpretation that the enhanced corotation torque arising during rapid gas accretion can temporarily overcome the (partially saturated) Lindblad torque and reverse the migration
direction.

\begin{figure*}[!htbp]
\centering
\includegraphics[width=0.75\textwidth]{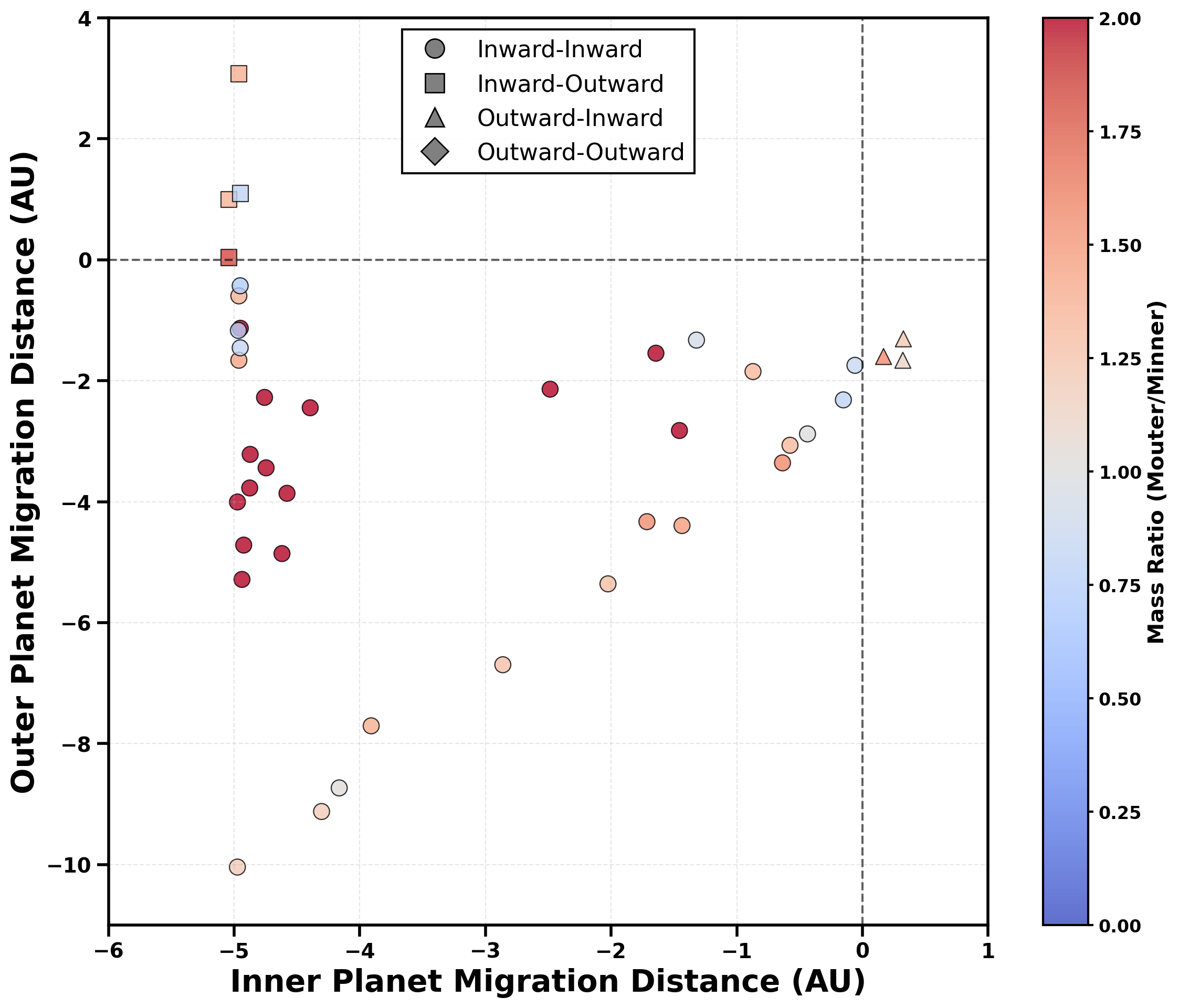}
\caption{Distribution of migration distances for the inner and outer planet pairs across varying mass ratios. The x-axis and y-axis denote the migration distance for the inner and outer planets, respectively. The color scale represents the mass ratio ($M_{outer}/M_{inner}$), with redder hues indicating a more massive outer planet relative to the inner one. Dashed lines indicate the zero-migration axes. The clustering at $-5$ AU highlights a sensitive dynamical threshold where outer planet trajectories diverge significantly, while the diagonal trend suggests coupled migration regimes. Systems with the extreme inward remain dynamically safe during the gas-disk phase either through MMR capture (convergent cases)or through an increasing period ratio (divergent cases).}\label{fig8}
\end{figure*}

Figure \ref{fig8} illustrates the relationship between the inner and outer planet migration distance. Each data point is color-coded according to the mass ratio ($M_{outer}/M_{inner}$). Over 85\% of systems exhibit inward migration for both the inner and outer planets, while no system in our simulation shows simultaneous outward migration of the two planetary bodies.

The extreme inward-migration cases in Figure~\ref{fig8} raise the question of whether such pairs remain from close encounters while gas is still present. In our sample, the answer depends on the migration. For convergent systems that occupy the strongly inward-migrating region of Figure~\ref{fig8}, capture into a low-order mean-motion resonance provides resonant protection. A minimum orbital separation is maintained once the resonant angles begin to librate and orbit crossing is prevented throughout the gas-disk phase. By contrast, many systems with comparably large inward are divergent and never enter an MMR. In those cases, the period ratio continues to increase, so close encounters are avoided by orbital divergence rather than by resonant locking. Thus, the extreme inward-migration cluster in Figure~\ref{fig8} does not imply imminent mergers that convergent members are safeguarded by resonance capture, while divergent members remain separated by an expanding period ratio.

\subsubsection{Final Mass Ratio and Resonance Capture}
\begin{figure*}[!htbp]
\centering
\includegraphics[width=0.75\textwidth]{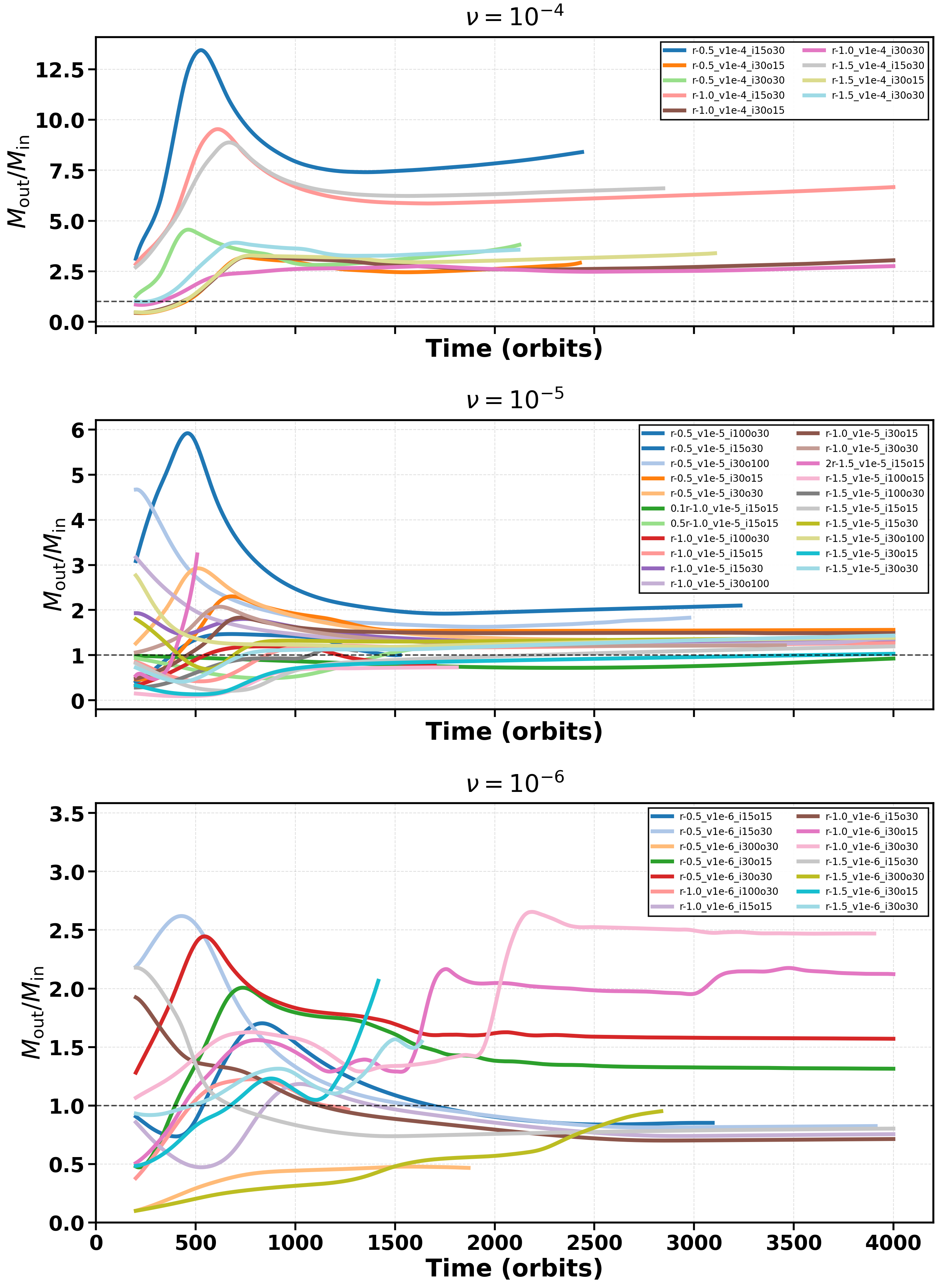}
\caption{Temporal evolution of the planet mass ratio across different viscosity regimes. The panels show the evolution of $M_{outer}/M_{inner}$ as a function of time (in orbits) for $\nu = 10^{-4}$ (top), $\nu = 10^{-5}$ (middle), and $\nu = 10^{-6}$ (bottom). In the high-viscosity regime, rapid differential accretion leads to early and sustained mass reversal. In the lower viscosity cases, the system exhibits more complex stochastic behavior during the first $1000$ orbits before settling toward a quasi-steady state, with the final mass ratio depending on the disk profile $\alpha$ and initial core-masses.}\label{fig9}
\end{figure*}	

\begin{table*}[htbp]
\centering
\caption{Summary of two planets simulations with $\nu = 10^{-4}$. $\Delta a_{\mathrm{outer}}$ and $\Delta a_{\mathrm{inner}}$ are the total migration distances. Negative (positive) values indicate inward (outward) migration. With $h = 0.04$, the kinematic viscosity $\nu = 10^{-4}$ corresponds to an effective Shakura--Sunyaev $\alpha \equiv \nu / h^2 \approx 0.0625.$}
\label{tab1}
\setlength{\tabcolsep}{2pt}
\begin{tabular}{c c c c c c c c}
\hline
{\bf Run} & {\bf $Core_{outer}/Core_{inner}$ ($M_\oplus$)} & {\bf $\Sigma$} & {\bf Migration} & {\bf MMR} & {\bf final $M_{outer}/M_{inner}$ ($M_\oplus$)} & {\bf $\Delta a_{outer}$ (AU)} & {\bf $\Delta a_{inner}$ (AU)} \\
\hline
1 & $15/30$ & $\Sigma_0 r^{-0.5}$ & Convergent & 3:1 & 1589:535 & $-5.29$ & $-4.94$ \\
2 & $30/15$ & $\Sigma_0 r^{-0.5}$ & Divergent & - & 2520:300 & $-3.77$ & $-4.88$ \\
3 & $30/30$ & $\Sigma_0 r^{-0.5}$ & Divergent & - & 1902:497 & $-4.00$ & $-4.97$ \\
\hline
4 & $15/30$ & $\Sigma_0 r^{-1.0}$ & Divergent & - & 1730:568 & $-4.72$ & $-4.92$ \\
5 & $30/15$ & $\Sigma_0 r^{-1.0}$ & Divergent & - & 2115:317 & $-3.86$ & $-4.58$ \\
6 & $30/30$ & $\Sigma_0 r^{-1.0}$ & Divergent & - & 1763:640 & $-3.44$ & $-4.75$ \\
\hline
7 & $15/30$ & $\Sigma_0 r^{-1.5}$ & Divergent & - & 1734:511 & $-3.22$ & $-4.87$ \\
8 & $30/15$ & $\Sigma_0 r^{-1.5}$ & Divergent & - & 1907:288 & $-2.45$ & $-4.39$ \\
9 & $30/30$ & $\Sigma_0 r^{-1.5}$ & Divergent & - & 1727:484 & $-2.28$ & $-4.76$ \\
\hline
\end{tabular}

\end{table*}

\begin{table*}[htbp]
\centering
\caption{Summary of two planets simulations with $\nu = 10^{-5}$. $\Delta a_{\mathrm{outer}}$ and $\Delta a_{\mathrm{inner}}$ are the total migration distances. Negative (positive) values indicate inward (outward) migration.}
\label{tab2}
\setlength{\tabcolsep}{2pt}
\begin{tabular}{c c c c c c c c}
\hline
{\bf Run} & {\bf $Core_{outer}/Core_{inner}$ ($M_\oplus$)} & {\bf $\Sigma$} & {\bf Migration} & {\bf MMR} & {\bf final $M_{outer}/M_{inner}$ ($M_\oplus$)} & {\bf $\Delta a_{outer}$ (AU)} & {\bf $\Delta a_{inner}$ (AU)} \\
\hline
1 & $15/30$ & $\Sigma_0 r^{-0.5}$ & Convergent & - & 798:511 & $-4.33$ & $-1.71$ \\
2 & $30/15$ & $\Sigma_0 r^{-0.5}$ & Divergent & - & 597:284 & $-1.13$ & $-4.95$ \\
3 & $30/30$ & $\Sigma_0 r^{-0.5}$ & Convergent & - & 795:593 & $-1.85$ & $-0.87$ \\
4 & $30/100$ & $\Sigma_0 r^{-0.5}$ & Convergent & 2:1 & 564:567 & $-2.88$ & $-0.44$ \\
5 & $100/30$ & $\Sigma_0 r^{-0.5}$ & Divergent & - & 796:435 & $0.04$ & $-5.04$ \\
\hline
6 & $15/15$ & $\Sigma_0 r^{-1.0}$ & Convergent & 2:1 & 288:226 & $-6.70$ & $-2.86$ \\
7 & $15/15$ & $0.5\,\Sigma_0 r^{-1.0}$ & Convergent & 2:1 & 184:142 & $-5.36$ & $-2.02$ \\
8 & $15/15$ & $0.1\,\Sigma_0 r^{-1.0}$ & Convergent & - & 37:40 & $-1.33$ & $-1.32$ \\
9 & $15/30$ & $\Sigma_0 r^{-1.0}$ & Convergent & 2:1 & 573:385 & $-4.39$ & $-1.44$ \\
10 & $30/15$ & $\Sigma_0 r^{-1.0}$ & Divergent & - & 454:335 & $-0.60$ & $-4.96$ \\
11 & $30/30$ & $\Sigma_0 r^{-1.0}$ & Convergent & -& 743:611 & $-1.31$ & $0.33$ \\
12 & $30/100$ & $\Sigma_0 r^{-1.0}$ & Convergent & 2:1 & 529:658 & $-2.32$ & $-0.15$ \\
13 & $100/30$ & $\Sigma_0 r^{-1.0}$ & Divergent & - & 637:467 & $1.00$ & $-5.04$ \\
\hline
14 & $15/15$ & $\Sigma_0 r^{-1.5}$ & Convergent & 2:1 & 251:210 & $-9.12$ & $-4.30$ \\
15 & $15/15$ & $2\,\Sigma_0 r^{-1.5}$ & Convergent & 3:1 & 1409:433 & $-4.86$ & $-4.62$ \\
16 & $15/30$ & $\Sigma_0 r^{-1.5}$ & Convergent & - & 269:261 & $-8.73$ & $-4.16$ \\
17 & $30/15$ & $\Sigma_0 r^{-1.5}$ & Convergent & - & 343:247 & $-7.71$ & $-3.91$ \\
18 & $30/30$ & $\Sigma_0 r^{-1.5}$ & Divergent & - & 478:333 & $-1.66$ & $-4.96$ \\
19 & $30/100$ & $\Sigma_0 r^{-1.5}$ & Convergent & 2:1 & 727:653 & $-1.66$ & $0.32$ \\
20 & $100/30$ & $\Sigma_0 r^{-1.5}$ & Divergent & - & 490:352 & $3.07$ & $-4.96$ \\
\hline
\end{tabular}
\end{table*}

\begin{table*}[htbp]
\centering
\caption{Summary of two planets simulations with $\nu = 10^{-6}$. $\Delta a_{\mathrm{outer}}$ and $\Delta a_{\mathrm{inner}}$ are the total migration distances. Negative (positive) values indicate inward (outward) migration.}
\label{tab3}
\setlength{\tabcolsep}{2pt}
\begin{tabular}{c c c c c c c c}
\hline
{\bf Run} & {\bf $Core_{outer}/Core_{inner}$ ($M_\oplus$)} & {\bf $\Sigma$} & {\bf Migration} & {\bf MMR} & {\bf final $M_{outer}/M_{inner}$ ($M_\oplus$)} & {\bf $\Delta a_{outer}$ (AU)} & {\bf $\Delta a_{inner}$ (AU)} \\
\hline
1 & $15/15$ & $\Sigma_0 r^{-0.5}$ & Convergent & - & 364:427 & $-1.75$ & $-0.06$ \\
2 & $15/30$ & $\Sigma_0 r^{-0.5}$ & Convergent & 2:1 & 774:589 & $-3.07$ & $-0.58$ \\
3 & $30/15$ & $\Sigma_0 r^{-0.5}$ & Divergent & - & 250:303 & $-1.46$ & $-4.95$ \\
4 & $30/30$ & $\Sigma_0 r^{-0.5}$ & Convergent & 2:1 & 833:530 & $-3.36$ & $-0.64$ \\
\hline
5 & $15/15$ & $\Sigma_0 r^{-1.0}$ & Divergent & - & 204:271 & $-1.17$ & $-4.97$ \\
6 & $15/30$ & $\Sigma_0 r^{-1.0}$ & Convergent & 2:1 & 1373:646 & $-2.82$ & $-1.46$ \\
7 & $30/15$ & $\Sigma_0 r^{-1.0}$ & Divergent & - & 205:288 & $-0.43$ & $-4.95$ \\
8 & $30/30$ & $\Sigma_0 r^{-1.0}$ & Convergent & 5:1 & 2013:815 & $-2.14$ & $-2.49$ \\
\hline
9 & $15/15$ & $\Sigma_0 r^{-1.5}$ & Convergent & 2:1 & 246:205 & $-10.04$ & $-4.97$ \\
10 & $15/30$ & $\Sigma_0 r^{-1.5}$ & Convergent & 2:1 & 1247:599 & $-1.55$ & $-1.64$ \\
11 & $30/15$ & $\Sigma_0 r^{-1.5}$ & Divergent & - & 166:206 & $1.10$ & $-4.95$ \\
12 & $30/30$ & $\Sigma_0 r^{-1.5}$ & Convergent & 2:1 & 785:501 & $-1.60$ & $0.17$ \\
\hline
\end{tabular}
\end{table*}

While the orbital evolution is relatively consistent across different runs, the mass evolution exhibits considerable variability depending on the disk's physical parameter. In most of our simulations, the outer planet ultimately accretes more gas than the inner one, even in cases where the inner planet initially had a larger core-mass. Figure \ref{fig9} shows the temporal evolution of the planetary mass ratio, which traces the emergence of the reversed mass configurations seen in mature planetary systems. The accretion initially exhibits a stochastic phase lasting approximately $1000$ orbits. During this stage, competitive gas accretion and dynamical interactions lead to frequent crossings of the unity threshold, particularly in the lower-viscosity cases ($\nu = 10^{-5}$ and $10^{-6}$). In contrast, in the high-viscosity regime ($\nu = 10^{-4}$), the system bypasses this prolonged competitive phase. The enhanced viscous mass transport from the outer disk enables the outer planet to rapidly overcome the gap-opening influence of the inner planet, resulting in an immediate and persistent divergence in their mass growth. This difference is primarily governed by a different gas-supply mechanism. The growth of the inner planet is limited by the rapid depletion of its local feeding zone, whereas the outer planet located near the pressure maximum at the outer edge of the common gap continues to efficiently accrete gas from the outer disk. As the system subsequently settles into a stable MMR and the coupled migration rate reaches a quasi-steady state after approximately $2000$ orbits, the planetary mass ratio stays robust at a value similar or even larger than unity. These results demonstrate that the final mass ratio is critically modulated by the $\nu$, $\alpha$ and initial core-masses.

The statistical results of our simulations are summarized in Table \ref{tab1} $-$ Table \ref{tab3}. The $\Sigma_0$ is the reference surface density (typically taken at 1 AU). The coefficients (e.g., $0.1$, $0.5$, $1$, $2$) represent an overall scaling factor of the surface density. For example, in Table \ref{tab2} Run 6 ($\Sigma_0$) enters 2:1 MMR and final mass is $1.27$. Run 7 ($0.5\Sigma_0)$ also enters 2:1 MMR but with lower final masses ($1.3$). Run 8 ($0.1\Sigma_0$) without a MMR and the final masses are much smaller ($0.93$). These results show that the scaling factor of the surface density directly controls both the efficiency of planetary growth and the dynamical evolution, making it a key parameter in the simulations.

In the high-viscosity regime ($\nu = 10^{-4}$, Table \ref{tab1}), the majority of cases result in divergent migration except for cases Run 1. As viscosity decreases to $\nu = 10^{-5}$ (Table \ref{tab2}), both convergent and divergent migrations occur and resonant capture emerges as a possible outcome under specific cases. The system exhibits a transition where the migration mode depends sensitively on the disk profile and initial mass ratio, allowing for the first instances of resonant capture. Finally, in the low viscosity limit ($\nu = 10^{-6}$, Table \ref{tab3}), convergent migration and stable resonance capture becomes the favored evolutionary pathway, occurring in the majority of simulated cases. Overall, the models suggest that disk density plays a key role in setting the efficiency of accretion and the stability of resonance capture, while asymmetric initial core-masses tend to amplify the outer planet's growth advantage.

\begin{figure*}[!htbp]
\centering
\includegraphics[width=0.9\textwidth]{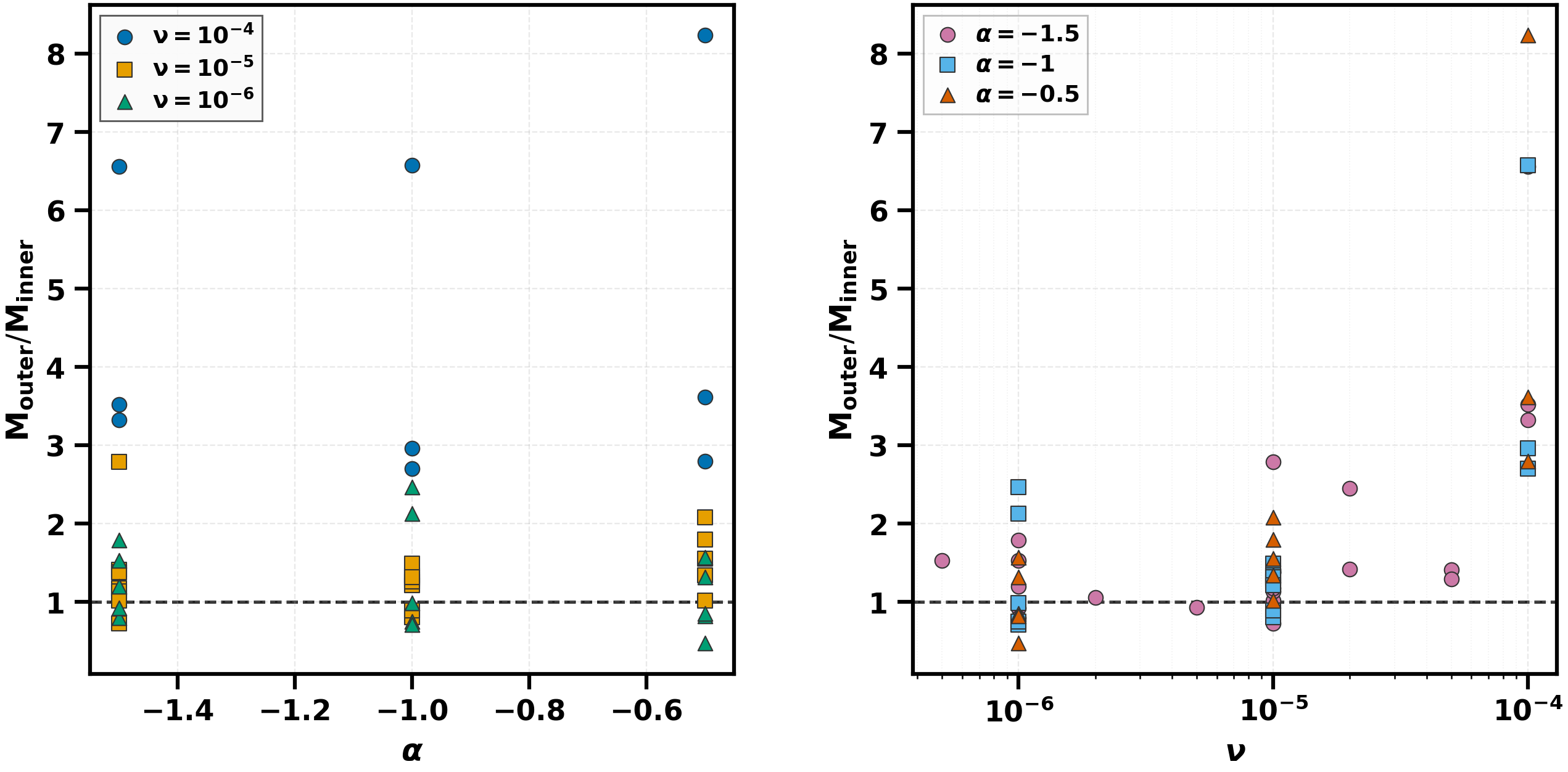}
\caption{The final mass ratio vs disk physical parameters. The panels illustrate the impact of $\nu$ and $\alpha$ on the differential gas accretion between the planet pair. Left panel: The final mass ratio as a function of viscosity $\nu$. In the high-viscosity regime ($\nu = 10^{-4}$), the outer planet consistently captures more mass flux, leading to significant mass reversal. In the low-viscosity limit ($\nu = 10^{-6}$), the system approaches a comparable mass configuration as convergent migration and stable resonance capture become the dominant evolutionary modes. Vertical spreads at specific parameter points (e.g., at $\nu = 10^{-5}$) reflects the variation in other grid parameters (surface density slope and initial mass ratio) for runs with the same viscosity}. Right panel: The final mass ratio $M_{\text{outer}}/M_{\text{inner}}$ plotted against the density slope $\alpha$. It suggests that steeper disk profiles may reduce the accretion advantage of the outer planet by modifying the gap structure. \label{fig10}
\end{figure*}

According to the simulation results, the final mass ratio $M_{\text{outer}}/M_{\text{inner}}$ exhibits a systematic dependence on the $\nu$ and $\alpha$ (see Figure \ref{fig10}). For $\nu = 10^{-4}$, the outer planet's growth is significantly accelerated as the increased mass from the outer disk bypasses the inner planet's gap more efficiently, leading to the most pronounced mass ratio reversals. As the system transitions toward the low-viscosity limit ($\nu = 10^{-6}$), the convergence of the final mass ratio toward unity suggests that the depletion of local gas supply becomes a shared constraint for both cores, despite the outer planet's strategic position. Furthermore, the observed negative correlation between the final mass ratio and the surface density slope $\alpha$ indicates that a steeper disk profile may partially mitigate the accretion asymmetry by altering the gap-opening threshold. Ultimately, these trends confirm that the outer-heavy configuration is a robust dynamical outcome of the interplay between gap-mediated accretion and resonant locking, particularly in disks with sufficient gas supply.

We observed migration in $41$ cases, $23$ of which show convergent migration. Of these convergent cases, $16$ are captured into mean-motion resonances (MMRs). Since we adopt a negative density profile, the inner planet should grow faster. However, the results show that the outer planets systematically achieve larger final masses. The initial density profile and mass ratio only affects the early stage of the growth (The panel b in Figure \ref{fig4}). Short explanations can be summarized as follows:

1. The inner core grows more quickly, opens gap earlier and faster, which suppresses its further growth (Panel a1 in Figure \ref{fig6}). In other words, gap opening self-limits the inner planet's gas accretion. 

2. Meanwhile, the outer core migrates into the out boundary of the gap where the disk density is highest. The tidal perturbation from the inner massive core also elongates the gap opening process of the outer core (Panel b1 in Figure \ref{fig6}).

3. The outer core can continue to accrete gas from the outer disk after the common gap has formed, whereas the inner core's growth is limited by the inner disk, which is quickly depleted (Panel c1 in Figure \ref{fig6}).

Furthermore, in nearly inviscid disks ($\nu \leq 10^{-6}$), the rapid formation of deep gaps can halt planetary growth at Neptune-like masses. This suggests that planets born in low-viscosity environments may not only exhibit reversed mass ratios but also fail to reach Jovian status, potentially providing a physical explanation for the moderate-mass planet desert observed at a few AUs.

\begin{figure*}[!htbp]
\centering
\includegraphics[width=0.4\textwidth]{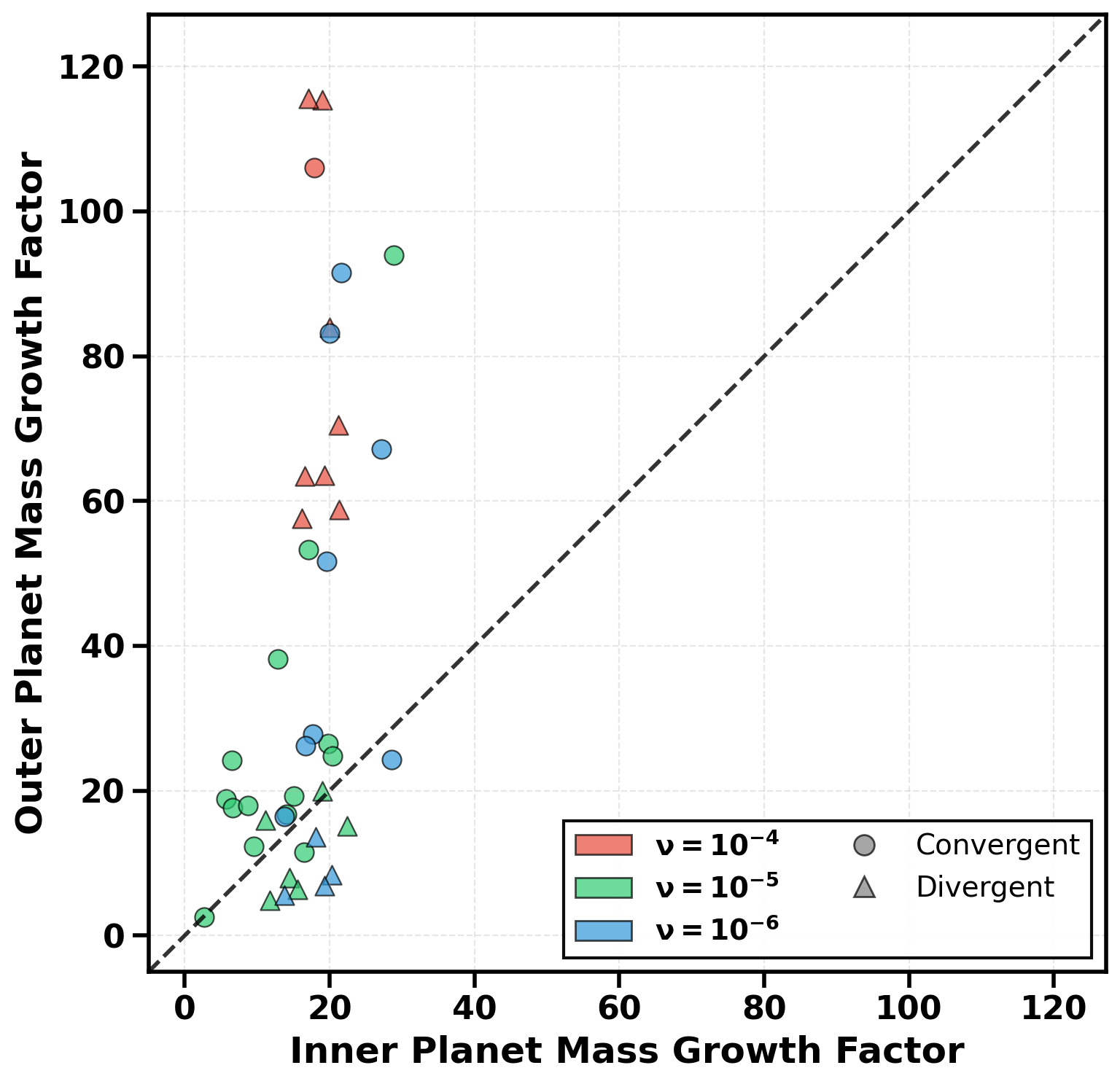}
\caption{Comparison of mass growth factors between inner and outer planets across different viscosity regimes. Colors represent viscosity. Marker shapes indicate orbital behavior. The dashed line marks the 1:1 equal growth relationship. Higher viscosity promotes divergent growth (outer planet grows faster than inner planet), while lower viscosity favors convergent growth.}\label{fig11}
\end{figure*}

Figure \ref{fig11} compared their mass growth factors across different viscosity regimes to clarify how disk viscosity influences the relative growth of inner and outer planets. The results reveal a clear viscosity-dependent trend. The efficiency of planetary gas accretion is governed by the disk viscosity. In particular, after gap opening, stronger viscous dissipation accelerates the inflow of gas into the gap. Consequently, a higher viscosity coefficient generally yields a larger final planetary mass. Nevertheless, the upper mass limit of accretion is tightly tied to the orbital location of each planet within the disk. Once the outer planet truncates the disk structure, the maximum accretable mass of the inner planet is constrained by the total mass of the inner disk. By contrast, the outer planet is fed by extensive outer disk material, allowing it to reach a much higher mass limit. Therefore, for outer planets, a higher viscosity systematically promotes a greater final mass. For inner planets, however, high viscosity does not necessarily correspond to a more massive final state, and the ultimate accretion outcome is strongly modulated by the specific orbital configuration of the planetary system.

\begin{figure*}[!htbp]
\centering
\includegraphics[width=0.9\textwidth]{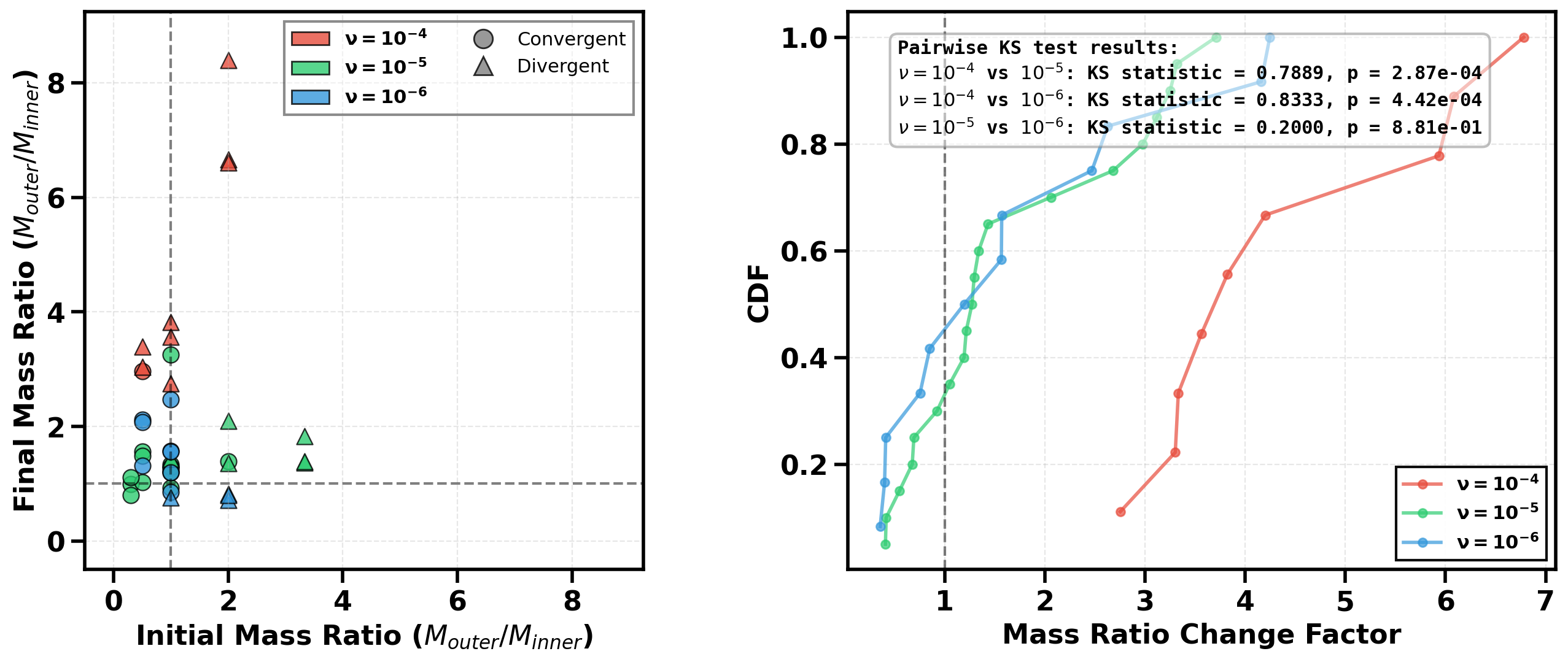}
\caption{Mass ratio evolution analysis showing the relationship between initial and final mass ratios (left) and cumulative distributions of the mass ratio change factor (right). In the left panel, colors denote viscosity red ($\nu = 10^{-4}$), green ($\nu = 10^{-5}$) and blue ($\nu = 10^{-6}$). Marker shapes indicate migration behavior. Circles for convergent systems and triangles for divergent systems. The right panel shows cumulative distribution functions with Kolmogorov-Smirnov test statistics comparing viscosity pairs. Higher viscosity leads to larger mass ratio changes, with $\nu = 10^{-4}$ showing significantly different distributions from lower viscosity cases.}\label{fig12}
\end{figure*}

We further analyzed the mass ratio evolution by comparing initial and final mass ratios for all cases (Figure \ref{fig12}). Convergent systems (circles) maintain their initial mass ratios, clustering tightly around the diagonal line where final ratios equal initial ratios. In contrast, divergent systems (triangles) show substantial deviations, particularly at high initial mass ratios and high viscosity. For example, at $\nu = 10^{-4}$ (red), initial mass ratios of 2 can evolve to final ratios exceeding 8, indicating extreme mass redistribution favoring the outer planet. The cumulative distribution functions (CDFs) of the mass ratio change factor (which is defined as the ratio of the final to initial mass ratio) quantify this effect. Kolmogorov-Smirnov tests shows that the $\nu = 10^{-4}$ distribution differs significantly from both $\nu = 10^{-5}$ ( ks statistic value D = 0.79, p = $2.87 \times 10^{-4}$) and $\nu = 10^{-6}$ (D = 0.83, p = $4.42 \times 10^{-4}$), while $\nu = 10^{-5}$ and $\nu = 10^{-6}$ show no significant difference (D = 0.2, p = 0.88). This suggests a critical viscosity threshold above which disk dynamics dramatically alter planetary mass ratios through competitive accretion.

\begin{figure*}[!htbp]
\centering
\includegraphics[width=0.9\textwidth]{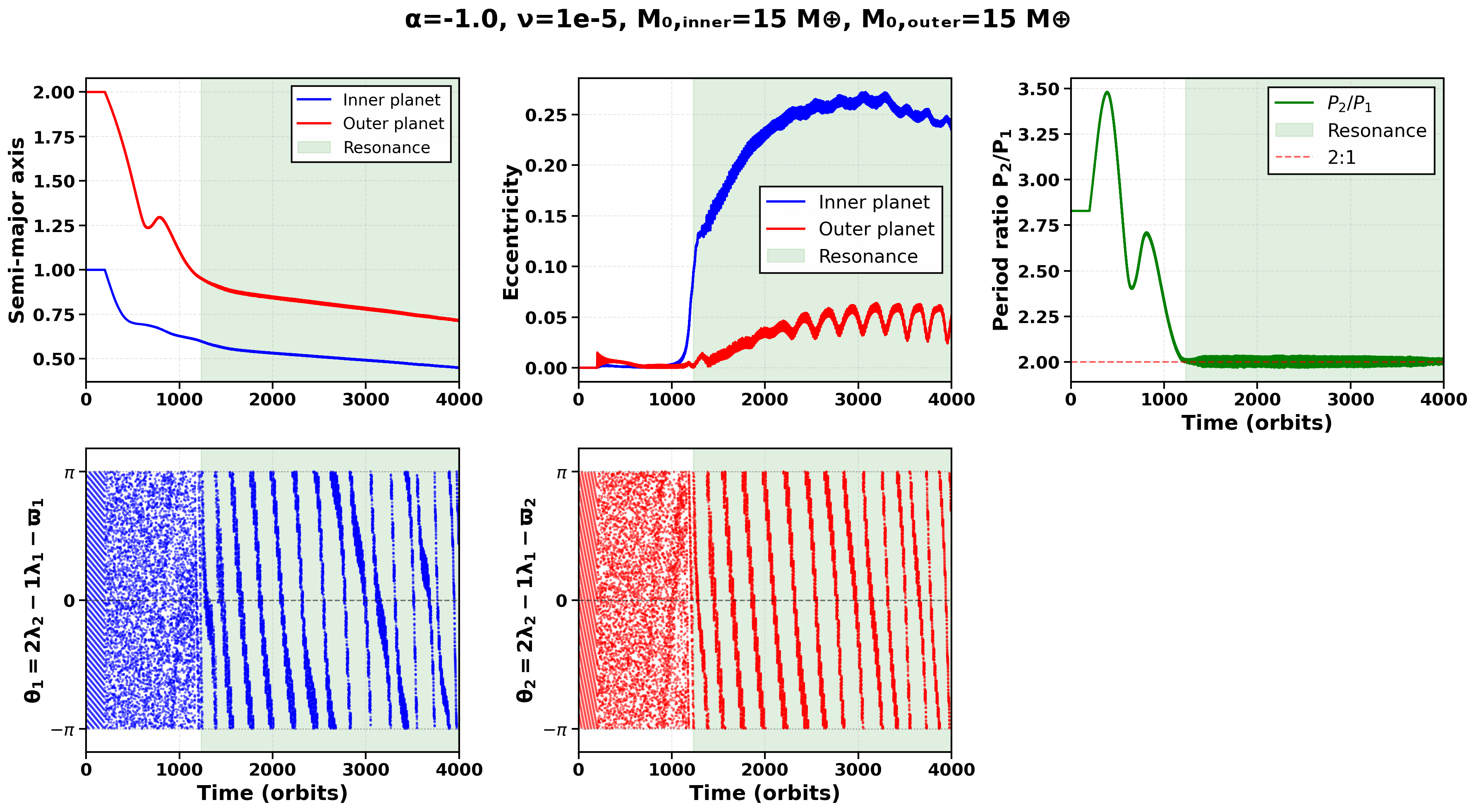}
\caption{Dynamical evolution of a two planets system captured into a 2:1 MMR. The top-left panel displays the inward migration of the semi-major axes for the inner (blue) and outer (red) planets. The top-middle panel illustrates the corresponding eccentricity evolution, showing the characteristic excitation of $e$ upon resonance entry. The top-right panel tracks the period ratio $P_2/P_1$, which stabilizes at the $2:1$ commensurability (dashed red line) at $T \approx 1200 P0$. The bottom panels show the evolution of the resonant angles $\theta_1 = 2\lambda_2 - \lambda_1 - \omega_1$ and $\theta_2 = 2\lambda_2 - \lambda_1 - \omega_2$. The shaded green region identifies the resonant state, characterized by the transition from circulation to libration of the resonant arguments.}\label{fig:2to1_69.2pct_r-1.0_v1e-5_i15o15}
\end{figure*}

Figure \ref{fig:2to1_69.2pct_r-1.0_v1e-5_i15o15} illustrates the evolution of a representative two-planet system exhibiting resonance capture. Initially ($T < 1200 P_0$), both planets experience inward migration driven by disk torques. The outer planet migrates at a higher rate relative to the inner planet, causing a rapid convergent evolution of their orbits. This differential migration leads to a steady decrease in the orbital period ratio. As the period ratio approaches the $2:1$ commensurability at $T \approx 1200 P0$, the system enters the resonance capture phase. At this critical junction, the resonant gravitational interactions between the planets counteract the disk-induced migration forces. The capture is marked by a sharp, non-linear excitation of the orbital eccentricities; notably, the inner planet's eccentricity surges to a peak of approximately $0.25$. In the subsequent phase (highlighted in green), the system achieves dynamical stability. The evolution of the resonant arguments, $\theta_1$ and $\theta_2$, transitions from a full $[-\pi, \pi]$ circulation to a steady libration around a fixed equilibrium point. This phase-locking ensures that despite the continued slow contraction of the semi-major axes, the period ratio remains strictly clamped at $2:1$. This robust configuration demonstrates how resonance capture can stabilize planetary architectures against further orbital decay or crossing.

\section{Discussion}\label{sec:discuss}

\subsection{Competitive or Cooperative Gas Accretion}

Our simulations show that once two giant planets form a common gap, gas accretion can proceed either competitively or cooperatively, depending on the disk conditions. In high-viscosity disks, efficient gas replenishment across the gap preferentially enhances accretion onto the outer planet, resulting in a strongly divergent growth pattern and a significantly increased final mass ratio. The inner planet is starved quickly because the inner disk is drained onto the star, while the outer planet monopolizes the gas inflow from the outer disk. This asymmetric access leads to a strong mass hierarchy and explains why many resonant pairs have an outer planet substantially heavier than the inner one. Disks with efficient angular momentum transport (e.g., due to turbulence) are thus more likely to produce nearly equal-mass resonant pairs, while laminar or weakly turbulent disks may preferentially yield systems with pronounced mass-ratio reversal.

\subsection{Final Mass Ratio}

The final mass ratio is also sensitive to the competition between disk viscosity and disk lifetime. High-viscosity promotes efficient gas replenishment, but if the disk disperses too quickly, both planets may be left at intermediate masses. Conversely, low-viscosity disks prolong the survival of gaps but suppress replenishment, favoring an outcome where the outer planet dominates. This interplay suggests that the diversity of mass ratios observed in resonant giant planet pairs may reflect not only differences in local disk conditions but also in the evolutionary timescales of their natal disks. Our results further imply that most resonant pairs with outer planets heavier than the inner ones should have formed in relatively low-viscosity environments with longer-lived gas. In contrast, systems where both giants have nearly equal masses likely formed in more turbulent or short-lived disks.

Because disk dispersal is not included in our simulations, the planetary masses measured at the end of each run should be interpreted as end-of-integration values rather than asymptotic post-dispersal masses. In most long-duration runs, however, $M_{\rm outer}/M_{\rm inner}$ evolves much more slowly at late times and approaches a quasi-steady value, although a few cases remain visibly rising and should be treated with caution. The young PDS 70 system provides an observational analogue \citep{Keppler-etal.2018,Muller-etal.2018}, its two actively accreting giant protoplanets are embedded in a gas-rich disk and have an orbital configuration consistent with, and likely close to, the 2:1 mean-motion resonance \citep{Bae-etal.2019,Wang-etal.2021,Benisty-etal.2021}. Thus our simulations characterize the relative growth and mass-ratio evolution of resonant or near-resonant giant-planet pairs during the gas-disk phase rather than their ultimate post-dispersal masses.

\begin{figure*}[!htbp]
\centering
\includegraphics[width=0.9\textwidth]{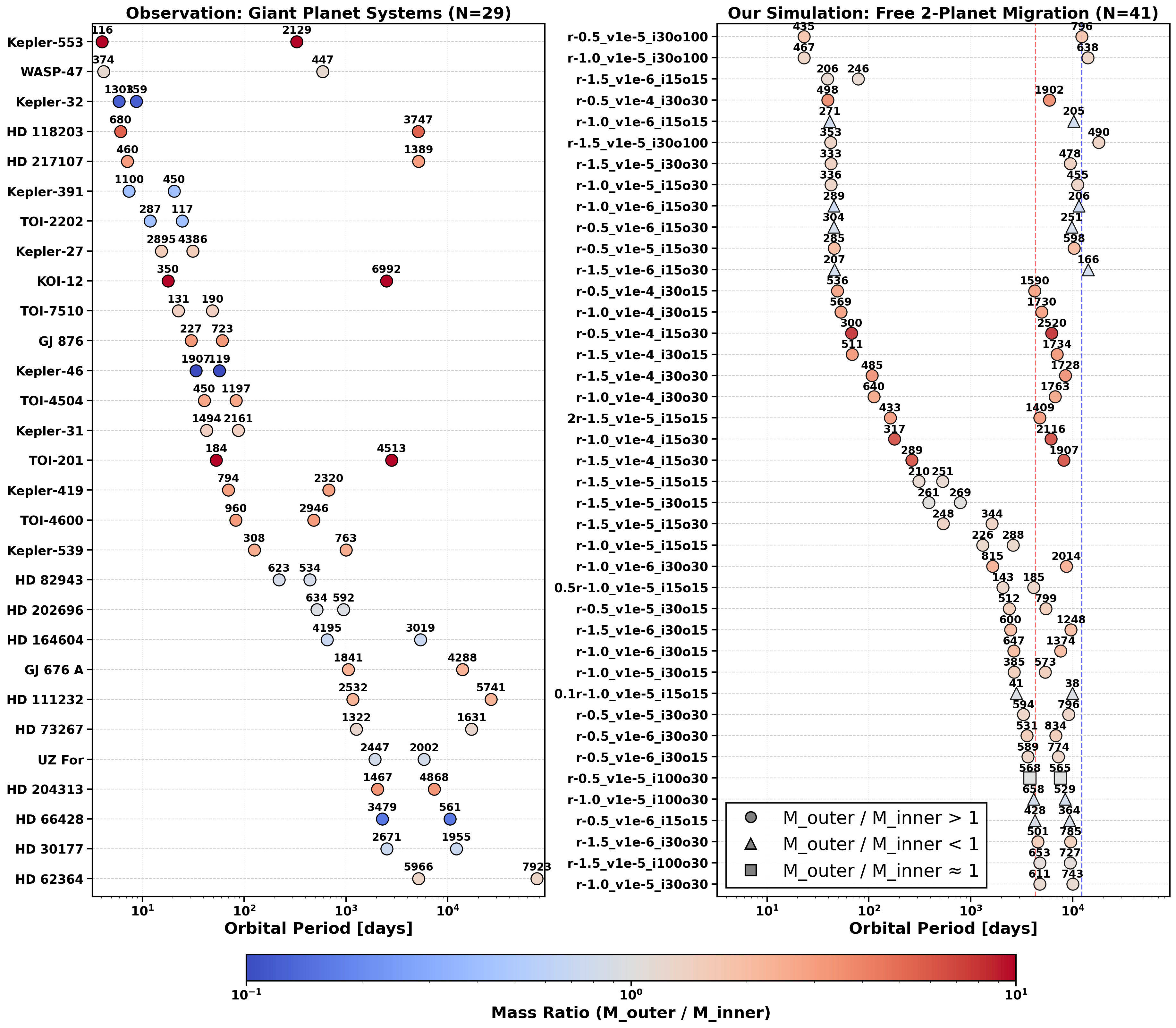}
\caption{Comparison of observed and simulated two-giant-planet system architectures. Left panel shows orbital periods for all known two planets systems with both companions $M > 100 M_\oplus$. Each row represents one system. The inner and outer planets are plotted individually. Marker color encodes $M_{\mathrm{outer}}/M_{\rm inner}$ on a logarithmic scale (blue is inner planet more massive and red is outer planet more massive). Numbers adjacent to each marker denote the masses in $M_\oplus$. Right panel equivalently represents for our simulation undergoing free two planets disk migration. Row labels encode the initial disk and planet parameters. Each row is labeled tith case-title such as $r$-$0.5\_v1e$-$5\_i30o15$, in which $r$ indicates profile index, $v$ denotes kinematic viscosity $\nu$ and i30o15 specifies the initial core-masses of the inner and outer planets (30 $M_\oplus$ and 15 $M_\oplus$, respectively). Vertical dashed lines mark the initial positions of the inner (red dashed) and outer (blue dashed) planet adopted uniformly across all runs. The scatter points represent the final orbital periods, illustrating the migration distance undergone by each system. Marker shape distinguishes outer-heavy (circle), inner-heavy (triangle) and comparable-mass (square) pairs in our simulation results.}\label{fig14}
\end{figure*}

\subsection{Comparison with Observations}
\subsubsection{Giant Planet Pair Architectures and Resonant Systems}
Observational studies (\citealt{Zhu2018, Zhu-Dong-2021, Liu2026}) have shown that outer cold Jupiters do not necessarily disrupt or clear out inner planet. Instead, multi-planet systems usually host inner planets coexisting with outer giant planets, suggesting that stable dynamical architectures can persist even in the presence of massive companions. Our hydrodynamical simulations provide a formation pathway consistent with those works. We find that when two planetary cores undergo simultaneous gas accretion and orbital migration, convergent migration naturally occurs over a broad range of disk conditions. The concurrent runaway gas accretion further enhances the mass growth of the planets during this phase. This process frequently leads to capture into mean-motion resonances, with the 2:1 resonance being particularly common. Resonant locking stabilizes the system during the gas-disk phase and prevents dynamical instability. During this evolution, gas accretion is strongly modulated by disk-planet interactions. The inner planet may experience reduced accretion due to depletion of nearby gas, while the outer planet continues to accrete efficiently from the more massive outer disk reservoir. As a result, comparable or even inverted mass hierarchies can naturally arise, with outer planets becoming as massive as or more massive than their inner companions.

A number of observed systems support this scenario. Compact resonant chains of giant planets, such as those reported in  \citealt{Almenara2025}, including TOI-7510 system, demonstrate that multi-giant resonant configurations can survive in tightly packed architectures. Well-studied resonant systems such as GJ 876 \citep{Bean2009} and HD 82943 \citep{Tan2013, Lee2006} also show that outer giants can be comparable in mass or more massive than inner companions while remaining locked in 2:1 resonance. In addition, transit timing variation (TTV) analyses of Kepler systems indicate that near-resonant giant pairs often exhibit outer planets with masses comparable to or exceeding those of inner planets \citep{Nesvorny2016, Xie2012}. TESS discoveries such as the TOI 216 \citep{Hirano2020, MacDonald2020} further support this trend, while young systems such as V1298 Tau \citep{David2019} suggest that such mass configurations may already be established at early evolutionary stages. Observations of young multi giant systems like V1298 $\tau$ further point toward the presence of external mass dominance already at early evolutionary stages (\citealt{David2019}).

Figure \ref{fig14} compares observed giant planet pairs with our simulation outcomes. The observational sample is from the NASA Exoplanet Archive (exoplanetarchive.ipac.caltech.edu, acquired in August of 2026). We retain only the recommended parameter set for each planet (\texttt{default\_flag}=1) and require a measured planetary mass (\texttt{pl\_masse}). Systems are included if they host exactly two giant planets with \(M_p\geq100\,M_\oplus\), both of which have a measured orbital period (\texttt{pl\_orbper}). Planets discovered by direct imaging are excluded. Masses estimated from planetary radii are not used for sample selection. The observed sample (left panel) are predominantly compact, with inner giants located well inside the snow line. In many cases, outer planets are more massive than their inner companions. Our simulations (right panel) reproduce these key features: planets initialized at wide separations undergo substantial inward migration, often by more than an order of magnitude in orbital period. The degree of inward migration, final orbital architecture, and mass ratios depend sensitively on disk properties, including surface density, viscosity, and initial core-masses. Systems with initially more massive outer cores tend to experience stronger convergent migration and are more likely to become locked into resonances, producing compact and outer-dominated configurations. In contrast, initially inner-dominated systems more often evolve toward wider separations or less compact architectures. Overall, the agreement between simulated and observed populations suggests that present-day orbital and mass architectures of giant planet pairs retain a direct imprint of disk-driven migration and early mass growth processes.

When comparing the simulation results with observed systems (Figure~\ref{fig1}), observational selection effects must be taken into account quantitatively. In our selected observational sample, the median inner and outer orbital periods are approximately \(53\) and \(950\)~days, respectively. About \(59\%\) of the inner planets satisfy \(P_{\mathrm{in}}<100\)~days, whereas only about \(38\%\) of the outer planets have periods exceeding \(3000\)~days. Roughly \(52\%\) of the planet pairs have period ratios \(P_{\mathrm{out}}/P_{\mathrm{in}}\leq 5\). By contrast, the final periods in the free two-planet simulations are overall longer, with median values of \(P_{\mathrm{in}}\sim260\)~days and \(P_{\mathrm{out}}\sim8200\)~days. This difference does not by itself invalidate the migration scenario. Transit and TTV surveys are strongly biased toward short orbital periods. Therefore, the concentration of the observational sample at \(P\lesssim 10^{3}\)~days (Figure~\ref{fig2} and~\ref{fig14}), together with the relative lack of counterparts at the long-period end of the simulation grid, can be explained at least in part by sample incompleteness and need not be interpreted as a failure of convergent migration. Selection effects also influence the mass-ratio trends shown in Figure~\ref{fig1} and Figure~\ref{fig2}. In the \(M_p\geq 100\,M_\oplus\) sample mentioned in Figure~\ref{fig1}, pairs with \(M_{\mathrm{out}}>M_{\mathrm{in}}\) account for about \(66\%\), while pairs with \(M_{\mathrm{out}}<M_{\mathrm{in}}\) account for about \(34\%\). Moreover, outer-heavier systems more often host short-period inner planets. The fraction with \(P_{\mathrm{in}}<100\)~days is about \(68\%\), compared with about \(40\%\) for inner-heavier systems. This is consistent with RV/transit detection sensitivity. Massive, short-period planets are more readily detected, whereas lower-mass outer companions at long periods are more easily missed, thereby suppressing the observed fraction of compact, outer-lighter systems. Planet pairs with smaller period ratios (\(P_{\mathrm{out}}/P_{\mathrm{in}}\leq 5\)) are the most complete observationally and also exhibit more nearly comparable masses, in agreement with Figure~\ref{fig1} panel b.

\subsubsection{The Super-Earth$-$Cold Jupiter Correlation}
Population analyses of exoplanetary systems have established a statiatical correlation between inner low-mass planets and outer gaint companions, commonly referred to as the Super-Earth$-$Cold Jupiter correlation. Radial velocity and transit surveys show that systems hosting inner super-Earths or sub-Neptunes exhibit a significantly enhanced occurrence rate of outer cold Jupiters compared to field star averages (\citealt{Zhu2018, Bryan2019, Schlecker2021, Zhu2024}). This trend is particularly prominent in systems containing inner sub-Neptunes and outer giant planets, suggesting a systematic architectural dichotomy in planetary systems that inner planets often remain in the super-Earth or sub-Neptune regime, while outer planets undergo runaway gas accretion to become gas giants. However, the physical origin of this correlation remains an open question. Our simulations suggest that competitive accretion during concurrent migration offers a plausible dynamical pathway. When two planetary embryos migrate through the disk while accreting gas, they can carve a shared gap structure, placing their growth in a competitive regime governed by disk-planet interactions. In this configuration, the growth of the inner planet becomes self-limiting. It rapidly depletes gas from the inner disk region and migrates toward the inner disk boundary, reducing its long-term accretion potential. Meanwhile, the outer planet continues to accrete efficiently from the more massive and less perturbed outer disk reservoir. This asymmetry naturally leads to divergent growth outcomes between inner and outer planets. We emphasize that our current simulations do not include an inner disk cavity, such as those produced by magnetospheric truncation or photoevaporation. As a result, gas accretion onto very close-in inner planets may be overestimated. Consequently, simulated inner planets can occasionally continue accreting after migrating to orbital distances of only $\sim 0.03$--$0.1\,{\rm AU}$, allowing some of them to reach masses exceeding $100\,M_{\oplus}$ or even approaching $1\,M_J$. In a more realistic disk containing an inner cavity, an initially low-mass inner planet (e.g., $\sim 15\,M_{\oplus}$) would likely migrate into the gas-depleted region, before undergoing substantial runaway gas accretion. In more realistic disks containing an inner cavity, inward-migrating inner planets would enter a gas-depleted region earlier, leading to a more abrupt termination of gas accretion. In this case, the inclusion of an inner cavity would further strengthen the proposed mechanism by truncating inner planet growth at an earlier stage. This would contribute to enhance the emergence of systems in which inner planets remain in the super-Earth or sub-Neptune regime while outer planets grow into cold Jupiters. Overall, this framework provides a physically motivated explanation for the observed Super-Earth$-$Cold Jupiter correlation and suggests that it may arise naturally from generic processes of concurrent migration and gas accretion in protoplanetary disks.

All simulations terminate while gas remains present. We therefore do not follow the long-term N-body evolution after disk dispersal, when hydrodynamical damping ceases and previously stable resonant pairs may become unstable on longer timescales. Our two-planet setup also omits inner super-Earths or terrestrial embryos interior to the migrating giants. The substantial inward migration of giant pairs found here would likely have strong dynamical consequences for such inner planets and may further shape the Super-Earth-Cold Jupiter architecture discussed above. A quantitative assessment of post-disk stability and the fate of inner low-mass planets requires dedicated N-body integrations after dispersal, which lies beyond the scope of this hydrodynamical study and prepare in the future work.

\section{Conclusions}\label{sec:conclusion}
We have investigated the evolutions of the orbits and mass ratio of two accreting planet cores. Our aim was to find out why reversed mass configurations are commonly observed in compact orbital systems. Our results show that if two planets cores accrete gas simultaneously, the outer one usually grows more massive, making the reversed mass configuration a natural outcome. The normal mass configuration would result from a different process. For example, the outer planet forming and being captured into resonance long after the inner giant planet had already formed.

1. Convergent migration and MMR capture occur frequently in our simulations. Resonance capture is achieved in $\sim 40\%$ of all migrating systems (16/41), and in $\sim 70\%$ of cases undergoing convergent migration (16/23), typically leading to long-term 2:1 MMR locking. Independent of initial core-masses and disk density profiles, two growing planets tend to migrate inward convergently due to differential migration rates and often captured into a 2:1 MMR. Once resonance locking occurs, the planets migrate inward as a pair and the opening of a common gap subsequently slows their joint migration.

2. Gap opening controls differential growth within resonant pairs. The inner planet typically accretes gas faster at early stages but opens a gap sooner, which limits its long-term growth by depleting the local gas supply. In contrast, the outer planet, residing near the density maximum at the outer edge of the common gap, continues to accrete efficiently from the outer disk. The resulting asymmetry in gas availability naturally drives a mass ratio reversal, with the outer planet eventually becoming more massive.

3. Reversed mass ratios are a natural outcome of coupled migration and accretion. Even if the inner core initially exceeds the outer one in mass, the interplay between convergent migration, resonance locking, and gap-mediated accretion typically leads to the outer planet overtaking in mass. This framework offers a coherent theoretical explanation for observed systems such as Kepler-553, Kepler-58 and HD217107, where outer planets are more massive.

4. Our simulations show that disk-driven accretion and migration shape giant planet pairs, with competitive accretion driving the Super-Earth$-$Cold-Jupiter correlation.


\section*{acknowledgements}
 We thank the anonymous referees for helpful comments. We are grateful to Robert A. Wittenmyer and Xianyu Wang for helpful discussions. This work was supported by the 100 Talents Programme of the Chinese Academy of Sciences. Y.W. acknowledges the EACOA Fellowship awarded by the East Asia Core Observatories Association. Y.P.L. was supported by the National Natural Science Foundation of China (grants 12373070, and 12192223). This work made use of the High Performance Computing Resource in the Core Facility for Advanced Research Computing at Shanghai Astronomical Observatory. 


\bibliography{zhao2026}

@ARTICLE{Bergez-Casalou-etal.2023,
       author = {{Bergez-Casalou}, C. and {Bitsch}, B. and {Raymond}, S.~N.},
        title = "{Simultaneous gas accretion onto a pair of giant planets: Impact on their final mass and on the protoplanetary disk structure}",
      journal = {\aap},
         year = 2023,
        month = jan,
       volume = {669},
          eid = {A129},
        pages = {A129},
          doi = {10.1051/0004-6361/202244988},
archivePrefix = {arXiv},
       eprint = {2211.16239},
 primaryClass = {astro-ph.EP},
       adsurl = {https://ui.adsabs.harvard.edu/abs/2023A&A...669A.129B}
}

@ARTICLE{Crida-etal.2007,
       author = {{Crida}, A. and {Morbidelli}, A. and {Masset}, F.},
        title = "{Simulating planet migration in globally evolving disks}",
      journal = {\aap},
         year = 2007,
        month = jan,
       volume = {461},
       number = {3},
        pages = {1173-1183},
          doi = {10.1051/0004-6361:20065870},
archivePrefix = {arXiv},
       eprint = {astro-ph/0608660},
 primaryClass = {astro-ph},
       adsurl = {https://ui.adsabs.harvard.edu/abs/2007A&A...461.1173C}
}

@ARTICLE{Zhu-Dong-2021,
       author = {{Zhu}, Wei and {Dong}, Subo},
        title = "{Exoplanet Statistics and Theoretical Implications}",
      journal = {\araa},
         year = 2021,
        month = sep,
       volume = {59},
        pages = {291-336},
          doi = {10.1146/annurev-astro-112420-020055},
archivePrefix = {arXiv},
       eprint = {2103.02127},
 primaryClass = {astro-ph.EP},
       adsurl = {https://ui.adsabs.harvard.edu/abs/2021ARA&A..59..291Z}
}

@ARTICLE{Masset-Snellgrove-2001,
       author = {{Masset}, F. and {Snellgrove}, M.},
        title = "{Reversing type II migration: resonance trapping of a lighter giant protoplanet}",
      journal = {\mnras},
         year = 2001,
        month = feb,
       volume = {320},
       number = {4},
        pages = {L55-L59},
          doi = {10.1046/j.1365-8711.2001.04159.x},
archivePrefix = {arXiv},
       eprint = {astro-ph/0003421},
 primaryClass = {astro-ph},
       adsurl = {https://ui.adsabs.harvard.edu/abs/2001MNRAS.320L..55M}
}

@ARTICLE{Morbidelli-Crida-2007,
       author = {{Morbidelli}, Alessandro and {Crida}, Aur{\'e}lien},
        title = "{The dynamics of Jupiter and Saturn in the gaseous protoplanetary disk}",
      journal = {\icarus},
         year = 2007,
        month = nov,
       volume = {191},
       number = {1},
        pages = {158-171},
          doi = {10.1016/j.icarus.2007.04.001},
       adsurl = {https://ui.adsabs.harvard.edu/abs/2007Icar..191..158M}
}

@ARTICLE{Zhang-Zhou-2010a,
       author = {{Zhang}, Hui and {Zhou}, Ji-Lin},
        title = "{On the Orbital Evolution of a Giant Planet Pair Embedded in a Gaseous Disk. I. Jupiter-Saturn Configuration}",
      journal = {\apj},
         year = 2010,
        month = may,
       volume = {714},
       number = {1},
        pages = {532-548},
          doi = {10.1088/0004-637X/714/1/532},
archivePrefix = {arXiv},
       eprint = {1002.2201},
 primaryClass = {astro-ph.EP},
       adsurl = {https://ui.adsabs.harvard.edu/abs/2010ApJ...714..532Z}
}

@ARTICLE{Zhang-Zhou-2010b,
       author = {{Zhang}, Hui and {Zhou}, Ji-Lin},
        title = "{On the Orbital Evolution of a Giant Planet Pair Embedded in a Gaseous Disk. II. A Saturn-Jupiter Configuration}",
      journal = {\apj},
         year = 2010,
        month = aug,
       volume = {719},
       number = {1},
        pages = {671-684},
          doi = {10.1088/0004-637X/719/1/671},
archivePrefix = {arXiv},
       eprint = {1007.4520},
 primaryClass = {astro-ph.EP},
       adsurl = {https://ui.adsabs.harvard.edu/abs/2010ApJ...719..671Z}
}

@ARTICLE{Baruteau-Papaloizou-2013,
       author = {{Baruteau}, Clement and {Papaloizou}, John C.~B.},
        title = "{Disk-Planets Interactions and the Diversity of Period Ratios in Kepler's Multi-planetary Systems}",
      journal = {\apj},
         year = 2013,
        month = nov,
       volume = {778},
       number = {1},
          eid = {7},
        pages = {7},
          doi = {10.1088/0004-637X/778/1/7},
archivePrefix = {arXiv},
       eprint = {1301.0779},
 primaryClass = {astro-ph.EP},
       adsurl = {https://ui.adsabs.harvard.edu/abs/2013ApJ...778....7B}
}

@ARTICLE{Bitsch-etal.2015,
       author = {{Bitsch}, Bertram and {Lambrechts}, Michiel and {Johansen}, Anders},
        title = "{The growth of planets by pebble accretion in evolving protoplanetary discs}",
      journal = {\aap},
         year = 2015,
        month = oct,
       volume = {582},
          eid = {A112},
        pages = {A112},
          doi = {10.1051/0004-6361/201526463},
archivePrefix = {arXiv},
       eprint = {1507.05209},
 primaryClass = {astro-ph.EP},
       adsurl = {https://ui.adsabs.harvard.edu/abs/2015A&A...582A.112B}
}

@ARTICLE{Pichierri-etal.2018,
       author = {{Pichierri}, Gabriele and {Morbidelli}, Alessandro and {Crida}, Aur{\'e}lien},
        title = "{Capture into first-order resonances and long-term stability of pairs of equal-mass planets}",
      journal = {Celestial Mechanics and Dynamical Astronomy},
         year = 2018,
        month = aug,
       volume = {130},
       number = {8},
          eid = {54},
        pages = {54},
          doi = {10.1007/s10569-018-9848-2},
archivePrefix = {arXiv},
       eprint = {1808.08059},
 primaryClass = {astro-ph.EP},
       adsurl = {https://ui.adsabs.harvard.edu/abs/2018CeMDA.130...54P}
}

@ARTICLE{Kley2000,
       author = {{Kley}, W.},
        title = "{On the migration of a system of protoplanets}",
      journal = {\mnras},
         year = 2000,
        month = apr,
       volume = {313},
       number = {4},
        pages = {L47-L51},
          doi = {10.1046/j.1365-8711.2000.03495.x},
archivePrefix = {arXiv},
       eprint = {astro-ph/9910155},
 primaryClass = {astro-ph},
       adsurl = {https://ui.adsabs.harvard.edu/abs/2000MNRAS.313L..47K}
}

@ARTICLE{Peplinski-etal.2008,
       author = {{Pepli{\'n}ski}, A. and {Artymowicz}, P. and {Mellema}, G.},
        title = "{Numerical simulations of type III planetary migration - III. Outward migration of massive planets}",
      journal = {\mnras},
         year = 2008,
        month = jul,
       volume = {387},
       number = {3},
        pages = {1063-1079},
          doi = {10.1111/j.1365-2966.2008.13339.x},
archivePrefix = {arXiv},
       eprint = {0802.2501},
 primaryClass = {astro-ph},
       adsurl = {https://ui.adsabs.harvard.edu/abs/2008MNRAS.387.1063P}
}

@ARTICLE{Bitsch-Kley-2011,
       author = {{Bitsch}, B. and {Kley}, W.},
        title = "{Range of outward migration and influence of the disc's mass on the migration of giant planet cores}",
      journal = {\aap},
         year = 2011,
        month = dec,
       volume = {536},
          eid = {A77},
        pages = {A77},
          doi = {10.1051/0004-6361/201117202},
archivePrefix = {arXiv},
       eprint = {1107.3844},
 primaryClass = {astro-ph.EP},
       adsurl = {https://ui.adsabs.harvard.edu/abs/2011A&A...536A..77B}
}

@ARTICLE{Crida-etal.2009,
       author = {{Crida}, Aur{\'e}lien and {Masset}, Fr{\'e}d{\'e}ric and {Morbidelli}, Alessandro},
        title = "{Long Range Outward Migration of Giant Planets, with Application to Fomalhaut b}",
      journal = {\apjl},
         year = 2009,
        month = nov,
       volume = {705},
       number = {2},
        pages = {L148-L152},
          doi = {10.1088/0004-637X/705/2/L148},
archivePrefix = {arXiv},
       eprint = {0910.1004},
 primaryClass = {astro-ph.EP},
       adsurl = {https://ui.adsabs.harvard.edu/abs/2009ApJ...705L.148C}
}

@ARTICLE{Lin-Papaloizou-2012,
       author = {{Lin}, Min-Kai and {Papaloizou}, John C.~B.},
        title = "{Outward migration of a giant planet with a gravitationally unstable gap edge}",
      journal = {\mnras},
         year = 2012,
        month = mar,
       volume = {421},
       number = {1},
        pages = {780-788},
          doi = {10.1111/j.1365-2966.2011.20352.x},
archivePrefix = {arXiv},
       eprint = {1112.1704},
 primaryClass = {astro-ph.EP},
       adsurl = {https://ui.adsabs.harvard.edu/abs/2012MNRAS.421..780L}
}

@ARTICLE{Wu-Chen-2025,
       author = {{Wu}, Yinhao and {Chen}, Yi-Xian},
        title = "{Planet migration in windy discs}",
      journal = {\mnras},
         year = 2025,
        month = jan,
       volume = {536},
       number = {1},
        pages = {L13-L18},
          doi = {10.1093/mnrasl/slae102},
archivePrefix = {arXiv},
       eprint = {2410.19731},
 primaryClass = {astro-ph.EP},
       adsurl = {https://ui.adsabs.harvard.edu/abs/2025MNRAS.536L..13W}
}

@ARTICLE{Kimmig-etal.2020,
       author = {{Kimmig}, C.~N. and {Dullemond}, C.~P. and {Kley}, W.},
        title = "{Effect of wind-driven accretion on planetary migration}",
      journal = {\aap},
         year = 2020,
        month = jan,
       volume = {633},
          eid = {A4},
        pages = {A4},
          doi = {10.1051/0004-6361/201936412},
archivePrefix = {arXiv},
       eprint = {1910.12889},
 primaryClass = {astro-ph.EP},
       adsurl = {https://ui.adsabs.harvard.edu/abs/2020A&A...633A...4K}
}

@ARTICLE{Andre-Papaloizou-2016,
       author = {{Andr{\'e}}, Q. and {Papaloizou}, J.~C.~B.},
        title = "{On the orbital evolution of a pair of giant planets in mean motion resonance}",
      journal = {\mnras},
         year = 2016,
        month = oct,
       volume = {461},
       number = {4},
        pages = {4406-4418},
          doi = {10.1093/mnras/stw1577},
archivePrefix = {arXiv},
       eprint = {1607.00628},
 primaryClass = {astro-ph.EP},
       adsurl = {https://ui.adsabs.harvard.edu/abs/2016MNRAS.461.4406A}
}

@ARTICLE{Crida-etal.2006,
       author = {{Crida}, A. and {Morbidelli}, A. and {Masset}, F.},
        title = "{On the width and shape of gaps in protoplanetary disks}",
      journal = {\icarus},
         year = 2006,
        month = apr,
       volume = {181},
       number = {2},
        pages = {587-604},
          doi = {10.1016/j.icarus.2005.10.007},
archivePrefix = {arXiv},
       eprint = {astro-ph/0511082},
 primaryClass = {astro-ph},
       adsurl = {https://ui.adsabs.harvard.edu/abs/2006Icar..181..587C}
}

@ARTICLE{Kanagawa-etal.2018,
       author = {{Kanagawa}, Kazuhiro D. and {Tanaka}, Hidekazu and {Szuszkiewicz}, Ewa},
        title = "{Radial Migration of Gap-opening Planets in Protoplanetary Disks. I. The Case of a Single Planet}",
      journal = {\apj},
         year = 2018,
        month = jul,
       volume = {861},
       number = {2},
          eid = {140},
        pages = {140},
          doi = {10.3847/1538-4357/aac8d9},
archivePrefix = {arXiv},
       eprint = {1805.11101},
 primaryClass = {astro-ph.EP},
       adsurl = {https://ui.adsabs.harvard.edu/abs/2018ApJ...861..140K}
}

@ARTICLE{Kanagawa-etal.2020,
       author = {{Kanagawa}, Kazuhiro D. and {Szuszkiewicz}, Ewa},
        title = "{Radial Migration of Gap-opening Planets in Protoplanetary Disks. II. The Case of a Planet Pair}",
      journal = {\apj},
         year = 2020,
        month = may,
       volume = {894},
       number = {1},
          eid = {59},
        pages = {59},
          doi = {10.3847/1538-4357/ab862f},
archivePrefix = {arXiv},
       eprint = {2004.01063},
 primaryClass = {astro-ph.EP},
       adsurl = {https://ui.adsabs.harvard.edu/abs/2020ApJ...894...59K}
}

@ARTICLE{Morbidelli-etal.2014,
       author = {{Morbidelli}, A. and {Szul{\'a}gyi}, J. and {Crida}, A. and {Lega}, E. and {Bitsch}, B. and {Tanigawa}, T. and {Kanagawa}, K.},
        title = "{Meridional circulation of gas into gaps opened by giant planets in three-dimensional low-viscosity disks}",
      journal = {\icarus},
         year = 2014,
        month = apr,
       volume = {232},
        pages = {266-270},
          doi = {10.1016/j.icarus.2014.01.010},
archivePrefix = {arXiv},
       eprint = {1401.2925},
 primaryClass = {astro-ph.EP},
       adsurl = {https://ui.adsabs.harvard.edu/abs/2014Icar..232..266M}
}

@ARTICLE{Ziampras-etal.2020,
       author = {{Ziampras}, Alexandros and {Kley}, Wilhelm and {Dullemond}, Cornelis P.},
        title = "{Importance of radiative effects in gap opening by planets in protoplanetary disks}",
      journal = {\aap},
         year = 2020,
        month = may,
       volume = {637},
          eid = {A50},
        pages = {A50},
          doi = {10.1051/0004-6361/201937048},
archivePrefix = {arXiv},
       eprint = {2003.02298},
 primaryClass = {astro-ph.EP},
       adsurl = {https://ui.adsabs.harvard.edu/abs/2020A&A...637A..50Z}
}

@ARTICLE{Sandor-Kley-2006,
       author = {{S{\'a}ndor}, Zs. and {Kley}, W.},
        title = "{On the evolution of the resonant planetary system HD 128311}",
      journal = {\aap},
         year = 2006,
        month = jun,
       volume = {451},
       number = {3},
        pages = {L31-L34},
          doi = {10.1051/0004-6361:20065196},
archivePrefix = {arXiv},
       eprint = {astro-ph/0603664},
 primaryClass = {astro-ph},
       adsurl = {https://ui.adsabs.harvard.edu/abs/2006A&A...451L..31S}
}

@ARTICLE{Rein-Papaloizou-2009,
       author = {{Rein}, H. and {Papaloizou}, J.~C.~B.},
        title = "{On the evolution of mean motion resonances through stochastic forcing: fast and slow libration modes and the origin of HD 128311}",
      journal = {\aap},
         year = 2009,
        month = apr,
       volume = {497},
       number = {2},
        pages = {595-609},
          doi = {10.1051/0004-6361/200811330},
archivePrefix = {arXiv},
       eprint = {0811.1813},
 primaryClass = {astro-ph},
       adsurl = {https://ui.adsabs.harvard.edu/abs/2009A&A...497..595R}
}

@ARTICLE{Steffen-etal.2012,
       author = {{Steffen}, Jason H. and {Ford}, Eric B. and {Rowe}, Jason F. and {Fabrycky}, Daniel C. and {Holman}, Matthew J. and {Welsh}, William F. and {Batalha}, Natalie M. and {Borucki}, William J. and {Bryson}, Steve and {Caldwell}, Douglas A. and {Ciardi}, David R. and {Jenkins}, Jon M. and {Kjeldsen}, Hans and {Koch}, David G. and {Pr{\v{s}}a}, Andrej and {Sanderfer}, Dwight T. and {Seader}, Shawn and {Twicken}, Joseph D.},
        title = "{Transit Timing Observations from Kepler. VI. Potentially Interesting Candidate Systems from Fourier-based Statistical Tests}",
      journal = {\apj},
         year = 2012,
        month = sep,
       volume = {756},
       number = {2},
          eid = {186},
        pages = {186},
          doi = {10.1088/0004-637X/756/2/186},
archivePrefix = {arXiv},
       eprint = {1201.1873},
 primaryClass = {astro-ph.EP},
       adsurl = {https://ui.adsabs.harvard.edu/abs/2012ApJ...756..186S}
}

@ARTICLE{Hadden-Lithwick-2014,
       author = {{Hadden}, Sam and {Lithwick}, Yoram},
        title = "{Densities and Eccentricities of 139 Kepler Planets from Transit Time Variations}",
      journal = {\apj},
         year = 2014,
        month = may,
       volume = {787},
       number = {1},
          eid = {80},
        pages = {80},
          doi = {10.1088/0004-637X/787/1/80},
archivePrefix = {arXiv},
       eprint = {1310.7942},
 primaryClass = {astro-ph.EP},
       adsurl = {https://ui.adsabs.harvard.edu/abs/2014ApJ...787...80H}
}

@ARTICLE{Almenara2025,
       author = {{Almenara}, J.~M. and {Leleu}, A. and {Guillot}, T. and others},
      journal = {\aap},
         year = 2025,
       volume = {aa57359-25},
       adsurl = {https://ui.adsabs.harvard.edu/abs/2025A&A...aa57359-25}
}

@ARTICLE{Bean2009,
       author = {{Bean}, J.~L. and {Seifahrt}, A.},
      journal = {\aap},
         year = 2009,
       volume = {496},
        pages = {249},
       adsurl = {https://ui.adsabs.harvard.edu/abs/2009A&A...496..249B}
}

@ARTICLE{Bryan2019,
       author = {{Bryan}, M.~L. and {Knutson}, H.~A. and {Lee}, E.~J. and others},
      journal = {\aj},
         year = 2019,
       volume = {157},
        pages = {52},
       adsurl = {https://ui.adsabs.harvard.edu/abs/2019AJ....157...52B}
}

@ARTICLE{ang2008,
       author = {{D'Angelo}, G. and {Lubow}, S.~H.},
      journal = {\apj},
         year = 2008,
       volume = {685},
        pages = {560},
       adsurl = {https://ui.adsabs.harvard.edu/abs/2008ApJ...685..560D}
}

@ARTICLE{David2019,
       author = {{David}, T.~J. and others},
      journal = {\apjl},
         year = 2019,
       volume = {885},
        pages = {L12},
       adsurl = {https://ui.adsabs.harvard.edu/abs/2019ApJ...885L..12D}
}

@ARTICLE{Fischer2016,
       author = {{Fischer}, D.~A. and {Anglada-Escud{\'e}}, G. and {Arriagada}, P. and others},
      journal = {\pasp},
         year = 2016,
       volume = {128},
        pages = {066001},
       adsurl = {https://ui.adsabs.harvard.edu/abs/2016PASP..128f6001F}
}

@ARTICLE{Hirano2020,
       author = {{Hirano}, T. and others},
      journal = {\mnras},
         year = 2020,
       volume = {493},
        pages = {5205},
       adsurl = {https://ui.adsabs.harvard.edu/abs/2020MNRAS.493.5205H}
}

@ARTICLE{Kley1999,
       author = {{Kley}, W.},
      journal = {\mnras},
         year = 1999,
       volume = {303},
        pages = {696},
       adsurl = {https://ui.adsabs.harvard.edu/abs/1999MNRAS.303..696K}
}

@ARTICLE{Kley2004,
       author = {{Kley}, W. and {Peitz}, J. and {Bryden}, G.},
      journal = {\aap},
         year = 2004,
       volume = {414},
        pages = {735},
       adsurl = {https://ui.adsabs.harvard.edu/abs/2004A&A...414..735K}
}

@ARTICLE{Laughlin2011,
       author = {{Laughlin}, G. and {Crismani}, M. and {Adams}, F.~C.},
      journal = {\apjl},
         year = 2011,
       volume = {729},
        pages = {L7},
       adsurl = {https://ui.adsabs.harvard.edu/abs/2011ApJ...729L...7L}
}

@ARTICLE{LeePeale2002,
       author = {{Lee}, M.~H. and {Peale}, S.~J.},
      journal = {\apj},
         year = 2002,
       volume = {567},
        pages = {596},
       adsurl = {https://ui.adsabs.harvard.edu/abs/2002ApJ...567..596L}
}

@ARTICLE{Lee2006,
       author = {{Lee}, M.~H. and {Peale}, S.~J.},
      journal = {\apj},
         year = 2006,
       volume = {641},
        pages = {1178},
       adsurl = {https://ui.adsabs.harvard.edu/abs/2006ApJ...641.1178L}
}

@ARTICLE{Li2024,
       author = {{Li}, Ya-Ping and {Chen}, Yi-Xian and {Lin}, Douglas N.~C.},
        title = "{Concurrent Accretion and Migration of Giant Planets in Their Natal Disks with Consistent Accretion Torque}",
      journal = {\apj},
         year = 2024,
        month = aug,
       volume = {971},
       number = {2},
        pages = {130},
          doi = {10.3847/1538-4357/ad5a06},
archivePrefix = {arXiv},
       eprint = {2406.12716},
 primaryClass = {astro-ph.EP},
       adsurl = {https://ui.adsabs.harvard.edu/abs/2024ApJ...971..130L}
}

@ARTICLE{MacDonald2020,
       author = {{MacDonald}, M.~G. and others},
      journal = {\aj},
         year = 2020,
       volume = {159},
        pages = {116},
       adsurl = {https://ui.adsabs.harvard.edu/abs/2020AJ....159..116M}
}

@ARTICLE{Mayor1995,
       author = {{Mayor}, M. and {Queloz}, D.},
        title = "{A Jupiter-mass companion to a solar-type star}",
      journal = {\nat},
         year = 1995,
       volume = {378},
        pages = {355},
          doi = {10.1038/378355a0},
       adsurl = {https://ui.adsabs.harvard.edu/abs/1995Natur.378..355M}
}

@ARTICLE{Militzer2013,
       author = {{Militzer}, B. and {Hubbard}, W.~B.},
      journal = {\apj},
         year = 2013,
       volume = {774},
        pages = {148},
       adsurl = {https://ui.adsabs.harvard.edu/abs/2013ApJ...774..148M}
}

@ARTICLE{Millholland2017,
       author = {{Millholland}, S. and {Wang}, S.~H. and {Laughlin}, G.},
      journal = {\apjl},
         year = 2017,
       volume = {849},
        pages = {L33},
       adsurl = {https://ui.adsabs.harvard.edu/abs/2017ApJ...849L..33M}
}

@ARTICLE{Nesvorny2016,
       author = {{Nesvorn{\'y}}, D. and others},
      journal = {\apj},
         year = 2016,
       volume = {827},
        pages = {66},
       adsurl = {https://ui.adsabs.harvard.edu/abs/2016ApJ...827...66N}
}

@ARTICLE{Otegi2022,
       author = {{Otegi}, J.~F. and {Dawson}, R.~I. and {Venturini}, J.},
      journal = {\aap},
         year = 2022,
       volume = {658},
          eid = {A107},
        pages = {A107},
       adsurl = {https://ui.adsabs.harvard.edu/abs/2022A&A...658A.107O}
}

@ARTICLE{Pan2026,
       author = {{Pan}, Junpeng and {Li}, Ya-Ping and {Chen}, Yi-Xian and {Ida}, Shigeru and {Lin}, Douglas N.~C.},
        title = "{Concurrent Accretion and Migration of Giant Planets in their Natal Disks with Consistent Accretion Torque. II. Parameter Survey and Condition for Outward Migration}",
      journal = {\apj},
         year = 2026,
        month = feb,
       volume = {997},
       number = {2},
        pages = {161},
          doi = {10.3847/1538-4357/ae261a},
archivePrefix = {arXiv},
       eprint = {2511.21323},
 primaryClass = {astro-ph.EP},
       adsurl = {https://ui.adsabs.harvard.edu/abs/2026ApJ...997..161P}
}

@ARTICLE{PapaloizouTerquem2006,
       author = {{Papaloizou}, J.~C.~B. and {Terquem}, C.},
      journal = {Reports on Progress in Physics},
         year = 2006,
       volume = {69},
        pages = {119},
       adsurl = {https://ui.adsabs.harvard.edu/abs/2006RPPh...69..119P}
}

@BOOK{Perryman2018,
       author = {{Perryman}, Michael},
        title = "{The Exoplanet Handbook}",
     publisher = {Cambridge University Press},
          year = 2018,
       edition = {2nd},
         address = {Cambridge}
}

@ARTICLE{Pollack1996,
       author = {{Pollack}, J.~B. and {Hubickyj}, O. and {Bodenheimer}, P. and others},
      journal = {\icarus},
         year = 1996,
       volume = {124},
        pages = {62},
       adsurl = {https://ui.adsabs.harvard.edu/abs/1996Icar..124...62P}
}

@ARTICLE{Schlecker2021,
       author = {{Schlecker}, M. and {Mordasini}, C. and {Emsenhuber}, A. and others},
      journal = {\aap},
         year = 2021,
       volume = {656},
          eid = {A71},
        pages = {A71},
       adsurl = {https://ui.adsabs.harvard.edu/abs/2021A&A...656A..71S}
}

@ARTICLE{Tan2013,
       author = {{Tan}, X. and others},
      journal = {arXiv e-prints},
         year = 2013,
archivePrefix = {arXiv},
       eprint = {1306.0687},
       adsurl = {https://ui.adsabs.harvard.edu/abs/2013arXiv1306.0687T}
}

@ARTICLE{Weiss2014,
       author = {{Weiss}, L.~M. and {Marcy}, G.~W.},
      journal = {\apjl},
         year = 2014,
       volume = {783},
        pages = {L6},
       adsurl = {https://ui.adsabs.harvard.edu/abs/2014ApJ...783L...6W}
}

@ARTICLE{Weiss2018,
       author = {{Weiss}, L.~M. and {Laughlin}, G. and {Teske}, J. and others},
      journal = {\aj},
         year = 2018,
       volume = {155},
        pages = {48},
       adsurl = {https://ui.adsabs.harvard.edu/abs/2018AJ....155...48W}
}

@ARTICLE{Wu2025,
       author = {{Wu}, Hening and {Li}, Ya-Ping},
        title = "{Effects of Thermodynamics on the Concurrent Accretion and Migration of Gas Giants in Protoplanetary Disks}",
      journal = {Universe},
         year = 2025,
        month = jan,
       volume = {11},
       number = {1},
        pages = {1},
          doi = {10.3390/universe11010001},
archivePrefix = {arXiv},
       eprint = {2501.15874},
 primaryClass = {astro-ph.EP},
       adsurl = {https://ui.adsabs.harvard.edu/abs/2025Univ...11....1W}
}

@ARTICLE{Xie2012,
       author = {{Xie}, J.~W.},
      journal = {\raa},
         year = 2012,
       volume = {12},
        pages = {1273},
       adsurl = {https://ui.adsabs.harvard.edu/abs/2012RAA....12.1273X}
}

@ARTICLE{Zhang2008,
       author = {{Zhang}, H. and {Yuan}, C. and {Lin}, D.~N.~C. and others},
      journal = {\apj},
         year = 2008,
       volume = {676},
        pages = {639},
       adsurl = {https://ui.adsabs.harvard.edu/abs/2008ApJ...676..639Z}
}

@ARTICLE{Zhu2018,
       author = {{Zhu}, W. and {Petrovich}, C. and {Wu}, Y. and others},
      journal = {\apj},
         year = 2018,
       volume = {860},
        pages = {101},
       adsurl = {https://ui.adsabs.harvard.edu/abs/2018ApJ...860..101P}
}

@ARTICLE{Zhu2024,
       author = {{Zhu}, W.},
      journal = {\raa},
         year = 2024,
       volume = {24},
        pages = {045013},
       adsurl = {https://ui.adsabs.harvard.edu/abs/2024RAA....24a45013Z}
}

@ARTICLE{Liu2026,
       author = {{Liu}, Beibei and {Baruteau}, Cl{\'e}ment and {Zhu}, Zhaohuan and {Li}, Ya-Ping and {Paardekooper}, Sijme-Jan and {Ida}, Shigeru},
        title = "{Hydrodynamical simulations of resonant breaking in multiplanet systems via rebound migration during disk dispersal}",
      journal = {\aap},
         year = 2026,
        month = jul,
       volume = {711},
          eid = {A160},
        pages = {A160},
          doi = {10.1051/0004-6361/202660554},
archivePrefix = {arXiv},
       eprint = {2606.11452},
 primaryClass = {astro-ph.EP},
       adsurl = {https://ui.adsabs.harvard.edu/abs/2026A&A...711A.160L}
}

@ARTICLE{Chen2020,
       author = {{Chen}, Yi-Xian and {Zhang}, Xiaojia and {Li}, Ya-Ping and {Li}, Hui and {Lin}, Douglas N.~C.},
        title = "{Retention of Long-period Gas Giant Planets: Type II Migration Revisited}",
      journal = {\apj},
         year = 2020,
        month = sep,
       volume = {900},
       number = {1},
          eid = {44},
        pages = {44},
          doi = {10.3847/1538-4357/abaab6},
archivePrefix = {arXiv},
       eprint = {2007.14905},
 primaryClass = {astro-ph.EP},
       adsurl = {https://ui.adsabs.harvard.edu/abs/2020ApJ...900...44C}
}

@ARTICLE{Laune2024,
       author = {{Laune}, JT and {Li}, Rixin and {Lai}, Dong},
        title = "{Migration of Accreting Planets and Black Holes in Disks}",
      journal = {\apj},
         year = 2024,
        month = nov,
       volume = {975},
       number = {2},
          eid = {296},
        pages = {296},
          doi = {10.3847/1538-4357/ad7117},
archivePrefix = {arXiv},
       eprint = {2405.00296},
 primaryClass = {astro-ph.EP},
       adsurl = {https://ui.adsabs.harvard.edu/abs/2024ApJ...975..296L}
}

@ARTICLE{Li2023,
       author = {{Li}, Ya-Ping and {Chen}, Yi-Xian and {Lin}, Douglas N.~C.},
        title = "{3D global simulations of accretion onto gap-opening planets: implications for circumplanetary disc structures and accretion rates}",
      journal = {\mnras},
         year = 2023,
        month = dec,
       volume = {526},
       number = {4},
        pages = {5346-5364},
          doi = {10.1093/mnras/stad3049},
archivePrefix = {arXiv},
       eprint = {2310.02822},
 primaryClass = {astro-ph.EP},
       adsurl = {https://ui.adsabs.harvard.edu/abs/2023MNRAS.526.5346L}
}

@ARTICLE{Machida2010,
       author = {{Machida}, Masahiro N. and {Kokubo}, Eiichiro and {Inutsuka}, Shu-Ichiro and {Matsumoto}, Tomoaki},
        title = "{Gas accretion onto a protoplanet and formation of a gas giant planet}",
      journal = {\mnras},
         year = 2010,
        month = jun,
       volume = {405},
       number = {2},
        pages = {1227-1243},
          doi = {10.1111/j.1365-2966.2010.16527.x},
archivePrefix = {arXiv},
       eprint = {1002.3002},
 primaryClass = {astro-ph.SR},
       adsurl = {https://ui.adsabs.harvard.edu/abs/2010MNRAS.405.1227M}
}

@ARTICLE{Choksi2023,
       author = {{Choksi}, Nick and {Chiang}, Eugene and {Fung}, Jeffrey and {Zhu}, Zhaohuan},
        title = "{The maximum accretion rate of a protoplanet: how fast can runaway be?}",
      journal = {\mnras},
         year = 2023,
        month = oct,
       volume = {525},
       number = {2},
        pages = {2806-2819},
          doi = {10.1093/mnras/stad2269},
archivePrefix = {arXiv},
       eprint = {2305.01684},
 primaryClass = {astro-ph.EP},
       adsurl = {https://ui.adsabs.harvard.edu/abs/2023MNRAS.525.2806C}
}

@ARTICLE{Li2021,
       author = {{Li}, Ya-Ping and {Chen}, Yi-Xian and {Lin}, Douglas N.~C. and {Zhang}, Xiaojia},
        title = "{Accretion of Gas Giants Constrained by the Tidal Barrier}",
      journal = {\apj},
         year = 2021,
        month = jan,
       volume = {906},
       number = {1},
          eid = {52},
        pages = {52},
          doi = {10.3847/1538-4357/abc88310.1002/essoar.10504346.1},
       adsurl = {https://ui.adsabs.harvard.edu/abs/2021ApJ...906...52L}
}

@ARTICLE{Tanigawa2016,
       author = {{Tanigawa}, Takayuki and {Tanaka}, Hidekazu},
        title = "{Final Masses of Giant Planets. II. Jupiter Formation in a Gas-depleted Disk}",
      journal = {\apj},
         year = 2016,
        month = may,
       volume = {823},
       number = {1},
          eid = {48},
        pages = {48},
          doi = {10.3847/0004-637X/823/1/48},
archivePrefix = {arXiv},
       eprint = {1510.06848},
 primaryClass = {astro-ph.EP},
       adsurl = {https://ui.adsabs.harvard.edu/abs/2016ApJ...823...48T}
}

@ARTICLE{Bodenheimer2013,
       author = {{Bodenheimer}, Peter and {D'Angelo}, Gennaro and {Lissauer}, Jack J. and {Fortney}, Jonathan J. and {Saumon}, Didier},
        title = "{Deuterium Burning in Massive Giant Planets and Low-mass Brown Dwarfs Formed by Core-nucleated Accretion}",
      journal = {\apj},
         year = 2013,
        month = jun,
       volume = {770},
       number = {2},
          eid = {120},
        pages = {120},
          doi = {10.1088/0004-637X/770/2/120},
archivePrefix = {arXiv},
       eprint = {1305.0980},
 primaryClass = {astro-ph.EP},
       adsurl = {https://ui.adsabs.harvard.edu/abs/2013ApJ...770..120B}
}

@ARTICLE{DAngelo2003,
       author = {{D'Angelo}, Gennaro and {Kley}, Willy and {Henning}, Thomas},
        title = "{Orbital Migration and Mass Accretion of Protoplanets in Three-dimensional Global Computations with Nested Grids}",
      journal = {\apj},
         year = 2003,
        month = mar,
       volume = {586},
       number = {1},
        pages = {540-561},
          doi = {10.1086/367555},
archivePrefix = {arXiv},
       eprint = {astro-ph/0308055},
 primaryClass = {astro-ph},
       adsurl = {https://ui.adsabs.harvard.edu/abs/2003ApJ...586..540D}
}

@ARTICLE{Ida2026,
       author = {{Ida}, Shigeru and {Li}, Ya-Ping and {Pan}, Jun-Peng and {Chen}, Yi-Xian and {Lin}, Douglas N.~C.},
        title = "{Outward Migration of a Gas Accreting Planet: A Semianalytical Formula}",
      journal = {\apj},
         year = 2026,
        month = feb,
       volume = {997},
       number = {2},
          eid = {160},
        pages = {160},
          doi = {10.3847/1538-4357/ae2612},
archivePrefix = {arXiv},
       eprint = {2512.00304},
 primaryClass = {astro-ph.EP},
       adsurl = {https://ui.adsabs.harvard.edu/abs/2026ApJ...997..160I}
}

@ARTICLE{Muller-etal.2018,
       author = {{M{\"u}ller}, A. and {Keppler}, M. and {Henning}, Th. and {Samland}, M. and {Chauvin}, G. and {Beust}, H. and {Maire}, A.-L. and {Molaverdikhani}, K. and {van Boekel}, R. and {Benisty}, M. and {Boccaletti}, A. and {Bonnefoy}, M. and {Cantalloube}, F. and {Charnay}, B. and {Baudino}, J.-L. and {Gennaro}, M. and {Long}, Z.~C. and {Cheetham}, A. and {Desidera}, S. and {Feldt}, M. and {Fusco}, T. and {Girard}, J. and {Gratton}, R. and {Hagelberg}, J. and {Janson}, M. and {Lagrange}, A.-M. and {Langlois}, M. and {Lazzoni}, C. and {Ligi}, R. and {M{\'e}nard}, F. and {Mesa}, D. and {Meyer}, M. and {Molli{\`e}re}, P. and {Mordasini}, C. and {Moulin}, T. and {Pavlov}, A. and {Pawellek}, N. and {Quanz}, S.~P. and {Ramos}, J. and {Rouan}, D. and {Sissa}, E. and {Stadler}, E. and {Vigan}, A. and {Wahhaj}, Z. and {Weber}, L. and {Zurlo}, A.},
        title = "{Orbital and atmospheric characterization of the planet within the gap of the PDS 70 transition disk}",
      journal = {\aap},
         year = 2018,
        month = sep,
       volume = {617},
          eid = {L2},
        pages = {L2},
          doi = {10.1051/0004-6361/201833584},
archivePrefix = {arXiv},
       eprint = {1806.11567},
 primaryClass = {astro-ph.EP},
       adsurl = {https://ui.adsabs.harvard.edu/abs/2018A&A...617L...2M}
}

@ARTICLE{Keppler-etal.2018,
       author = {{Keppler}, M. and {Benisty}, M. and {M{\"u}ller}, A. and {Henning}, Th. and {van Boekel}, R. and {Cantalloube}, F. and {Ginski}, C. and {van Holstein}, R.~G. and {Maire}, A.-L. and {Pohl}, A. and {Samland}, M. and {Avenhaus}, H. and {Baudino}, J.-L. and {Boccaletti}, A. and {de Boer}, J. and {Bonnefoy}, M. and {Chauvin}, G. and {Desidera}, S. and {Langlois}, M. and {Lazzoni}, C. and {Marleau}, G.-D. and {Mordasini}, C. and {Pawellek}, N. and {Stolker}, T. and {Vigan}, A. and {Zurlo}, A. and {Birnstiel}, T. and {Brandner}, W. and {Feldt}, M. and {Flock}, M. and {Girard}, J. and {Gratton}, R. and {Hagelberg}, J. and {Isella}, A. and {Janson}, M. and {Juhasz}, A. and {Kemmer}, J. and {Kral}, Q. and {Lagrange}, A.-M. and {Launhardt}, R. and {Matter}, A. and {M{\'e}nard}, F. and {Milli}, J. and {Molli{\`e}re}, P. and {Olofsson}, J. and {P{\'e}rez}, L. and {Pinilla}, P. and {Pinte}, C. and {Quanz}, S.~P. and {Schmidt}, T. and {Udry}, S. and {Wahhaj}, Z. and {Williams}, J.~P. and {Buenzli}, E. and {Cudel}, M. and {Dominik}, C. and {Galicher}, R. and {Kasper}, M. and {Lannier}, J. and {Mesa}, D. and {Mouillet}, D. and {Peretti}, S. and {Perrot}, C. and {Salter}, G. and {Sissa}, E. and {Wildi}, F. and {Abe}, L. and {Antichi}, J. and {Augereau}, J.-C. and {Baruffolo}, A. and {Baudoz}, P. and {Bazzon}, A. and {Beuzit}, J.-L. and {Blanchard}, P. and {Brems}, S.~S. and {Buey}, T. and {De Caprio}, V. and {Carbillet}, M. and {Carle}, M. and {Cascone}, E. and {Cheetham}, A. and {Claudi}, R. and {Costille}, A. and {Delboulb{\'e}}, A. and {Dohlen}, K. and {Fantinel}, D. and {Feautrier}, P. and {Fusco}, T. and {Giro}, E. and {Gluck}, L. and {Gry}, C. and {Hubin}, N. and {Hugot}, E. and {Jaquet}, M. and {Le Mignant}, D. and {Llored}, M. and {Madec}, F. and {Magnard}, Y. and {Martinez}, P. and {Maurel}, D. and {Meyer}, M. and {M{\"o}ller-Nilsson}, O. and {Moulin}, T. and {Mugnier}, L. and {Orign{\'e}}, A. and {Pavlov}, A. and {Perret}, D. and {Petit}, C. and {Pragt}, J. and {Puget}, P. and {Rabou}, P. and {Ramos}, J. and {Rigal}, F. and {Rochat}, S. and {Roelfsema}, R. and {Rousset}, G. and {Roux}, A. and {Salasnich}, B. and {Sauvage}, J.-F. and {Sevin}, A. and {Soenke}, C. and {Stadler}, E. and {Suarez}, M. and {Turatto}, M. and {Weber}, L.},
        title = "{Discovery of a planetary-mass companion within the gap of the transition disk around PDS 70}",
      journal = {\aap},
         year = 2018,
        month = sep,
       volume = {617},
          eid = {A44},
        pages = {A44},
          doi = {10.1051/0004-6361/201832957},
archivePrefix = {arXiv},
       eprint = {1806.11568},
 primaryClass = {astro-ph.EP},
       adsurl = {https://ui.adsabs.harvard.edu/abs/2018A&A...617A..44K}
}

@ARTICLE{Benisty-etal.2021,
       author = {{Benisty}, Myriam and {Bae}, Jaehan and {Facchini}, Stefano and {Keppler}, Miriam and {Teague}, Richard and {Isella}, Andrea and {Kurtovic}, Nicolas T. and {P{\'e}rez}, Laura M. and {Sierra}, Anibal and {Andrews}, Sean M. and {Carpenter}, John and {Czekala}, Ian and {Dominik}, Carsten and {Henning}, Thomas and {Menard}, Francois and {Pinilla}, Paola and {Zurlo}, Alice},
        title = "{A Circumplanetary Disk around PDS70c}",
      journal = {\apjl},
         year = 2021,
        month = jul,
       volume = {916},
       number = {1},
          eid = {L2},
        pages = {L2},
          doi = {10.3847/2041-8213/ac0f83},
archivePrefix = {arXiv},
       eprint = {2108.07123},
 primaryClass = {astro-ph.EP},
       adsurl = {https://ui.adsabs.harvard.edu/abs/2021ApJ...916L...2B}
}

@ARTICLE{Wang-etal.2021,
       author = {{Wang}, J.~J. and {Vigan}, A. and {Lacour}, S. and {Nowak}, M. and {Stolker}, T. and {De Rosa}, R.~J. and {Ginzburg}, S. and {Gao}, P. and {Abuter}, R. and {Amorim}, A. and {Asensio-Torres}, R. and {Baub{\"o}ck}, M. and {Benisty}, M. and {Berger}, J.~P. and {Beust}, H. and {Beuzit}, J.-L. and {Blunt}, S. and {Boccaletti}, A. and {Bohn}, A. and {Bonnefoy}, M. and {Bonnet}, H. and {Brandner}, W. and {Cantalloube}, F. and {Caselli}, P. and {Charnay}, B. and {Chauvin}, G. and {Choquet}, E. and {Christiaens}, V. and {Cl{\'e}net}, Y. and {Coud{\'e} Du Foresto}, V. and {Cridland}, A. and {de Zeeuw}, P.~T. and {Dembet}, R. and {Dexter}, J. and {Drescher}, A. and {Duvert}, G. and {Eckart}, A. and {Eisenhauer}, F. and {Facchini}, S. and {Gao}, F. and {Garcia}, P. and {Garcia Lopez}, R. and {Gardner}, T. and {Gendron}, E. and {Genzel}, R. and {Gillessen}, S. and {Girard}, J. and {Haubois}, X. and {Hei{\ss}el}, G. and {Henning}, T. and {Hinkley}, S. and {Hippler}, S. and {Horrobin}, M. and {Houll{\'e}}, M. and {Hubert}, Z. and {Jim{\'e}nez-Rosales}, A. and {Jocou}, L. and {Kammerer}, J. and {Keppler}, M. and {Kervella}, P. and {Meyer}, M. and {Kreidberg}, L. and {Lagrange}, A.-M. and {Lapeyr{\`e}re}, V. and {Le Bouquin}, J.-B. and {L{\'e}na}, P. and {Lutz}, D. and {Maire}, A.-L. and {M{\'e}nard}, F. and {M{\'e}rand}, A. and {Molli{\`e}re}, P. and {Monnier}, J.~D. and {Mouillet}, D. and {M{\"u}ller}, A. and {Nasedkin}, E. and {Ott}, T. and {Otten}, G.~P.~P.~L. and {Paladini}, C. and {Paumard}, T. and {Perraut}, K. and {Perrin}, G. and {Pfuhl}, O. and {Pueyo}, L. and {Rameau}, J. and {Rodet}, L. and {Rodr{\'\i}guez-Coira}, G. and {Rousset}, G. and {Scheithauer}, S. and {Shangguan}, J. and {Shimizu}, T. and {Stadler}, J. and {Straub}, O. and {Straubmeier}, C. and {Sturm}, E. and {Tacconi}, L.~J. and {van Dishoeck}, E.~F. and {Vincent}, F. and {von Fellenberg}, S.~D. and {Ward-Duong}, K. and {Widmann}, F. and {Wieprecht}, E. and {Wiezorrek}, E. and {Woillez}, J. and {Gravity Collaboration}},
        title = "{Constraining the Nature of the PDS 70 Protoplanets with VLTI/GRAVITY}",
      journal = {\aj},
         year = 2021,
        month = mar,
       volume = {161},
       number = {3},
          eid = {148},
        pages = {148},
          doi = {10.3847/1538-3881/abdb2d},
archivePrefix = {arXiv},
       eprint = {2101.04187},
 primaryClass = {astro-ph.EP},
       adsurl = {https://ui.adsabs.harvard.edu/abs/2021AJ....161..148W}
}

@ARTICLE{Bae-etal.2019,
       author = {{Bae}, Jaehan and {Zhu}, Zhaohuan and {Baruteau}, Cl{\'e}ment and {Benisty}, Myriam and {Dullemond}, Cornelis P. and {Facchini}, Stefano and {Isella}, Andrea and {Keppler}, Miriam and {P{\'e}rez}, Laura M. and {Teague}, Richard},
        title = "{An Ideal Testbed for Planet-Disk Interaction: Two Giant Protoplanets in Resonance Shaping the PDS 70 Protoplanetary Disk}",
      journal = {\apjl},
         year = 2019,
        month = oct,
       volume = {884},
       number = {2},
          eid = {L41},
        pages = {L41},
          doi = {10.3847/2041-8213/ab46b0},
archivePrefix = {arXiv},
       eprint = {1909.09476},
 primaryClass = {astro-ph.EP},
       adsurl = {https://ui.adsabs.harvard.edu/abs/2019ApJ...884L..41B}
}

@ARTICLE{Wu-etal.2023,
       author = {{Wu}, Yinhao and {Baruteau}, Cl{\'e}ment and {Nayakshin}, Sergei},
        title = "{Using planet migration and dust drift to weigh protoplanetary discs}",
      journal = {\mnras},
         year = 2023,
        month = aug,
       volume = {523},
       number = {4},
        pages = {4869-4882},
          doi = {10.1093/mnras/stad1791},
archivePrefix = {arXiv},
       eprint = {2305.01493},
 primaryClass = {astro-ph.EP},
       adsurl = {https://ui.adsabs.harvard.edu/abs/2023MNRAS.523.4869W}
}
\bibliographystyle{aasjournal}
\end{CJK*}
\end{document}